\documentclass[universe,review,accept,moreauthors,pdftex,10pt,a4paper]{mdpi}
\usepackage{amsmath}
\usepackage{hyperref}
\usepackage{CJK}
\usepackage{color}

\usepackage{natbib}
\usepackage{hyperref}
\usepackage{cleveref}
\newcommand{\cQMsq}{c^2_{\rm QM}}

\newcommand{\De}{\Delta}
\newcommand{\ep}{\varepsilon}

\newcommand{\Msun}{\,\text{M}_{\odot}}

\firstpage{1}
\articlenumber{x}
\doinum{10.3390/10.3390/universe7060182}
\pubvolume{xx}
\pubyear{2021}
\copyrightyear{2021}
\history{Received: May 10, 2021, accepted May 31, 2021, published June 4, 2021}
\Title{Progress in Constraining Nuclear Symmetry Energy Using Neutron Star Observables since GW170817}

\Author{{Bao-An Li}$^{1}$\orcidA{}*,
	Bao-Jun Cai$^{2}$, Wen-Jie Xie$^{3}$, Nai-Bo Zhang$^{4}$}
\AuthorNames{Bao-An Li, Bao-Jun Cai, Wen-Jie Xie and Nai-Bo Zhang}

\address{%
	$^{1}$ \quad Department of Physics and Astronomy, Texas A$\&$M
University-Commerce, Commerce, TX 75429-3011, USA\\

	$^{2}$ \quad Quantum Machine Learning Laboratory, Shadow Creator Inc., Shanghai 201208, China\\
	$^{3}$ \quad Department of Physics, Yuncheng University, Yuncheng 044000, China\\
	$^{4}$ \quad School of Physics, Southeast University, Nanjing 211189, China}

\corres{\bf Correspondence: Bao-An.Li@TAMUC.edu}

\abstract{The density dependence of nuclear symmetry energy is among the most uncertain parts of the Equation of State (EOS) of dense neutron-rich nuclear matter.  It is currently poorly known especially at suprasaturation densities partially because of our poor knowledge about isovector nuclear interactions at short distances. Because of its broad impacts on many interesting issues, to pin down the density dependence of nuclear symmetry energy has been a longstanding and shared goal of both astrophysics and nuclear physics. New observational data of neutron stars including their masses, radii, and tidal deformations since GW170817 have helped improve our knowledge about nuclear symmetry energy especially at high densities. Based on various model analyses of these new data by many people in the nuclear astrophysics community, while our brief review might be incomplete and biased unintentionally, we have learned particularly:
(1) The slope parameter $L$ of nuclear symmetry energy at saturation density $\rho_0$ of nuclear matter from 24 new analyses of neutron star observables is about $L\approx 57.7\pm 19$ MeV at 68\% confidence level consistent with its fiducial value from surveys of over 50 earlier analyses of both terrestrial and astrophysical data within error bars, (2) The curvature $K_{\rm{sym}}$ of nuclear symmetry energy at $\rho_0$ from 16 new analyses of neutron star observables is about $K_{\rm{sym}}\approx -107\pm 88$ MeV at 68\% confidence level in very good agreement with the systematics of earlier analyses, (3) The magnitude of nuclear symmetry energy at $2\rho_0$, i.e.  $E_{\rm{sym}}(2\rho_0)\approx 51\pm 13$ MeV at 68\% confidence level, has been extracted from 9 new analyses of neutron star observables consistent with results from earlier analyses of heavy-ion reactions and the latest predictions of the  state-of-the-art nuclear many-body theories, (4) while the available data from canonical neutron stars do not provide tight constraints on nuclear symmetry energy at densities above about $2\rho_0$, the lower radius boundary $R_{2.01}=12.2$ km from NICER's very recent observation of PSR J0740+6620 of mass $2.08\pm 0.07$ $M_{\odot}$ and radius $R=12.2-16.3$ km at 68\% confidence level sets a tight lower limit for nuclear symmetry energy at densities above $2\rho_0$, (5) Bayesian inferences of nuclear symmetry energy using models encapsulating a first-order hadron-quark phase transition from observables of canonical neutron stars indicate that the phase transition shifts appreciably both the $L$ and $K_{\rm{sym}}$ to higher values but with larger uncertainties compared to analyses assuming no such phase transition, (6) The high-density behavior of nuclear symmetry energy affects significantly the minimum frequency necessary to rotationally support GW190814's secondary component of mass (2.50-2.67) $M_{\odot}$ as the fastest and most massive pulsar discovered so far. Overall, thanks to the hard work of many people in the astrophysics and nuclear physics community, new data of neutron star observations since the discovery of GW170817 have enriched significantly our knowledge about the symmetry energy of dense neutron-rich nuclear matter.
	}


\usepackage{lipsum}% this package is included to get dummy paragraphs for sample purpose.
\usepackage{graphicx,pstricks,epsfig,rotating,amsmath,amssymb}

\newcommand{\bea}{\begin{eqnarray}}
\newcommand{\eea}{\end{eqnarray}}
\renewcommand{\d}{\mathrm{d}}

\firstpage{1} 
\pubvolume{7}
\issuenum{1}
\articlenumber{182}
\pubyear{2021}
\copyrightyear{2021}
\externaleditor{Academic Editor: Ignazio Bombaci; Rosa Poggiani}
\begin{document}
\tableofcontents
\newpage
\section{Introduction}
Understanding the nature and constrain the Equation of State (EOS) of dense neutron-rich nuclear matter is a major science goal~\cite{NAP2011,NAP2012,LRP2015,NuPECC} shared by many other astrophysical observations (see, e.g.,~the analyses and reviews in~\citep{LP01,Lat16,Watts16,Oz16a,Oertel17,Baiotti,BALI19,Web,Mark,Capano20,David,Kat20,Annala2018,Kievsky2018,Landry2020,Dietrich2020,ALi20}) and terrestrial nuclear experiments (see, e.g.,~\cite{Dan02,ditoro,Steiner05,LCK08,Chen11,Trau12,Tsang12,Tesym,Bal16,Li1,PPNP-Li,burg}). However, realizing this goal is very challenging for many scientific and technical reasons. The~EOS
of Asymmetric Nuclear Matter (ANM) at nucleon density $\rho=\rho_n+\rho_p$ and isospin asymmetry $\delta\equiv (\rho_n-\rho_p)/\rho$
can be expressed in terms of the nuclear pressure $P(\rho, \delta)=\rho^2\frac{d\epsilon(\rho,\delta)/\rho}{d\rho}$ as a function of nucleon energy density $\epsilon(\rho,\delta)=\rho E(\rho,\delta)+\rho M$
where $E(\rho ,\delta )$ and $M$ are the average nucleon energy and mass, respectively.~$E(\rho ,\delta )$ can be well approximated by~\cite{Bom91}:
\begin{equation}
E(\rho ,\delta )=E_0(\rho)+E_{\rm{sym}}(\rho )\cdot \delta ^{2} +\mathcal{O}(\delta^4)
\end{equation}
\textls[-33]{according to essentially all existing nuclear many-body theories. The~first term \mbox{$E_0(\rho)\equiv E_{\rm{SNM}}(\rho)$}} is the nucleon energy in Symmetric Nuclear Matter (SNM), having equal numbers of neutrons and protons, while the symmetry energy $E_{\rm{sym}}(\rho )$ quantifies the energy needed to make nuclear matter more neutron rich. Since the pressure in ANM can be written as:
\begin{equation}\label{pressure}
 P(\rho, \delta)=\rho^2\frac{dE(\rho,\delta)}{d\rho}=\rho^2[\frac{dE_{\rm{SNM}}(\rho)}{d\rho}+\frac{dE_{\rm{sym}}(\rho)}{d\rho}\delta^2]=P_{\rm{SNM}}(\rho)+\rho^2\frac{dE_{\rm{sym}}(\rho)}{d\rho}\delta^2
\end{equation}
where $P_{\rm{SNM}}(\rho)\equiv \rho^2\frac{dE_{\rm{SNM}}(\rho)}{d\rho}$ is the pressure in SNM, the~pressure in Pure Neutron Matter (PNM) $P_{\rm{PNM}}(\rho)\equiv P(\rho,\delta=1)$ can be written as:
\begin{equation}
 P_{\rm{PNM}}(\rho)=P_{\rm{SNM}}(\rho)+\rho^2\frac{dE_{\rm{sym}}(\rho)}{d\rho}.
\end{equation}
%MDPI: Is the noindent necessary? NO
By inverting the above equation, one can express the symmetry energy $E_{\rm{sym}}(\rho )$ as~\cite{Li19-PLB}:
\begin{equation}\label{esym}
 E_{\textrm{sym}}(\rho)=E_{\textrm{sym}}(\rho_i)+\int_{\rho_i}^{\rho}\frac{P_{\rm{PNM}}(\rho_v)-P_{\rm{SNM}}(\rho_v)}{\rho_v^2}d\rho_v
\end{equation}
where $\rho_i$ is a reference density. While many observables in terrestrial nuclear experiments have provided us much useful information about the pressure $P_{\rm{SNM}}$ in SNM over a broad density range~\cite{Dan02}, neutron star observables are messengers of nuclear pressure in neutron-rich matter towards PNM. Combining all knowledge from analyzing the observables of both neutron stars and their mergers, as well as terrestrial nuclear experiments holds the promise of ultimately determining the density dependence of nuclear symmetry energy $E_{\rm{sym}}(\rho )$.

The nuclear symmetry energy $E_{\rm{sym}}(\rho )$ has broad ramifications for many properties of both isolated Neutron Stars (NSs) and gravitational waves from their mergers. For~example, the~density profile $\delta(\rho)$ of isospin asymmetry in NSs at $\beta$-equilibrium is uniquely determined by the $E_{\rm{sym}}(\rho )$ through the $\beta$-equilibrium and charge neutrality conditions. Once the $\delta(\rho)$ is determined by the $E_{\rm{sym}}(\rho )$, both the pressure $p(\rho, \delta)$ and energy density $\epsilon(\rho,\delta)$ reduce to functions of nucleon density only. Their relation $p(\epsilon)$ can then be used in solving the TOV equation to study NS structures and/or simulators of their mergers. It is well known that the critical nucleon density $\rho_c$ above which the fast cooling of protoneutron stars by neutrino emissions through the direct URCA process can occur, the~crust-core transition density, and~pressure in NSs all depend sensitively on $E_{\rm{sym}}(\rho )$. Moreover, the~frequencies and damping times of various oscillations, especially the f- and g-modes of the core, as well as the torsional mode of the crust, quadrupole deformations of isolated NSs, and~the tidal deformability of NSs in inspiralling binaries all depend significantly on $E_{\rm{sym}}(\rho )$. 

{Furthermore, the~binding energy and spacetime curvature of NSs also depend on the symmetry energy~\cite{Newton09,He15}. In~fact, there is a degeneracy between the EOS of super-dense neutron-rich matter and the strong-field gravity in determining the maximum mass of NSs. As was pointed out already~\cite{Psa08}, NSs are among the densest objects with the strongest gravity in the Universe, making them ideal places to test Einstein's General Relativity (GR) in regions where it has not been fully tested yet. One reason for testing GR in the extremely strong-gravity region is that there is no fundamental reason to choose Einstein's GR over other alternatives, and~it is known that GR may break down at the limit of very strong gravitational fields. It was pointed out that the EOS--gravity degeneracy is tied to the fundamental Strong Equivalence Principle and can only be broken by using at least two independent observables~\citep{Yun10}. It has also been shown that the variation of the mass--radius relation, especially the maximum NS mass an EOS can support, due to the variation of gravity theory (with respect to GR predictions), is much larger than that due to the uncertainty of the poorly known EOS of dense neutron-rich matter~\cite{Ded03}. In~fact, the~EOS--gravity degeneracy has promoted some people to ask the question: Can the maximum mass of neutron stars rule out any EOS of dense stellar matter before the strong-field gravity is well understood? One possible answer is no~\cite{WenLiChen}. One possible way out of the EOS--gravity degeneracy is to simultaneously determine both the strong-field gravity and the EOS of superdense matter using massive NSs. In~this sense, the~recent discovery of GW190814 with its secondary component in the NS--black hole mass gap is particularly interesting.} Furthermore, the~minimum frequency for GW190814's secondary component of mass (2.50--2.67)~$M_{\odot}$ to be a supermassive and superfast pulsar that is r-mode stable against run-away gravitational radiations depends critically on the high-density $E_{\rm{sym}}(\rho )$~\cite{Zhang20b,Zhou21}. Thus, a~precise determination of $E_{\rm{sym}}(\rho )$ has broad impacts in many areas of astrophysics, cosmology, and~nuclear physics. In~turn, many astrophysical observables from various compact objects and/or processes may carry useful information about nuclear symmetry energy. Indeed, various data of several observables, e.g.,~radii, masses, and~tidal deformations of canonical neutron stars with masses around 1.4~$M_\odot$, have been analyzed within different model frameworks to extract the symmetry energy and the EOS of SNM. Despite the vast diversity of approaches used, as~we shall show, rather consistent results on the characteristics of symmetry energy around $\rho_0$ have been extracted, albeit within still relatively large error~bars.

The symmetry energy $E_{\rm{sym}}(\rho )$ at suprasaturation densities and the possible hadron--quark phase transition are among the most uncertain parts of the EOS of dense neutron-rich matter~\cite{Web,Mark,David,Tesym}.
Moreover, the~appearance of new particles, such as $\Delta(1232)$ resonances and various hyperons, also depends strongly on the high-density behavior of nuclear symmetry energy~\cite{Drago14,Cai15,Zhu16,Sahoo18,Li18,Ribes19,Li19,Raduta20,Raduta21,Thapa21,Sen21,Jiang12,Pro19,Vidana18,Choi20,Fortin21}. Since the nuclear symmetry energy will lose its physical meaning above the hadron--quark transition density, it is imperative to determine both the high-density $E_{\rm{sym}}(\rho )$ and the properties of the hadron--quark phase transition simultaneously by using combined data from astrophysical observations and nuclear experiments. Since many existing studies assume that neutron stars are made of nucleons and leptons only and no hadron--quark phase transition is considered, it is thus interesting to compare the $E_{\rm{sym}}(\rho )$ extracted from neutron stars with and without considering the hadron--quark phase~transition.

Thanks to the great efforts of many people in both astrophysics and nuclear physics over the last two decades, significant progress has been made in constraining the symmetry energy $E_{\rm{sym}}(\rho )$, especially around and below the saturation density of nuclear matter $\rho_0$. Compared to terrestrial experiments, neutron stars are particularly useful for probing the symmetry energy at suprasaturation densities.
While still much less is known about $E_{\rm{sym}}(\rho )$ at suprasaturation densities, astrophysical data since the detection of GW170817 have indeed stimulated many interesting new studies on nuclear symmetry energy in a broad density range. Results of these new studies together with those from earlier analyses of mostly terrestrial experiments have certainly improved our knowledge about the density dependence of nuclear symmetry energy. One of the main purposes of this brief review was to give an update on the systematics of the slope $L$ and curvature $K_{\rm{sym}}$ of $E_{\rm{sym}}(\rho )$ at $\rho_0$, as well as its magnitude at twice the saturation density, i.e.,~$E_{\rm{sym}}(2\rho_0 )$, based on recently published results from analyzing astrophysical data by the community since GW170817 was discovered. Moreover, as~examples of how the nuclear symmetry energy affects astrophysical observables, we also briefly reviewed the following issues mostly based on our own recent~work:
\begin{enumerate}
\item What have we learned about the symmetry energy from the tidal deformation of canonical neutron stars from GW170817, the mass of PSR J0740+6620, and NICER's simultaneous observation of mass and radius of PSR J0030+0451 and PSR J0740+6620?
\item How do the symmetry energy parameters extracted from recent observations of neutron stars compare with what we knew before the discovery of GW170817 that were mostly from terrestrial experiments?
\item What can we learn about the high-density symmetry energy from future, more precise radius measurement of massive neutron stars?
\item What are the effects of hadron--quark phase transition on extracting the symmetry energy from neutron star observables? How does the symmetry energy affect the fraction and size of quark cores in hybrid stars?
\item What are the effects of symmetry energy on the nature of GW190814's second component of mass (2.50--2.67) $M_{\odot}$? %MDPI: Please check if it should be italic or normal, and unify the whole tex. Normalized to italic
\item If all the characteristics of nuclear symmetry energy at saturation density $\rho_0$, e.g.,~its slope $L$, curvature $K_{\rm{sym}}$, and~skewness $J_{\rm{sym}}$, are precisely determined by the astrophysical observations and/or terrestrial experiments, how do we use them to predict the symmetry energy at suprasaturation densities, such as $2-3 \rho_0$? Nuclear symmetry energy $E_{\rm{sym}}(\rho)$ is normally expanded or simply parameterized as a function of $\chi=(\rho-\rho_0)/3\rho_0$ in the form of $E_{\rm{sym}}(\rho)\approx S+L\chi+2^{-1}K_{\rm{sym}}\chi^2+6^{-1}J_{\rm{sym}}\chi^3+\cdots$. However, such kinds of expansions/parameterizations do not converge at suprasaturation densities where $\chi$ is not small enough, hindering an accurate determination of high-density $E_{\rm{sym}}(\rho)$.
Is there a better way that one can predict accurately the symmetry energy at high densities using $L$, $K_{\rm{sym}}$, and~$J_{\rm{sym}}$?

\end{enumerate}

Answers to these questions are expected to be useful for further understanding the nature and EOS of density neutron-rich nuclear matter using high-precision data especially from more advanced X-ray observatories and gravitational wave detectors, as well as terrestrial experiments, especially those at high-energy rare isotope beam laboratories being built around the~world. 

\section{What Have We Learned about the Symmetry Energy from the Tidal Deformation of Canonical Neutron Stars from GW170817 as Well as NICER's Simultaneous Observations of Mass and Radius of PSR J0030+0451 and PSR J0740+6620? How Do They Compare with What We Knew before GW170817?}\label{ASTR}
The detection of GW170817 marked the beginning of a new era in astronomy in particular and physics in general. The~nuclear astrophysics community has studied many aspects of isolated neutron stars and their mergers.
Various interesting physics have been extracted from analyzing the GW170817 event as documented in the literature. Here, we briefly summarize what we have learned about the symmetry energy
from studying the tidal deformation of canonical neutron stars observed by LIGO/VIRGO together with the recent mass and radius measurements from other observatories. While our brief review here is probably not complete, we tried to be unbiased and hope to provide a useful picture for the community on this particular~issue.

\subsection{Updated Systematics of Symmetry Energy Parameters at $\rho_0$ after Incorporating the Results of Recent Analyses of Neutron Star Observables since~GW170817}
Before reaching the hadron--quark phase transition density, the~SNM EOS $E_{0}(\rho)$ and symmetry energy $E_{\rm{sym}}(\rho)$ can be parameterized as:
\begin{eqnarray}
E_{0}(\rho)&=&E_0(\rho_0)+\frac{K_0}{2}(\frac{\rho-\rho_0}{3\rho_0})^2+\frac{J_0}{6}(\frac{\rho-\rho_0}{3\rho_0})^3,\\
E_{\rm{sym}}(\rho)&=&E_{\rm{sym}}(\rho_0)+L(\frac{\rho-\rho_0}{3\rho_0})+\frac{K_{\rm{sym}}}{2}(\frac{\rho-\rho_0}{3\rho_0})^2+\frac{J_{\rm{sym}}}{6}(\frac{\rho-\rho_0}{3\rho_0})^3
\end{eqnarray}
in solving the inverse-structure problems of neutron stars, such as Bayesian analysis. In~this case, the~posterior Probability Distribution Functions (PDFs) of the EOS parameters and their correlations are inferred directly from observational data. Conventionally, the $E_{0}(\rho)$ and $E_{\rm{sym}}(\rho)$ predicted by nuclear many-body theories can be Taylor expanded around $\rho_0$ in the same form as above. In~this case, the~coefficients
in the Taylor expansions are obtained from the predicted energy density functionals. The~symmetry energy is then characterized by its magnitude $S\equiv E_{\rm{sym}}(\rho_0)$,
slope $L=[3\rho\d E_{\rm{sym}}/\d\rho]_{\rho_0}$, curvature $K_{\rm{sym}}=[9\rho^2\d^2 E_{\rm{sym}}/\d\rho^2]_{\rho_0}$, and~skewness $J_{\rm{sym}}=[27\rho^3\d^3 E_{\rm{sym}}/\d\rho^3]_{\rho_0}$ at saturation density. Similarly, the~SNM EOS is characterized by the binding energy $E_0(\rho_0)$, incompressibility $K_0$, and~skewness $J_0$ at $\rho_0$. It is worth noting that while the parameterizations can have as many high-order terms as one wishes as long as one can find enough data to fix them, the~Taylor expansions have the general issue of convergence at high densities, as discussed in detail recently in~\cite{CL21}. The~above parameterizations are often used in metamodeling of neutron star EOSs in Bayesian inferences. On~the other hand, even when the exact theoretical energy density functionals are used in either the inversion processes or directly comparing forward model predictions with observational data (mostly in the latter case), often, only the first few coefficients, e.g.,~$E_{\rm{sym}}(\rho_0)$, $L$, and $K_{\rm{sym}}$, are reported and compared with those from other analyses. This practice is normally appropriate as the parameterizations themselves can be considered as energy density functionals. However, one should be cautious that simply plugging the extracted coefficients from comparing model predictions with data into the above parameterizations may not reproduce the underlying symmetry energy at high densities due to the convergence issue~\cite{CL21}.

Before the discovery of GW170817, much effort was devoted to extracting $E_{\rm{sym}}(\rho_0)$ and $L$, as well as their correlation using mostly data from terrestrial experiments.
For example, a~survey of 29 analyses performed before 2013 found the fiducial of $E_{\rm{sym}}(\rho_0)=31.6\pm 2.7$ MeV and $L=58.9\pm 16$ MeV\,\cite{LiBA13}. These values were changed to $E_{\rm{sym}}(\rho_0)=31.7 \pm 3.2$~MeV and $L = 58.7 \pm 28.1$~MeV in the 2016 survey of 53 analyses~\cite{Oertel17}. Interestingly, as~more diverse approaches were used in analyzing some of the same data, the~uncertainty ranges increased somewhat, while the mean values remained the same. During~the same time, microscopic and ab~initio theories made more accurate predictions. For~instance, using a novel Bayesian approach to quantify the truncation errors in chiral Effective Field Theory (EFT) predictions for pure neutron matter and many-body perturbation theory with consistent nucleon--nucleon and three-nucleon interactions up to fourth-order in the EFT expansion,~$E_{\rm{sym}}(\rho_0)$ and $L$ were recently predicted to be $E_{\rm{sym}}(\rho_0)=31.7 \pm 1.1$~MeV and $L = 59.8 \pm 4.1$~MeV~\cite{Ohio20}, respectively. The~mean values of these predictions were in excellent agreement with the fiducial values found~earlier.

The discovery of GW170817 has triggered many analyses of neutron star observables, mostly the tidal deformability and radii, using various models. Most of these new analyses have actually used the existing fiducial values of $E_{\rm{sym}}(\rho_0)$ and L in setting their prior ranges, albeit using various prior PDFs. To~the best of our knowledge, the~resulting posterior means and 68\% confidence intervals of $E_{\rm{sym}}(\rho_0)$ are not much different from the fiducial values given above as most of the neutron star observable studies are not really sensitive to the variation of $E_{\rm{sym}}(\rho_0)$. However, new values of $L$ and $K_{\rm{sym}}$ have been extracted in many~studies.

\begin{figure}[H]
%\widefigure
\includegraphics[width=1\linewidth]{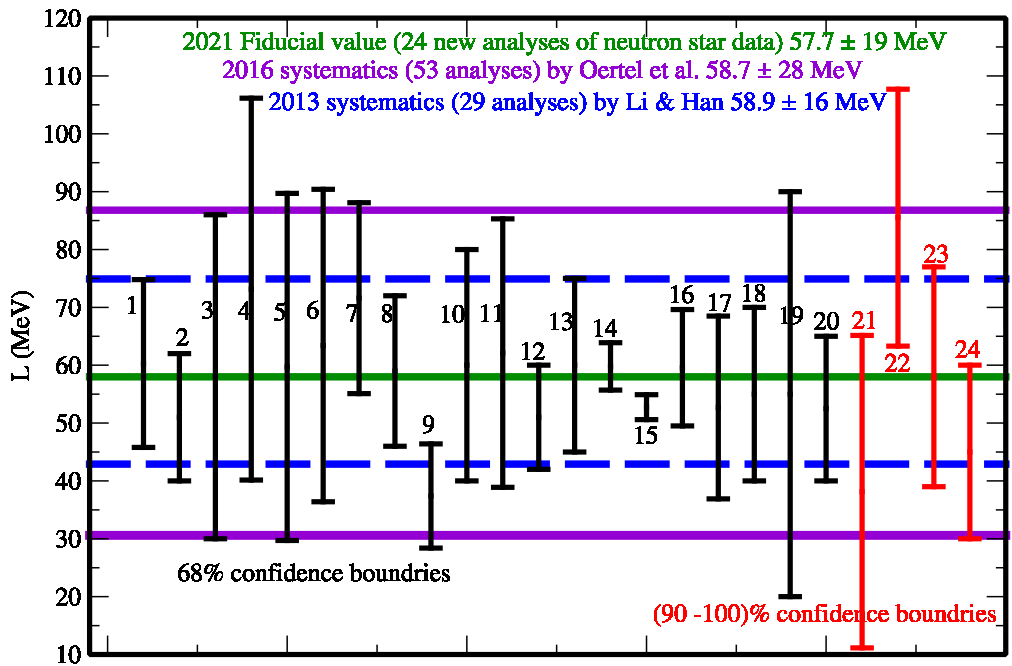}
%MDPI: Please replace with a sharper image.: For the first 3 figures, I tried to sharpen the images. However, the software XMGRACE I used to make these plots use a dash for a minus sign. 
\caption{The slope parameter $L$ of nuclear symmetry from the (1) 2013 survey of 29 analyses of terrestrial experiments and astrophysical observations (between the dashed blue lines)~\cite{LiBA13}, (2) 2016 surveys of 53 analyses (between the violet lines)~\cite{Oertel17}, and (3) 24 new analyses of neutron star observables since GW170817 (see the detailed list in the text). The~green line is the average value of $L$ from these 24 new~analyses.}
\label{Lnew}
\end{figure}

Shown in Figure~\ref{Lnew} are $L$ values from 24 new analyses of neutron star observables in comparison with those from the 2013 and 2016 surveys. The~results collected here are likely incomplete, and purely theoretical predictions are not included unless they are explicitly compared with new data of neutron star observables since GW170817. The~results displayed randomly from the left are from (1) Mondal~et~al.~\cite{Mondal17}, (2) and (3) \mbox{Malik~et~al.~\cite{Malik18}},
(4), (5), and~(6) Biswas~et~al.~\cite{Biswas20}, (7) Tsang~et~al.~\cite{Tsang20}, (8) Xie and Li~\cite{Xie20}, (9) Baillot d'Etivaux~et~al.~\cite{France1}, (10) Malik~et~al.~\cite{Malik20}, (11) Zhao and Lattimer~\cite{Zhao18}, (12) Lim and Holt~\cite{Lim19}, (13) Margueron~et~al.~\cite{Margueron18}, (14) Drischler~et~al.~\cite{Ohio20}, (15) Tan~et~al.~\cite{Tan20}, (16) Zhang~et~al.~\cite{XZhang20}, (17) Chamel~et~al.~\cite{Chamel19}, (18) Huang~et~al.~\cite{Huang20}, (19) Tews~et~al.~\cite{Tews19}, (20) Gil~et~al.~\cite{Gil21}, (21) Raithel and {\"O}zel~\cite{Raithel19}, (22) Yue~et~al.~\cite{Yue21}, (23) Essick~et~al.~\cite{Essick21}, and~(24) Li~et~al.~\cite{ALi20}.

Interestingly, they are rather consistent with each other within the error bars. While it is beyond our ability to discuss the detailed differences among these analyses using vastly different models, obviously, reducing the error bars is one of the future tasks. A~major contribution to the error bars is the correlation between $L$ and the even less constrained $K_{\rm{sym}}$ parameter~\cite{LiM}. The~average value of $L$ from these 24 new analyses of neutron star observables was about $L\approx 57.7\pm 19$ MeV at a 68\% confidence level, as indicated by the green line. Not so surprisingly, this was quite consistent with the fiducial values from both the 2013 and 2016 surveys within the error bars. If~one considers all results equally reliable within the published error bars, the~fiducial value of $L$ remains about 58 MeV from, now in total, about 80 independent analyses of various observables of NSs and nuclear experiments, while there is not much reduction of its error bar. In~fact, we noticed that the estimation of the error bars for the fiducial value of $L$ was not scientifically very rigorous as the nature, approach, and~data used in the vastly different analyses were not completely transparent and~compatible.

While the focus of this review was on the progress in constraining nuclear symmetry energy using NS observables since GW170817, it is worth noting that continuous efforts have been made in nuclear physics to constrain the symmetry energy. Indeed, there are many interesting results in the literature. In~particular, we noticed that the fiducial value of $L\approx 57.7\pm 19$ MeV at a 68\% confidence level from the 24 new analyses of NS observables since GW170817, as well as the $L=58.9\pm 16$ MeV from the 2013 survey of 29 analyses and the
$L = 58.7 \pm 28.1$~MeV from the 2016 survey of 53 analyses were all consistent with the latest report of $L$ between 42 and 117 MeV from studying the pion spectrum ratio in heavy-ion collision in an experiment performed at RIKEN~\cite{Estee21}, but in serious tension with the implications of both the PREX-I and PREX-II experiments measuring the size of neutron skin in $^{208}$Pb using parity violating electron scatterings. The~PREX-II experiment found very recently a neutron skin in $^{208}$Pb of size $R_n-R_p = 0.283 \pm 0.071$ fm. This implies a value of \mbox{$L = 106 \pm 37$ MeV~\cite{Prex2}} based on the Relativistic Mean Field (RMF) model calculations~\cite{Chuck21}. 

On the other hand, to~be consistent with the results from other nuclear experiments including the sizes of neutron skins in several Sn isotopes, neutron star observables, as well as the state-of-the-art chiral EFT prediction, the~neutron skin in $^{208}$Pb was predicted to be 0.17--0.18 fm based on Bayesian analyses using mocked data before the PREX-II result was announced~\cite{Xu20}. Using a similar approach with essentially the same data sets of NS observables and neutron skin in Sn isotopes, a~more recent analysis~\cite{Essick21} \mbox{found
$L=58 \pm 19$ MeV} and a neutron skin of $0.19^{+0.03}_{-0.04}$ fm for $^{208}$Pb, in good agreement with that found in~\cite{Xu20} and the systematics discussed above. These interesting agreements and disagreements require further studies by the community. While finishing up this review, we learned about the very recent work of Biswas, who just performed a comprehensive Bayesian
analysis using the latest NS observational data available (GW170817, NICER, and the revised mass measurement of PSR J0740+6620) and the PREX-II result~\cite{Biswas21}. Before~adding the PREX-II results, the~bound on $L$ was $61^{+17}_{-16}$ MeV at a $1 \sigma$ confidence level, consistent with the systematics discussed above. After~including the PREX-II data, $L$ becomes $69^{+16}_{-16}$ MeV. This is not much different from the result obtained from using only the astrophysics data, which dominate the whole data set used and have relatively smaller errors (4$\sim$7\%) for the NS radii compared to the uncertainty ($\sim$25\%) for the neutron skin for $^{208}$Pb from PREX-II. Moreover, the~inferred posterior value of neutron skin $R_{\rm skin}^{208} = 0.20_{-0.04}^{+0.04}$ fm is significantly smaller than the PREX-II measured value, but consistent with that found in~\cite{Xu20,Essick21} using similar approaches. Furthermore, the~curvature of symmetry energy $K_{\rm sym}$ inferred is $K_{\rm sym}=-163^{+123}_{-107}$ MeV, consistent with its fiducial value, which we shall discuss~next.

\begin{figure}[H]
%\widefigure
\includegraphics[width=1\linewidth]{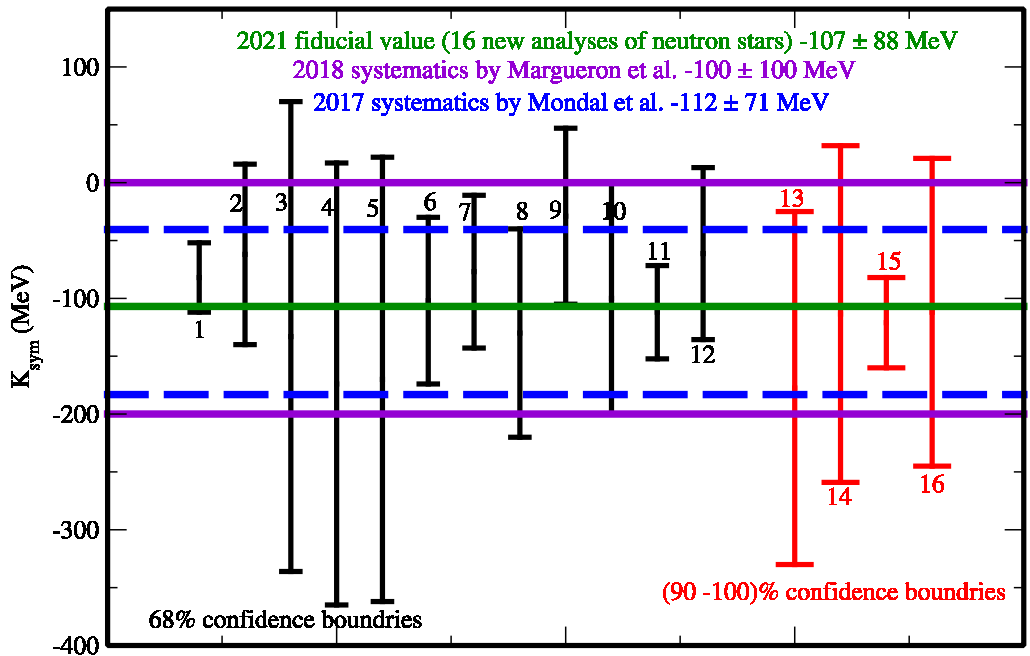}
%MDPI: Please replace with a sharper image.
%MDPI: Please change hyphen to minus sign in the image.
\caption{\textls[-5]{The curvature parameter $K_{\rm{sym}}$ of nuclear symmetry from the (1) 2017 survey of 500 energy density functional predictions constrained by available terrestrial experiments and astrophysical observations (between the dashed blue lines)~\cite{Mondal17}, (2) 2018 surveys based on a metamodel of nuclear EOS under similar constraints (between the violet lines)~\cite{Margueron18}, and (3) 16 new analyses of neutron star observables since GW170817 (see the detailed list in the text). The~green line is the average value of $K_{\rm{sym}}$ from these 16 new analyses.}}
\label{Knew}
\end{figure}

Shown in Figure~\ref{Knew} is a comparison of $K_{\rm{sym}}$ from 16 new analyses of neutron star observables since GW170817 with respect to (1) $K_{\rm{sym}}\approx -112\pm 71$ MeV from the 2017 systematics by Mondal~et~al.~\cite{Mondal17} from analyzing the predictions of over 500 energy density functionals under the constraints of both terrestrial experiments and astrophysical observations available at the time and (2) $K_{\rm{sym}}\approx -100\pm 100$ MeV from the 2018 systematics by Margueron~et~al.~\cite{Margueron18} from
a metamodeling of nuclear EOS under similar constraints, respectively. Most of the 16 analyses are the same ones used in Figure~\ref{Lnew}, but~not all analyses gave simultaneously both the $L$ and $K_{\rm{sym}}$ values by marginalizing one of them. More specifically, (1) and (2) were from Malik~et~al.~\cite{Malik18} and (3), (4), and~(5) from Biswas~et~al.~\cite{Biswas20}. The~rest were from (6) Zimmerman~et~al.~\cite{Zimmerman20},
(7) Tsang~et~al.~\cite{Tsang20}, (8) Xie and Li~\cite{Xie20}, (9) Baillot d'Etivaux~et~al.~\cite{France1}, (10) Malik~et~al.~\cite{Malik20}, (11) Lim and Holt~\cite{Lim19}, (12) Chamel~et~al.~\cite{Chamel19}, (13) Raithel and {\"O}zel~\cite{Raithel19}, (14) Carson~et~al.~\cite{Carson19}, (15) Yue~et~al.~\cite{Yue21}, and~(16) Essick~et~al.~\cite{Essick21}. The~average of these 16 new analyses was about $K_{\rm{sym}}\approx -107\pm 88$ MeV at a 68\% confidence level. Obviously, this was in very good agreement with both the 2017 and 2018 systematics shown in the same~plot. 

It is understood that, within~the errors,~$L$ and $K_{\rm{sym}}$ are generally correlated. Their correlations have significant effects on a number of neutron star properties~\cite{BALI19}. While the individual values of $L$ and $K_{\rm{sym}}$ shown in Figures~\ref{Lnew} and \ref{Knew} are useful, future constraints on their correlations will also be important~\cite{LiM,Tews0,NBZ0,Jeremy}.

\subsection{Symmetry Energy at $2\rho_0$ Extracted from Neutron Star~Observables}
\begin{figure}[H]
%\widefigure
\includegraphics[width=1\linewidth]{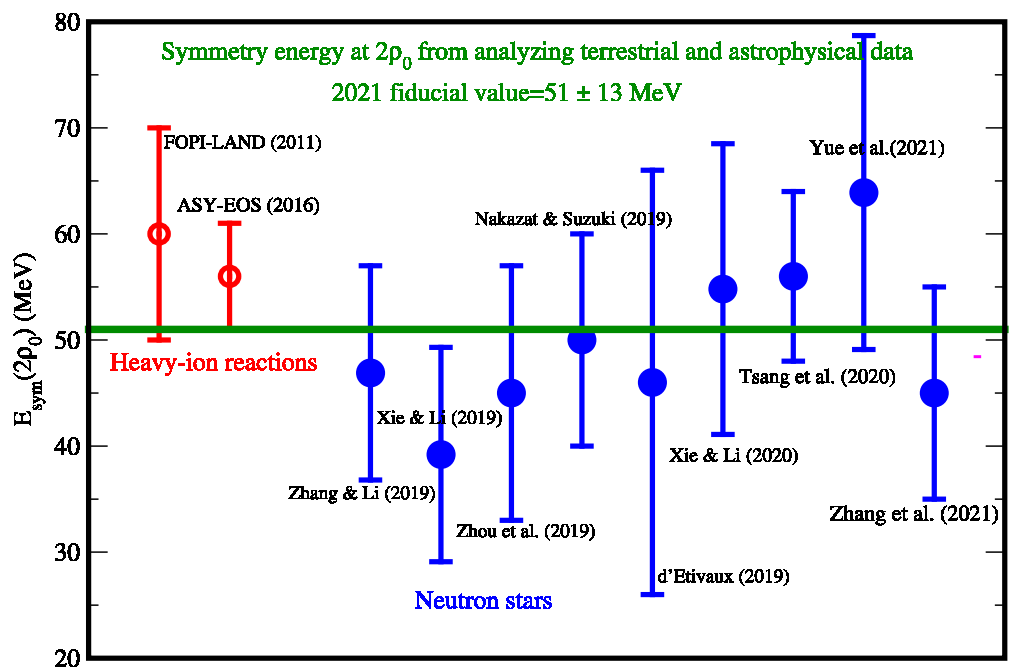}
%MDPI: Please replace with a sharper image.
\caption{Nuclear symmetry $E_{\rm{sym}}(2\rho_0)$ at twice the saturation density of nuclear matter extracted from earlier heavy-ion reaction experiments in terrestrial nuclear laboratories (red) and 9 recent analyses of neutron star observables (see the text for the detailed list). The~green line serving as the latest fiducial value of $E_{\rm{sym}}(2\rho_0)$ is the global average of all points~shown.}\label{Esym2}%in the figures throughout, change & to and
\end{figure}

Besides constraining the characteristics of the symmetry energy at $\rho_0$ discussed above, recent neutron star observations have also been used to constrain explicitly the symmetry energy at suprasaturation densities.
 As an example, shown in Figure~\ref{Esym2} is the magnitude of symmetry energy at twice the saturation density of nuclear matter from nine recent analyses of neutron stars in comparison with the two results from earlier heavy-ion reaction experiments (from the FOPI-LAND~\cite{Rus11} and the ASY-EOS~\cite{Rus16} Collaborations by analyzing the relative flows and yields of light mirror nuclei, as well as neutrons and protons in heavy-ion collisions at beam energies of 400 MeV/nucleon). More specifically, the~nine analyses were from (1) (Zhang and Li 2019) directly inverting observed NS radii, tidal deformability, and~maximum mass in the high-density EOS space~\cite{Zhang18,Zhang19epj,Zhang19apj}, (2) (Xie and Li 2019) a Bayesian inference from the radii of canonical NSs observed by using Chandra X-rays and gravitational waves from GW170817~\cite{Xie19}, (3) (Zhou~et~al. 2019) analyses of NS radii, tidal deformability, and~maximum mass within an extended Skyrme--Hartree--Fock approach (eSHF)~\cite{YZhou19a}, (4) (Nakazato and Suzuki 2019) analyzing cooling timescales of protoneutron stars, as well as the radius and tidal deformability of GW170817~\cite{Nakazato19}, (5) (d'Etivaux~et~al. 2019) a Bayesian inference directly from the X-ray data of seven quiescent low-mass X-ray binaries in globular clusters~\cite{France1}, (6) (Xie and Li 2020) a Bayesian inference from the radii of NSs observed by NICER and LIGO/VIRGO~\cite{Xie20}, (7) (Tsang~et~al. 2020) Bayesian analyses of tidal deformation of canonical NSs from LIGO/VIRGO~\cite{Tsang20}, (8) (Yue~et~al. 2021) eSHF analyses of tidal deformation from GW170817 and radii from NICER~\cite{Yue21}, and (9) (Zhang~et~al. 2021) Skyrme--Hartree--Fock predictions with interaction parameters constrained by heavy-ion reaction experiments, the neutron skin of heavy nuclei, as well as the tidal deformation and radii of neutron stars from LIGO/VIRGO~\cite{XZhang20}.
 
While certainly the model dependence and the error bars are still relatively large, all results from both heavy-ion reactions and neutron stars scatter around an overall mean of $E_{\rm{sym}}(2\rho_0)\approx 51\pm 13$ MeV at a 68\% confidence level, as indicated by the green line. The~symmetry energy around $2\rho_0$ is particularly interesting because the pressure around this density determines
the radii of canonical NSs~\cite{LP01}. Moreover, around \mbox{$(1-2)\rho_0$,} the symmetry energy contribution to the NS pressure competes strongly with that from SNM~\cite{LCK08}. This is also the density range where
the current chiral EFT and other ab~initio theories are still applicable. It is thus interesting to note that the latest many-body perturbation theory calculations with consistent nucleon--nucleon and three-nucleon interactions up to fourth-order in the chiral EFT expansion predicted a value of $E_{\rm{sym}}(2\rho_0) \approx 45 \pm 3$ MeV~\citep{Ohio20}. Similarly, the~latest quantum Monte Carlo calculations using local interactions derived from the chiral EFT up to next-to-next-to-leading order predicted a value of $E_{\rm{sym}}(2\rho_0) \approx 46 \pm 4$ MeV~\citep{Diego}. Obviously, the~fiducial value of $E_{\rm{sym}}(2\rho_0)$ from the analyses listed above was in good agreement with these state-of-the-art nuclear theory predictions, albeit with a significantly larger error~bar.

\subsection{Directly Solving Neutron Star Inverse-Structure Problems in the High-Density EOS Parameter~Space}
How are the constraints on $E_{\rm{sym}}(2\rho_0)$ or generally on the density dependence of nuclear symmetry energy $E_{\rm{sym}}(\rho )$ at suprasaturation densities extracted from neutron star observables?
To answer this question, we provide two examples from our own recent work. The~first example presented in the following is the direct inversion of several NS observables in a three-dimensional high-density EOS parameter space~\cite{Zhang20b}. Another example to be presented in the next subsection is from Bayesian statistical analyses of the same NS~observables.

Since the magnitude and slope of symmetry energy, as well as the binding energy and incompressibility of SNM at $\rho_0$ are relatively well determined, as we discussed above, they can be fixed at their currently known most probable values given above. One can then metamodel NS EOSs in the three-dimensional $J_0-K_{\rm{sym}}-J_{\rm{sym}}$ high-density EOS parameter space. Given an NS observable, one can find all points (EOSs) necessary to reproduce the observable, thus numerically solving the NS inverse-structure problem. Such an approach has been found very successful in several applications~\cite{Zhang18,Zhang19epj,Zhang19jpg,Zhang19apj}.

\begin{figure}[H]
%\widefigure
\includegraphics[width=.87\linewidth]{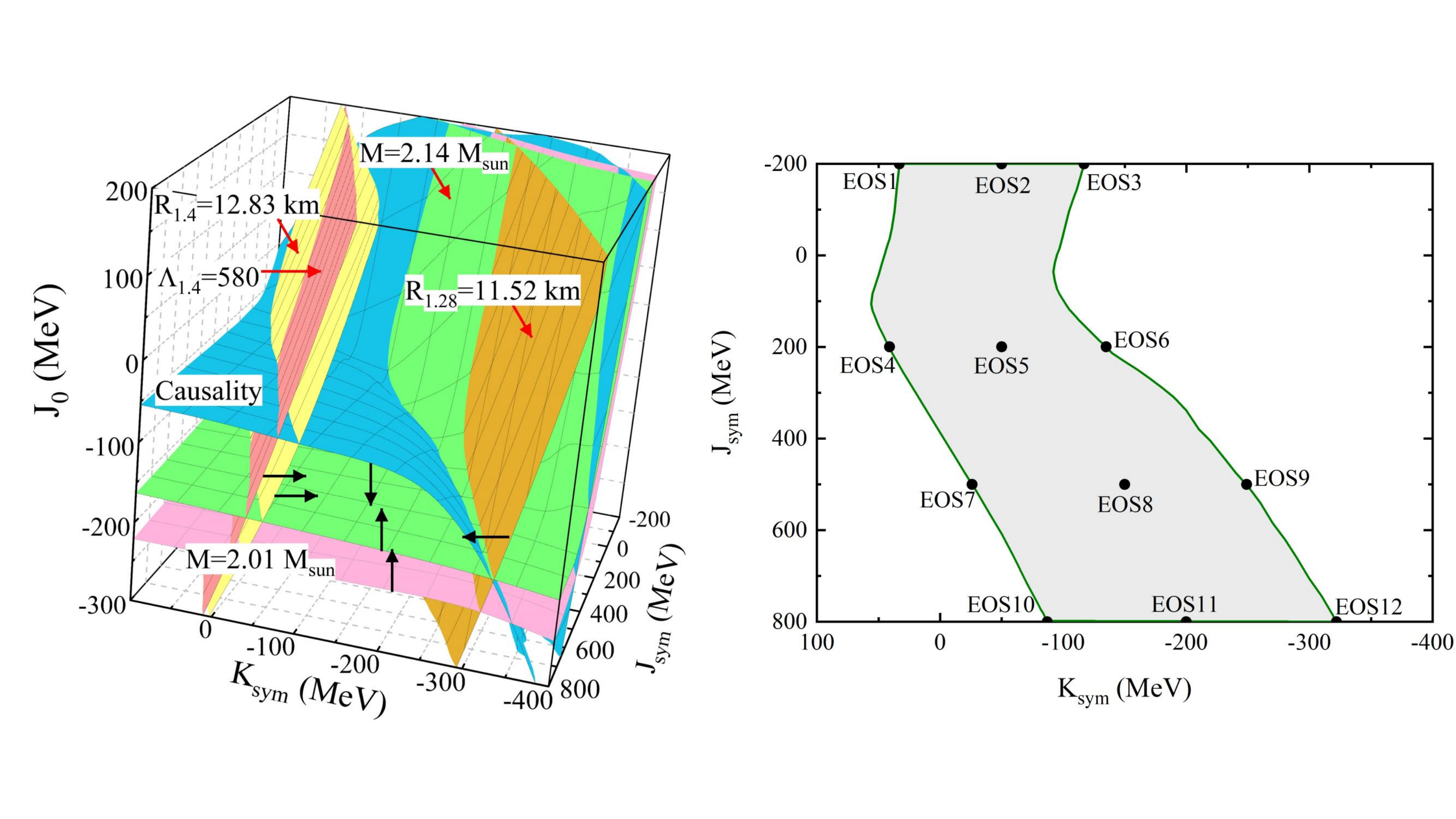}
%MDPI: Please change hyphen to minus sign in the image.
\caption{({Left}) The constant surfaces of M = $2.01$~$M_\odot$ (pink surface), M = $2.14$~$M_\odot$ (green surface), $R_{1.4} = 12.83$ km (yellow surface), $\Lambda_{1.4} = 580$ (red surface), $R_{1.28} = 11.52$ km (orange surface), and~causality condition (blue surface) in the 3-dimensional parameter space of $K_{\rm sym}-J_{\rm sym}-J_0$. The~black arrows show the directions supporting the corresponding constraints, and the red arrows direct to the corresponding surfaces. ({Right}) Boundaries of the symmetry energy in the $K_{\rm{sym}}-J_{\rm{sym}}$ plane.
 Modified from the figures originally published in~\cite{Zhang20,Zhang20b}.}
%MDPI: Please make sure that permission has been obtained and there is no copyright issue. Yes, we own the copy right of all figures used.
\label{Constraints}%in the figures throughout, use () not [] around units and space before ( . Please carefully check all figures.
\end{figure}

As an example, the~left window of Figure~\ref{Constraints} illustrates how neutron star radii and the tidal deformation $\Lambda_{1.4}$ of canonical NSs from Chandra, GW170817~\citep{LIGO18}, and~NICER~\citep{Miller19}, as well as the presently observed NS maximum mass~$M_{\rm{max}}$ and causality condition together can constrain the high-density EOS parameter space spanned in $J_0-K_{\rm{sym}}-J_{\rm{sym}}$ using $E_{\rm{sym}}(\rho_0)=31.6$ MeV and $L=58.9$ MeV. On~each surface, the~indicated NS observable is a constant, while on the causality surface, {the central density of the maximum mass NS is equal to the critical density satisfying the causality condition (the speed of sound is equal to that of light)~\cite{Zhang19epj}}. It is seen that while the skewness $J_0$ of SNM significantly affects the NS maximum mass (notice its change when the~$M_{\rm{max}}$ changes from 2.01 to 2.14~$M_{\odot}$), it has little effects on the radius or the tidal deformation, as indicated by the vertical surfaces of constant radii and tidal deformation. On~the other hand, all of the observables shown and the causality surface depend sensitively on the high-density $E_{\rm{sym}}(\rho )$ parameters $K_{\rm{sym}}$ and $J_{\rm{sym}}$. The~crosslines between the constant surfaces set upper and lower limits for $E_{\rm{sym}}(\rho )$ and $J_0$. For~example, the~crossline between the surface of %please ensure that the original meaning is retained. Yes
radius $R_{1.4}=12.83$ km for a 1.4~$M_{\odot}$ NS (it is the 68\% confidence upper boundary from an earlier analysis of Chandra data~\cite{Lattimer14} and very close to the $\Lambda_{1.4}= 580$ surface from LIGO/VIRGO) and the causality surface sets an upper boundary for $E_{\rm{sym}}(\rho )$, while the crossline between the causality surface and the surface of the previously reported NS maximum mass of 2.14~$M_\odot$ from PSR~J0740+6620~\citep{Mmax} sets approximately the lower limit for $E_{\rm{sym}}(\rho )$ {(note that the mass of PSR~J0740+6620 was reduced from the previously published $M=2.14~M_\odot$
to 2.08 $\pm$ 0.07~$M_\odot$ recently~\cite{Fonseca21,Miller21,Riley21}; for~this review, we still used some of our previous results obtained from using $M=2.14~M_\odot$)}. The~projections of the above two crosslines onto the $J_{\rm{sym}}-K_{\rm{sym}}$ plane are shown in the right window of Figure~\ref{Constraints}. It is seen that these projections stringently limit the range of $K_{\rm{sym}}$ compared to its prior range, but do not narrow down the range of $J_{\rm{sym}}$, thus limited constraints on the symmetry energy at densities above about $2\rho_0$.

% start a new page without indent 4.6cm
\clearpage
%\end{paracol}
\nointerlineskip
%\begin{paracol}{2}
%\linenumbers
%\switchcolumn

Moreover, comparing the crossline between the causality surface and the surface of the NS maximum mass of 2.14~$M_\odot$ (previously reported mass~\cite{Mmax}) from PSR~J0740+6620 with that between the causality surface and the previously known NS maximum mass of 2.01~$M_\odot$ from PSR J0348+0432, it is seen that the NS maximum mass used clearly affects the lower boundary of symmetry energy at suprasaturation densities. Detailed analyses~\cite{Zhang19apj,YZhou19b} indicated that the change of the observed NS maximum mass from 2.01 to 2.14~$M_{\odot}$ mostly affects the $E_{\rm{sym}}(\rho )$ above about $2\rho_0$, and with~2.14~$M_{\odot}$, the high-density symmetry energy becomes appreciably stiffer. It is also interesting to note that the $R_{1.28}=11.52$ km surface for a 1.28~$M_{\odot}$ NS from NICER's observation of PSR J0030+0451~\cite{Riley19,Miller19} is slightly outside the crossline between the causality surface and the surface of NS of a maximum mass of 2.14~$M_\odot$. Thus, this lower limit of radius for the light NS does not really provide additional constraints on the symmetry~energy.

Very interestingly, the NICER Collaboration has just announced their radius measurement for PSR J0740+6620, which is the currently observed most massive NS. To~avoid confusion in the following discussions, we noticed again here that this NS was reported earlier to have a mass of $2.14_{-0.09}^{+0.10} M_{\odot}$~\cite{Mmax}. Some of the results from our earlier publications were obtained by using 2.14~$M_{\odot}$ as the minimum maximum mass of NSs. 
Now, it has an updated mass of $2.08\pm 0.07$~$M_{\odot}$~\cite{Fonseca21,Miller21a,Miller21,Riley21}. Miller~et~al. reported a radius of $13.7^{+2.6}_{-1.5}$ km (68\%)~\citep{Miller21a,Miller21} 
for PSR J0740+6620. Consistently, an~independent analysis of the NICER data by Riley~et~al. found a radius of $12.39_{-0.98}^{+1.30}$~km~\citep{Riley21}. Our discussions here used the data from Miller~et~al. as we first learned the NICER results from Miller's seminar earlier~\cite{Miller21a}. Results from using the data reported by Riley~et~al. were approximately the same within the remaining uncertainties of the $L$ parameter~\cite{NBZ21}. 

\begin{figure}[H]
%\widefigure
\includegraphics[width=1.0\linewidth]{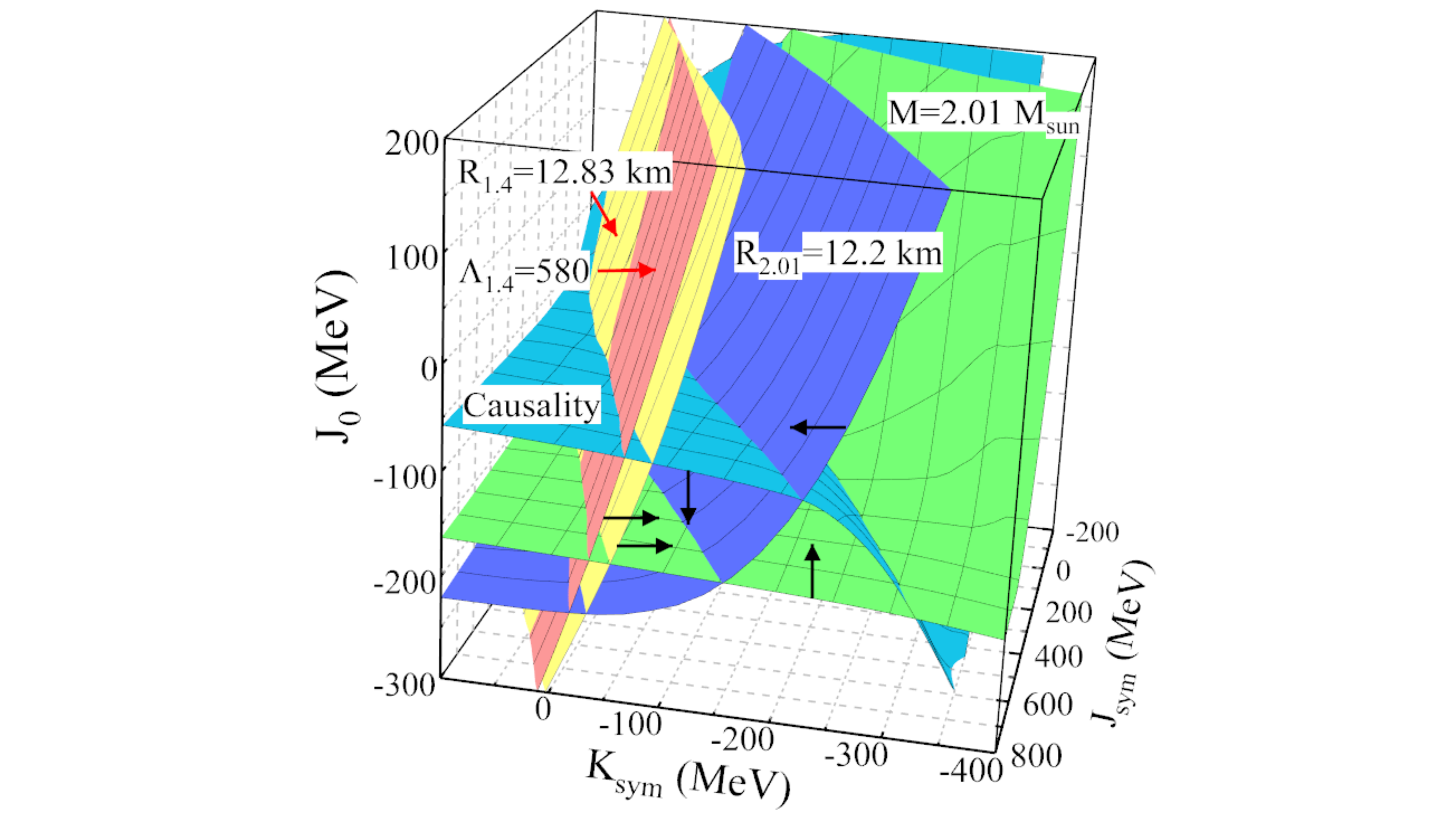}
%MDPI: Please change hyphen to minus sign in the image.
\caption{\textls[-25]{Similar to Figure~\ref{Constraints}, but with the constant surface of the minimum radius $R=12.2$~km} at a 68\% confidence level for PSR J0740+6620 with a mass of $2.08\pm 0.07$~$M_{\odot}$~\cite{Miller21}.}
\label{NICER2}
\end{figure}

While the upper radius limit from Miller~et~al. or Riley~et~al. did not provide additional constraints on the upper boundary of symmetry energy compared to the upper radius limit of $R_{1.4}=12.83$ km from the earlier analysis of Chandra data~\cite{Lattimer14} or the upper limit of $\Lambda_{1.4}= 580$ from LIGO/VIRGO, the~lower radius limit of $R_{2.01}=12.2$ km for PSR J0740+6620 as indicated by the blue surface in Figure~\ref{NICER2} provides an even more stringent limit on the lower boundary of high-density symmetry energy compared to that discussed above. Moreover, comparing the $R_{2.01}=12.2$ km surface with the minimum NS maximum mass surface of 2.01~$M_{\odot}$ shown in Figure~\ref{Constraints}, one sees clearly the power of measuring both the mass and radius simultaneously of heavy NSs for the purposes of limiting the high-density EOS parameter space. There is a large gap in the direction of $J_0$ between the two surfaces in the region where both $K_{\rm{sym}}$ and $J_{\rm{sym}}$ are small (correspondingly, the symmetry energy is very soft). 
More specifically, the~left side of the $R_{2.01}=12.2$ km surface is the allowed high-density EOS parameter space. Before~this new radius observation, the~lower boundary of $E_{\rm{sym}}(\rho )$ was determined by the crossline between the causality surface and the surface of the NS maximum mass of 2.14~$M_\odot$ shown in Figure~\ref{Constraints}. It was now determined by the crossline between the causality surface and the surface of $R_{2.01}=12.2$ km. It was at significantly higher $K_{\rm{sym}}$ values, moving the lower limit of $E_{\rm{sym}}(\rho )$ upward. Moreover, if~only the mass of $2.08\pm 0.07$~$M_{\odot}$ were measured for PSR J0740+6620, then the 
lower boundary of the high-density symmetry energy would be determined by the crossline between the causality surface and the constant surface of 2.01~$M_{\odot}$. It is on the right of the crossline between the causality and M = 2.14~$M_\odot$ surfaces, providing an even looser lower boundary for the high-density symmetry energy. 

\vspace{-7pt}

\begin{figure}[H]
%\widefigure
\includegraphics[width=.4\linewidth,angle=0]{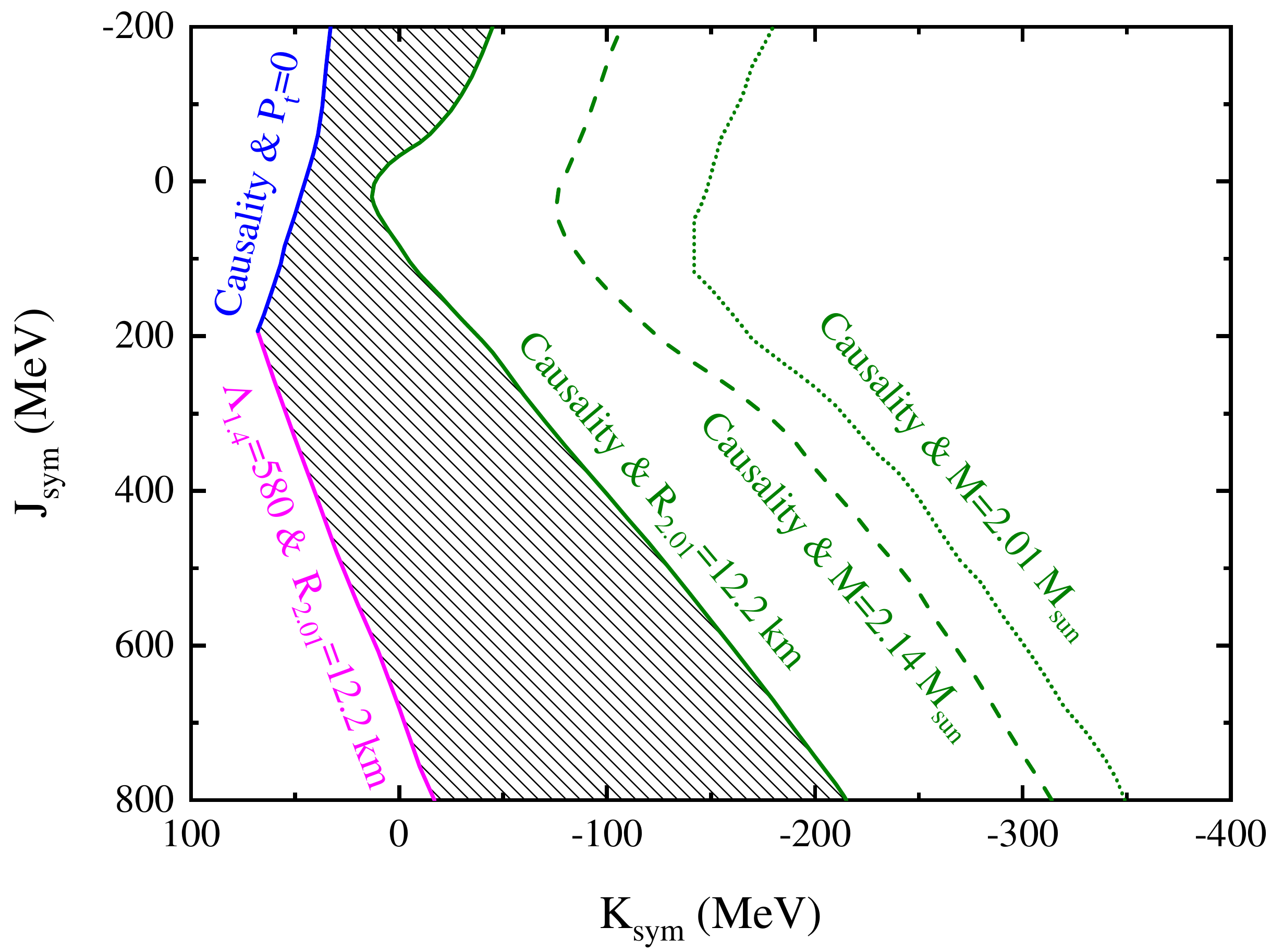}
\includegraphics[width=.4\linewidth,angle=0]{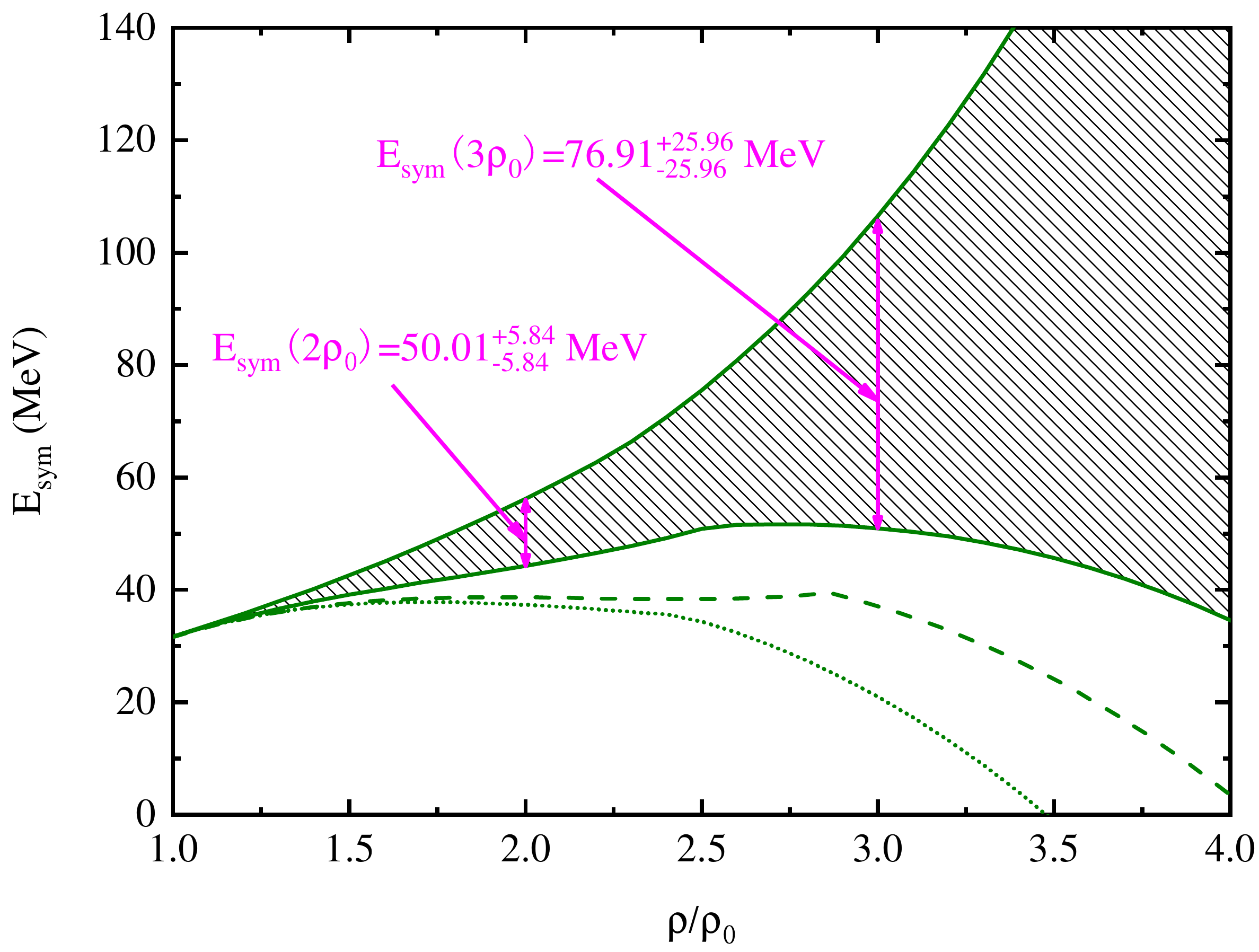}
%MDPI: Please change hyphen to minus sign in the image.
\caption{({Left}) Projections of the indicated crosslines of constant surfaces of NS observables to the $K_{\rm sym}-J_{\rm sym}$ plane for $L=58.7$ MeV. ({Right}) The constraints on symmetry energy extracted from the combinations of $K_{\rm sym}-J_{\rm sym}$ in the left plot.~$E_{\rm sym}(2\rho_0)$ and $E_{\rm sym}(3\rho_0)$ are also extracted and~labeled.}
\label{KsymJsymL60}
\end{figure}

To~see more quantitatively the relative locations of these boundaries and their constraints on the high-density symmetry energy, shown in the left window of Figure~\ref{KsymJsymL60} are projections of the indicated crosslines of constant surfaces of NS observables to the $K_{\rm sym}-J_{\rm sym}$ plane with $L$ set at $L=58.7$ MeV. The~right window shows the resulting constraints on the symmetry energy. For~quantitative comparisons with the systematics discussed earlier, the~new bounds on $E_{\rm sym}(2\rho_0)$ and $E_{\rm sym}(3\rho_0)$ were also extracted and labeled. Clearly, the~radius measurement for PSR J0740+6620 had a significant effect on refining the constraint on the lower boundary of high-density symmetry energy. This has very interesting implications to several predicted effects associated with the super-soft (decreasing symmetry energy with increasing density), e.g.,~the formation of proton polarons, kaon condensation and isospin separation instability in the cores of NSs; see~\cite{NBZ21}
for more \mbox{detailed discussions}.

% start a new page without indent 4.6cm
\clearpage
%\end{paracol}
\nointerlineskip
%\begin{paracol}{2}
%\linenumbers
%\switchcolumn

\textls[-11]{Furthermore, from~the separations between the constant surfaces of \mbox{$R_{1.4}=12.83$ km} and $R_{1.28}=11.52$ km shown in the left window of Figure~\ref{Constraints}, as well as \mbox{$R_{2.01}=12.2$ km} for PSR J0740+6620 shown in Figure~\ref{NICER2}, one can clearly see the importance of precise measurements of NS masses and radii for determining the symmetry energy. To~see more clearly the effects of symmetry energy on the radius and tidal deformability, one can examine the constant surfaces of radius and tidal polarizability in the three-dimensional symmetry energy parameter space of $L-K_{\rm{sym}}-J_{\rm{sym}}$ by setting the skewness of SNM $J_0$ to a constant, as it has little effect on these two~observables. }

\begin{figure}[H]
%\widefigure
\includegraphics[width=.85\linewidth,angle=0]{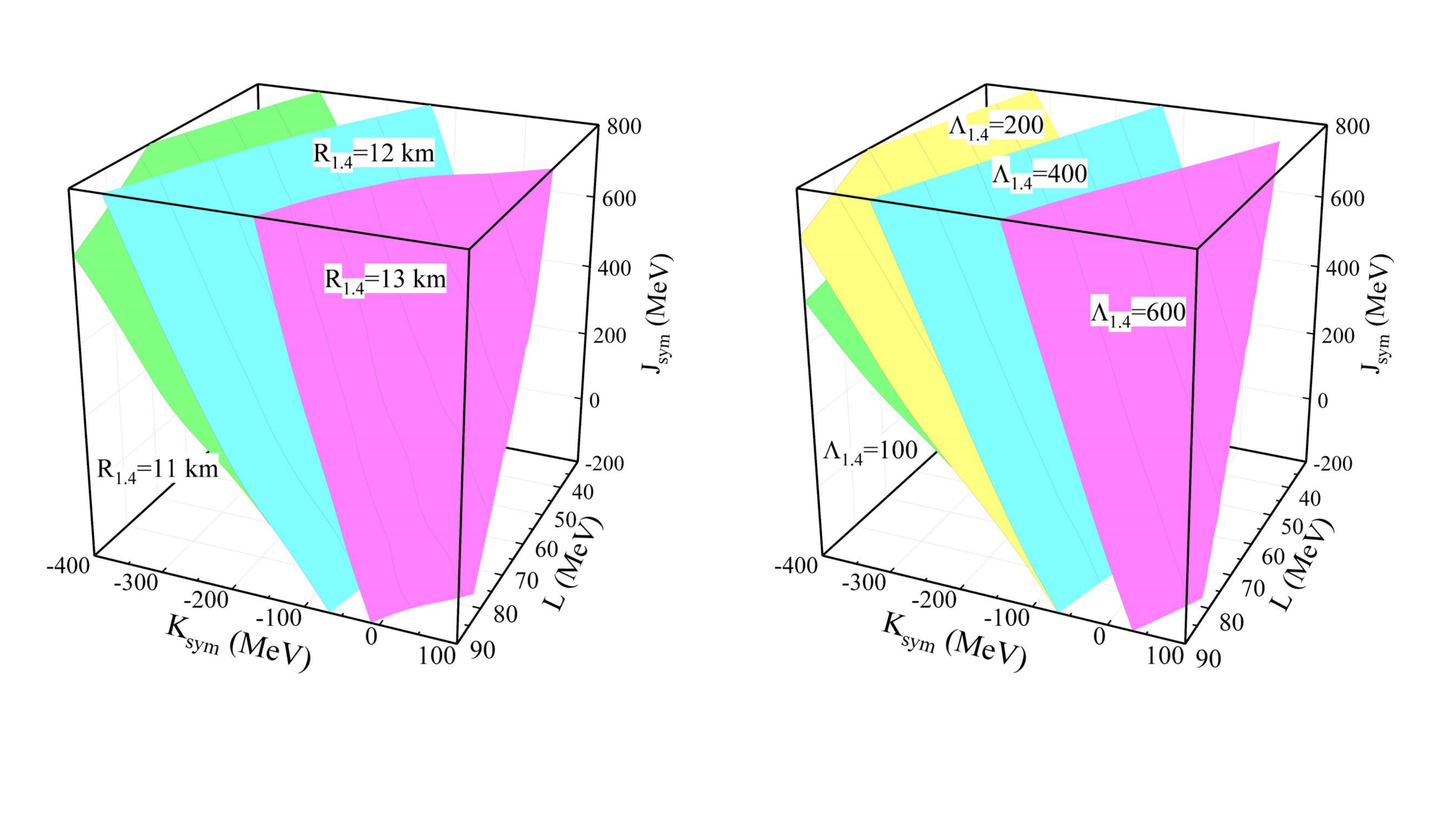}
%MDPI: Please change hyphen to minus sign in the image.
\vspace{-1cm}
\caption{The constant surfaces of radius ({left}) and tidal polarizability ({right}) in the symmetry energy parameter space of $L-K_{\rm{sym}}-J_{\rm{sym}}$, respectively. Taken from~\cite{Zhang19jpg}.}
%MDPI: Please make sure that permission has been obtained and there is no copyright issue.
\label{NBZ-jpg}
\end{figure}

Shown in Figure~\ref{NBZ-jpg} is an example for this purpose in the three-dimensional space within the known uncertainties of the three symmetry energy parameters by setting $J_0=-180$ MeV. It is seen again that only one observable, either $R_{\rm{1.4}}$ or $\Lambda_{1.4}$, is insufficient to completely determine the three parameter,s but provides a strong constraint on their correlations. As~mentioned earlier, since the average density reached in canonical NSs is not very high, except~when $K_{\rm{sym}}$ is very small (having large negative values), the high-density parameter $J_{\rm{sym}}$ plays a small role in determining the radii or tidal deformations of these NSs, as indicated by their largely vertical surfaces. Interestingly, but not surprisingly, while $L$ dominates,~$K_{\rm{sym}}$ has an appreciable role in determining the radius and/or tidal deformation. This explains why the community has extracted both $L$ and $K_{\rm{sym}}$ from analyzing GW170817, but not $J_{\rm{sym}}$ yet, as we summarized in the previous section. It also indicates that it is insufficient to adjust $L$ in models or simply report $L$ without giving any information
about the $K_{\rm{sym}}$ parameter. For~example, for~$R_{\rm{1.4}}$ = 12 km, it can be obtained with a large $L$, but small $K_{\rm{sym}}$ or a small $L$, but larger $K_{\rm{sym}}$. Therefore, a~precise measurement of $R_{\rm{1.4}}$ or $\Lambda_{1.4}$ alone is not sufficient to precisely fix $L$ and $K_{\rm{sym}}$ individually. This is partially responsible for the large error bars of both of them shown in Figures~\ref{Lnew} and \ref{Knew} in the previous section. Moreover, the~fact that $R_{\rm{1.4}}$ and $\Lambda_{1.4}$ do not provide a stringent constraint on $J_{\rm{sym}}$, as shown in Figure~\ref{NBZ-jpg} (which is also verified by the Bayesian analysis to be discussed next), implies that these two observables do not constrain the symmetry energy at very high densities. We discuss this issue more quantitatively~next.

% start a new page without indent 4.6cm
\clearpage
%\end{paracol}
\nointerlineskip
%\begin{paracol}{2}
%\linenumbers
%\switchcolumn

\begin{figure}[H]
%\widefigure
\includegraphics[width=.97\linewidth]{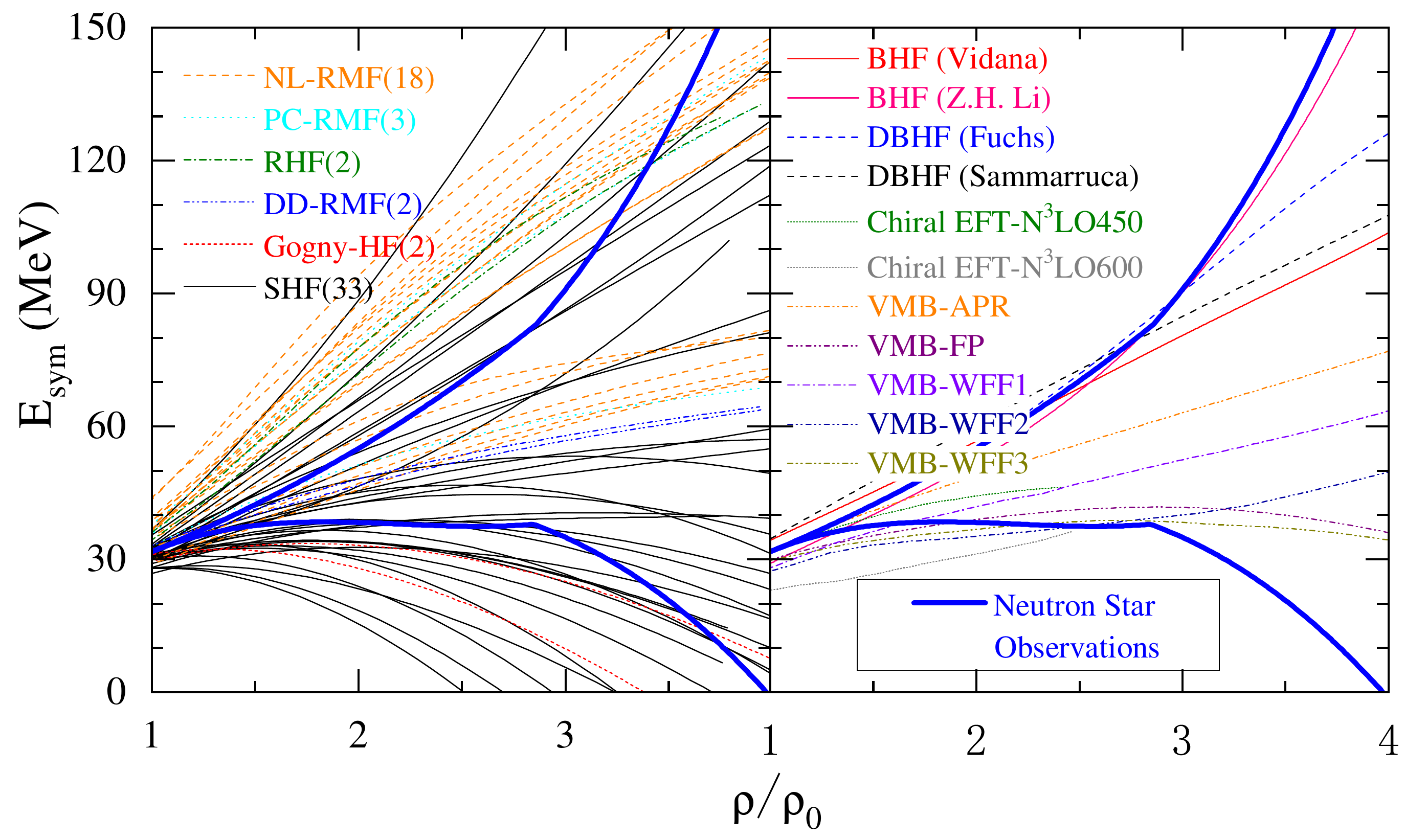}
\caption{Examples of the density dependence of nuclear symmetry energy predicted by nuclear many-body theories using different interactions, energy density functionals, and/or techniques
in comparison with the constraining boundaries extracted from studying the properties of neutron stars. Modified from the figures originally published in~\cite{Zhang20,Zhang20b,Chen15}.}
\label{Esymcon}
\end{figure}

The resulting boundaries of $E_{\rm{sym}}(\rho )$ from the crosslines of neutron star observables shown in Figure~\ref{Constraints} and discussed above are shown as thick blue lines in Figure~\ref{Esymcon} in comparisons with the predictions of phenomenological models (left) and microscopic theories (right). Essentially, all existing nuclear many-body theories using all available nuclear forces have been used to predict the density dependence of nuclear symmetry energy $E_{\rm{sym}}(\rho )$. Shown in the left window are sixty examples selected from six classes of over five-hundred twenty {\it phenomenological} models and/or energy density functional theories, while the right window shows eleven examples from {\it microscopic} and/or ab~initio theories. Mostly by design, they all agree with existing constraints within error bars available around and below the saturation density $\rho_0$. However, at~suprasaturation densities, their predictions diverge very~broadly. 

The fundamental reason for the very uncertain high-density $E_{\rm{sym}}(\rho )$ is our poor knowledge about the relatively weak isospin dependence (i.e., the~difference between neutron--proton interactions in isotriplet and isosinglet channels and their differences from neutron--neutron and proton--proton interactions) of the two-body force, as well as the spin--isospin dependence of the three-body and tensor forces at short distances in dense neutron-rich nuclear matter.
While the astrophysical constraints discussed above can rule out many model predictions up to about $2\rho_0$, huge uncertainties remain at higher densities. This is mainly because the radii and/or tidal deformations of canonical NSs are mostly determined by the pressure around the average density $2\rho_0$ in these NSs~\citep{LP01}. To~constrain $E_{\rm{sym}}(\rho )$ at higher densities, one thus has to use radii of more massive NSs, or~messengers directly from the core of isolated NSs, e.g.,~neutrinos, or~high-frequency gravitational waves from the post merger phase of colliding NSs. Moreover, many theories predict that at densities higher than about $(2\sim 4)\rho_0$, a~hadron--quark phase transition will occur. Since $E_{\rm{sym}}(\rho )$ will lose its physical meaning once the hadron--quark phase transition happens, one thus has to extract the high-density $E_{\rm{sym}}(\rho )$ from astrophysical observables using NS models that properly consider the hadron--quark phase~transition.

\begin{figure}[H]
%\widefigure
\hspace{-.8mm}\includegraphics[width=1.\linewidth,angle=0]{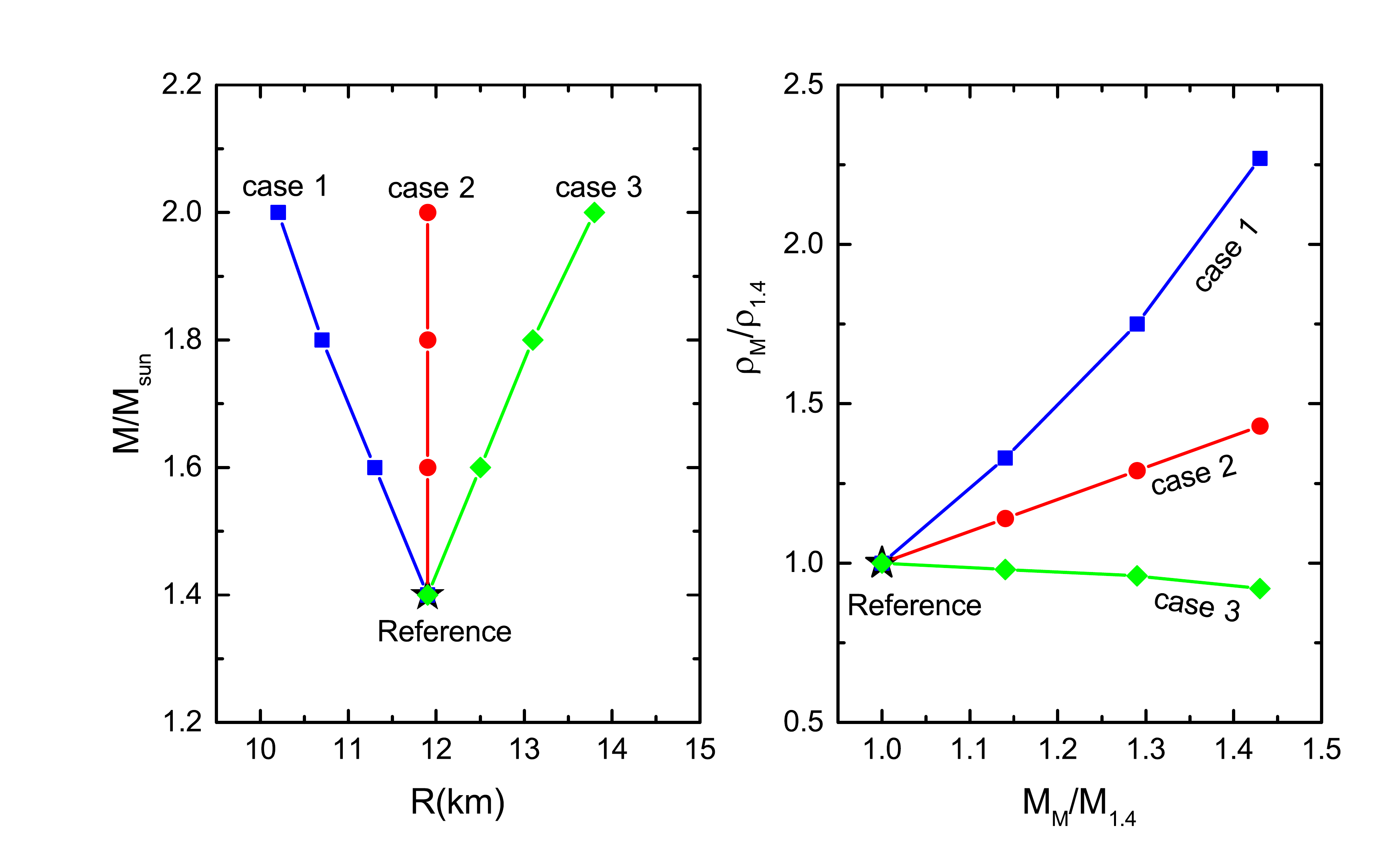}
\caption{({Left}) Representative mocked mass--radius correlations considered for massive NSs with respect to the reference of $R_{1.4}=11.9\pm 1.4$ km at a 90\% confidence level for canonical NSs from GW170817. ({Right}) The corresponding average density in NSs of mass M scaled by that of canonical NSs as a function of the mass ratio $M/M_{1.4}$. Taken from~\cite{Xie20}.}
%MDPI: Please make sure that permission has been obtained and there is no copyright issue.
\label{MR-model}
\end{figure}%in the figures throughout, change case-1, -2, -3 to Case 1, 2, 3
\vspace{-8pt}
%MDPI: Figures should be cited in sequential numerical order. Please confirm if it’s invalid citation, if no, please revise the order. To solve this problem, we moved the original Fig. 11 and its description to here
\subsection{Bayesian Inference of Symmetry Energy Parameters from the Radii of Canonical Neutron~Stars}
\begin{figure}[H]
%\widefigure
\vspace{-1cm}
\hspace{-1.5mm}\includegraphics[width=1.2\linewidth,angle=0]{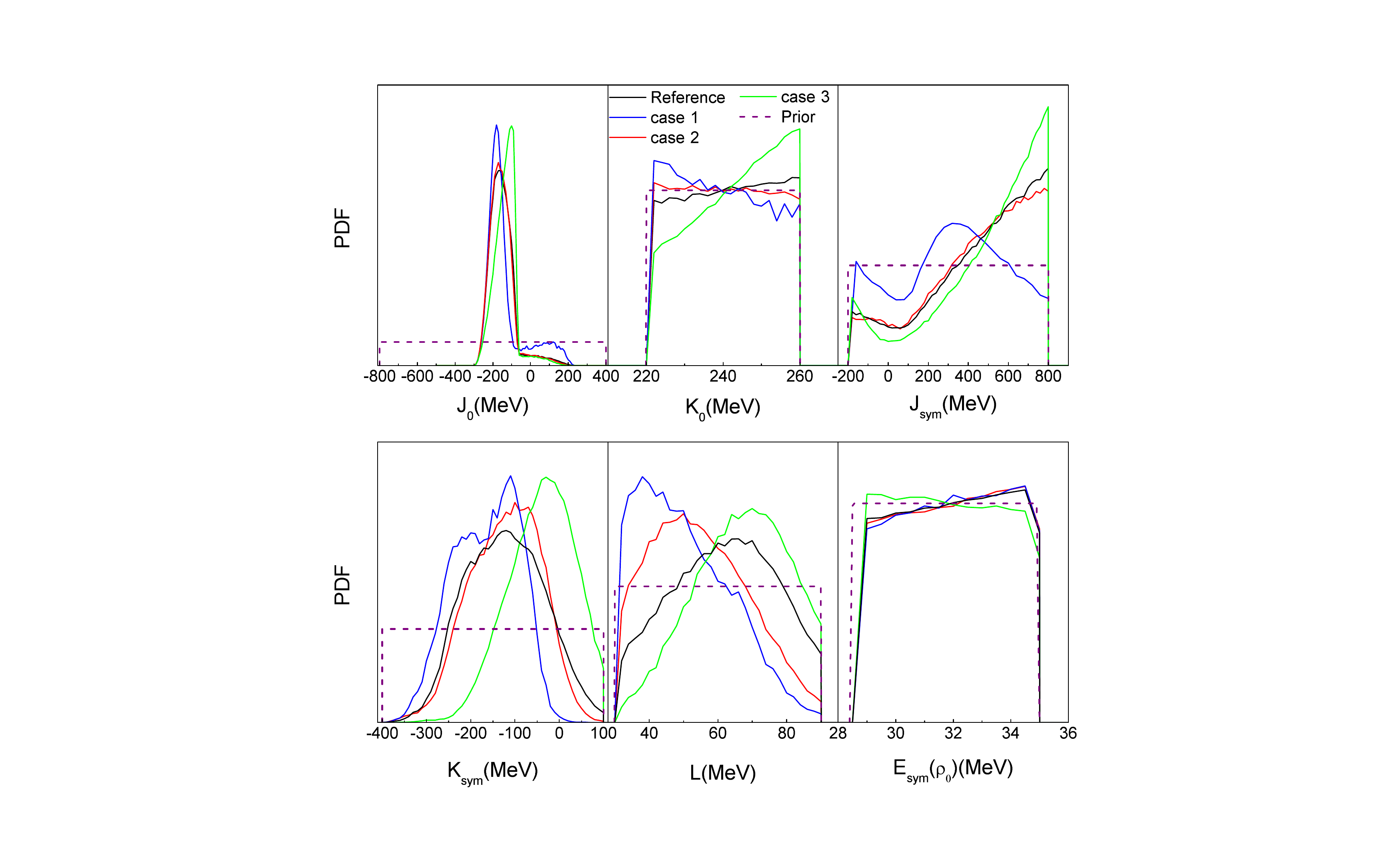}
%MDPI: Please change hyphen to minus sign in the image.
\caption{The posterior PDFs of the 6 EOS parameters in comparison with their prior PDFs for the three cases using the mass--radius data shown in Figure~\ref{MR-model}. Taken from~\cite{Xie20}.}
%MDPI: Figures should be cited in sequential numerical order. Please confirm if it’s invalid citation, if no, please revise the order. Valid, in this REVIEW we are using figures from several of our earlier publications.
%Sometimes, we need to use information from different figures to make a point.
\label{Xie-PDF}
\end{figure}

\begin{figure}[H]
%\widefigure
\includegraphics[width=1.\linewidth,angle=0]{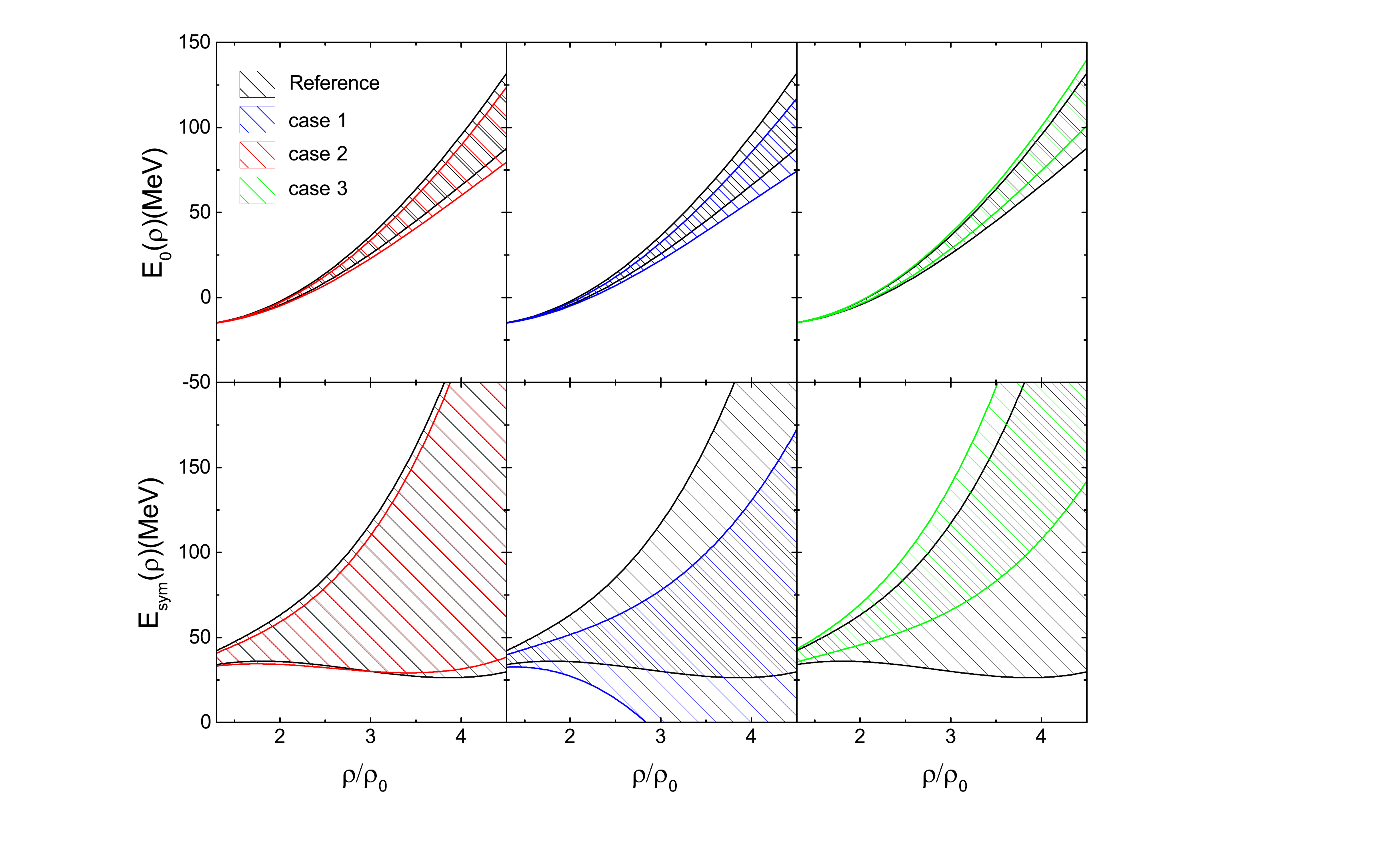}
%MDPI: Please change hyphen to minus sign in the image.
\caption{The 68\% boundaries of nuclear symmetry symmetry (bottom) and SNM EOS (top) inferred from the neutron star radius data shown in Figure~\ref{MR-model}. Taken from~\cite{Xie20}.}
%MDPI: Please make sure that permission has been obtained and there is no copyright issue.
\label{Xie-Esym}
\end{figure}

While the direct inversion technique used in the examples given above has the advantage of enabling us to visualize how each observable may help
constrain one or more high-density EOS parameters, it is limited to the three-dimensional space. It is well known that the general technique of multidimensional inversion is the Bayesian statistical inference. It has been widely used in
analyzing various data within different EOS model frameworks. For~example, albeit giving quantitatively slightly different results, several Bayesian analyses of the GW17817 data have indicated that the radii of canonical NSs are approximately mass independent~\cite{LIGO18,De18,Capano20}. For~example, the~principal NS in GW170817 has a mass between 1.36 and 1.58~$M_{\odot}$, while its secondary has a mass between 1.18 and 1.36~$M_{\odot}$~\citep{LIGO18}. Assuming initially their radii are different in their model analyses, the LIGO/VIRGO Collaborations found a common radius $R_{1.4}=11.9\pm 1.4$ km for the two NSs involved. Using the later as the radius data (labeled as Reference in Figures~\ref{MR-model}, \ref{Xie-PDF} and \ref{Xie-Esym}), together with other general constraints, such as the minimum NS maximum mass of 1.97~$M_{\odot}$ and causality condition, in~a Bayesian analysis using the metamodel of NS EOSs~\cite{Xie20}, the~PDFs of the six EOS parameters were inferred and shown as the black curves in Figure~\ref{Xie-PDF}.  The other curves labeled as Case 1, 2, 3 are results of using mocked mass-radius relations shown in Figure~\ref{MR-model}. They will be discussed in detail in Section \ref{future}. 

It is seen that the PDFs of $L$ and $K_{\rm{sym}}$ are strongly peaked compared to their uniform prior PDFs, leading to the extraction of $L=66^{+12}_{-20}$ MeV and $K_{\rm{sym}}=-120^{+80}_{-100}$ MeV at a 68\% confidence level. As~discussed in detail in Reference~\cite{Xie20}, the~peak of $J_0$ is mainly due to the requirement to support NSs with masses at least as massive as 1.97~$M_{\odot}$,
while the considered NS data have little effect on constraining $K_0$ or $E_{\rm{sym}}(\rho_0)$, indicated by their very similar prior and posterior PDFs. This is because these two parameters characterize only the properties
of neutron-rich matter at saturation density. On~the other hand, the~$J_{\rm{sym}}$ parameter characterizes the behavior of $E_{\rm{sym}}(\rho)$ at densities above $2\rho_0$. It is seen that the posterior PDF of $J_{\rm{sym}}$ peaks at its upper boundary and would keep changing as its prior range changes, indicating that the NS data used do not constrain this parameter (thus the behavior of $E_{\rm{sym}}(\rho)$ at densities higher than about $2\rho_0$, consistent with the findings from the direction inversion approach shown in Figure~\ref{Constraints}). Again, this is mainly because the radii of canonical NSs are determined mainly by the nuclear pressure around $2\rho_0$. They are
not sensitive to the symmetry energy at higher densities, while the NS mass is mostly determined by the SNM EOS, unless the symmetry energy becomes super-soft, as we shall~discuss.

\begin{figure}[H]
%\widefigure
\includegraphics[width=.485\linewidth]{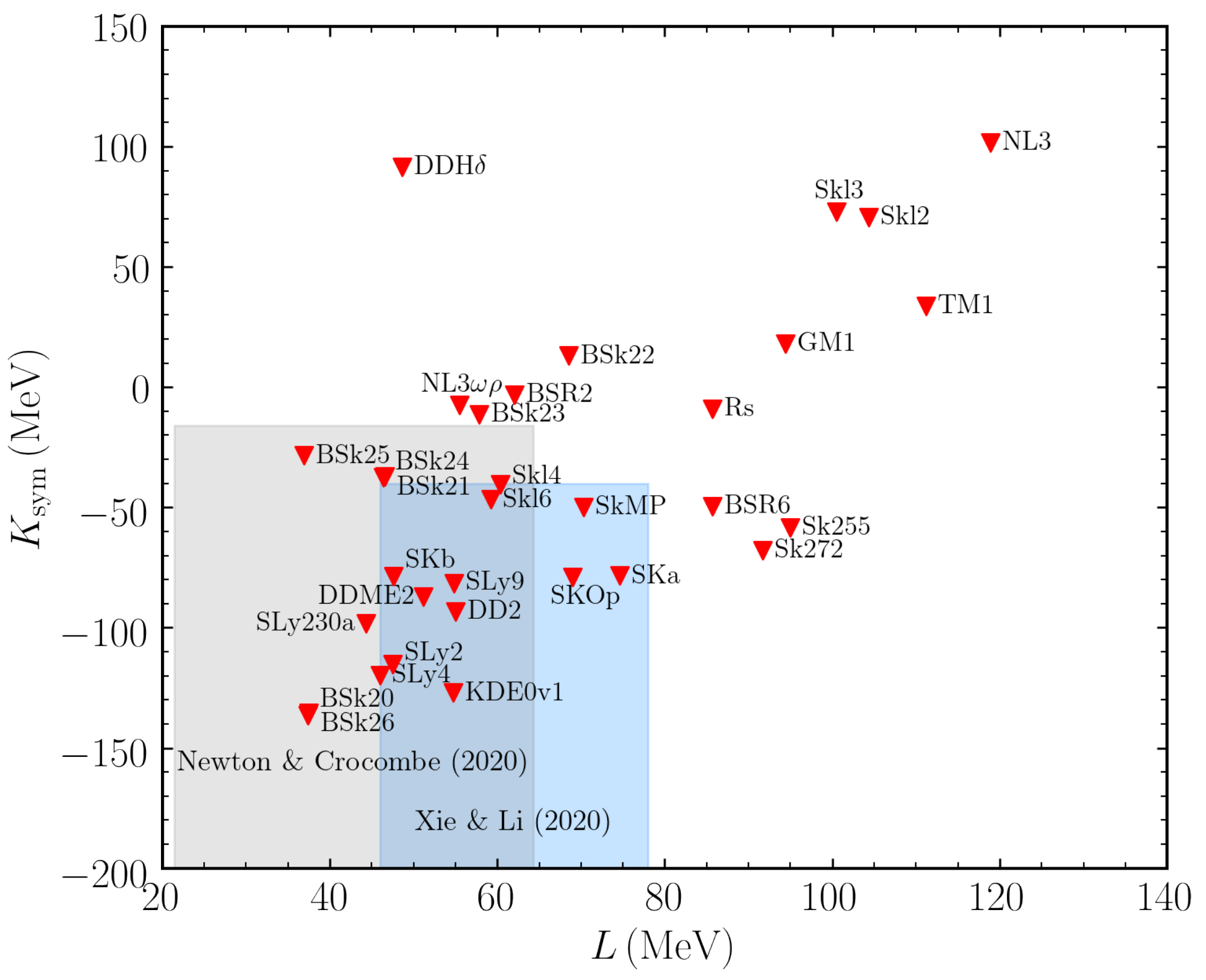}
\includegraphics[width=.48\linewidth]{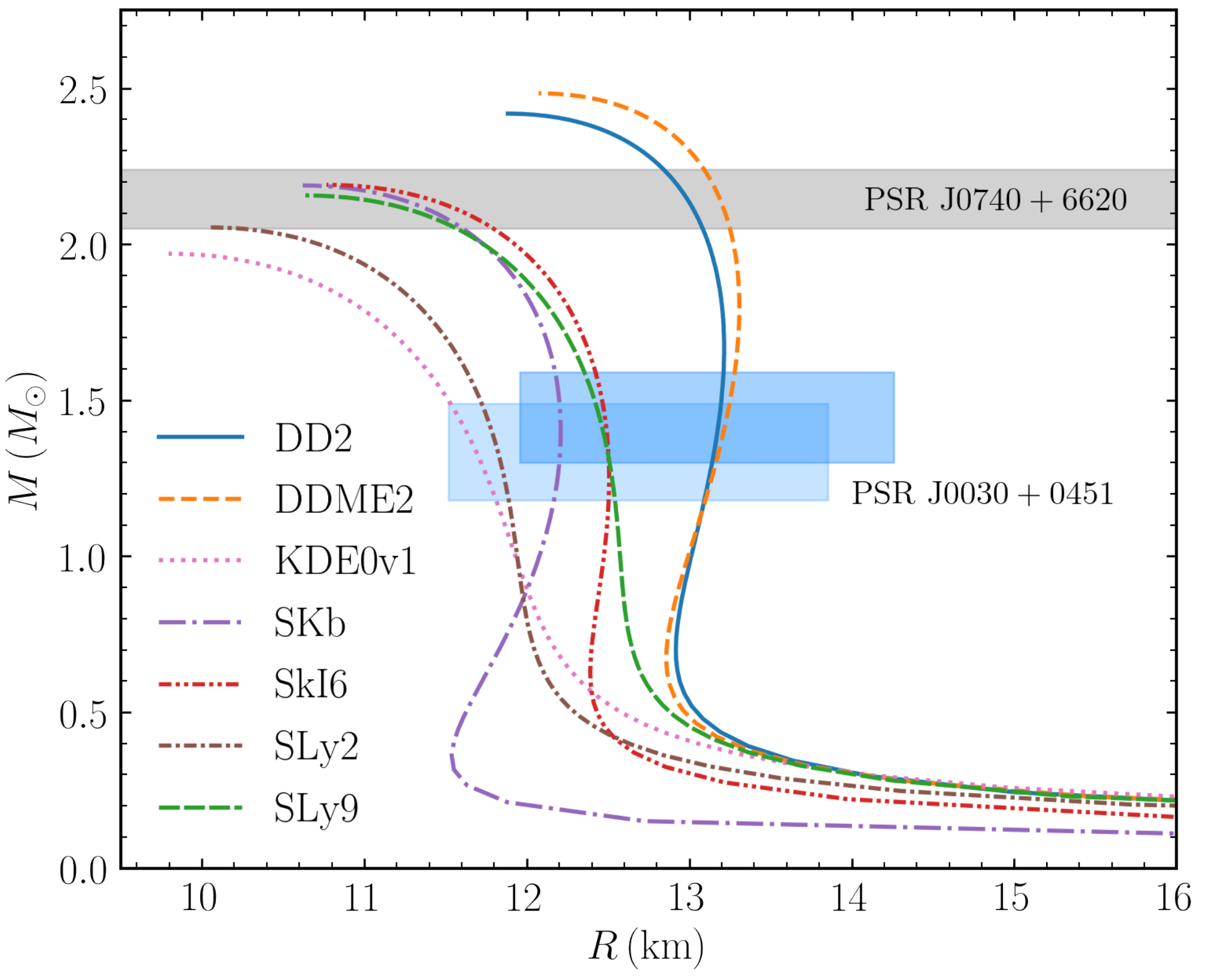}
\caption{({Left}) $K_{\rm sym}$ versus $L$ for 33 unified neutron star EOSs from \citet{Fortin2016} in comparison with
their $68\%$ confidence boundaries from analyzing neutron observables (Xie and Li)~\cite{Xie20}, as well as terrestrial experiments and theoretical calculations for the PNM EOS (Newton and Crocombe)~\cite{Newton20}. ({Right}) Mass--radius correlations of the EOSs filtered by the $K_{\rm sym}$ versus $L$ constraints in comparison with the observational constraints. Taken from~\cite{Zhou21}.}
%MDPI: Please make sure that permission has been obtained and there is no copyright issue.
\label{Zhou}
\end{figure}

To see the impacts of the constraints on $L$ and $K_{\rm sym}$ from the above analyses, shown on the left of Figure~\ref{Zhou} is
$K_{\rm sym}$ as a function of $L$ for 33 unified neutron star EOSs from \mbox{\citet{Fortin2016}} in comparison with
their $68\%$ confidence boundaries from analyzing neutron observables (Xie and Li)~\cite{Xie20}, as well as terrestrial experiments and theoretical calculations for the PNM EOS (Newton and Crocombe)~\cite{Newton20}. The~33 unified EOSs were derived by \mbox{\citet{Fortin2016}} within the Skyrme--Hartree--Fock and the RMF models for the core and the Thomas--Fermi model for the crust using the same interactions. Only seven of them are in the overlapping area of the two constraints used. Shown in the right window are the
mass--radius correlations predicted by these seven EOSs in comparison with the latest observational constraints.
It is seen that the KDE0v1 EOS is further excluded by the mass 2.14~$M_{\odot}$ of MSP J0740+6620, as Xie and Li's
constraint was derived by using an NS minimum mass of 1.97~$M_{\odot}$, as in the original LIGO/VIRGO data analysis of GW170817.
Thus, it is clear that the constraints on $L$ and $K_{\rm sym}$ from analyzing the observables of neutron stars help greatly in
screening theoretical predictions for the EOS of dense neutron-rich~matter.

% start a new page without indent 4.6cm
%\clearpage
%\end{paracol}
%\newpage
\nointerlineskip
%\begin{paracol}{2}
%\linenumbers
%\switchcolumn
%\vspace{-9pt} 

\subsection{Future Radius Measurements of Massive Neutron Stars and Their Constrains on High-Density Nuclear Symmetry~Energy}\label{future}
Will future high-precision radius measurements of more massive NSs help constrain the symmetry energy at densities above $2\rho_0$? This question was studied in~\cite{Xie20}. In~the following,
we summarize the main findings with a few illustrations. To~answer this question, three sets of mock mass--radius correlations were used in a comprehensive Bayesian analysis, as shown in Figure~\ref{MR-model}.
They all start from the same reference point of $R_{1.4}=11.9\pm 1.4$ km and use the same uncertainty of 1.4 km at a 90\% confidence level for all data points. The~scaled average densities in NSs for these three typical mass--radius relations are shown in the right window of Figure~\ref{MR-model}, assuming NSs are made of neutrons, protons, electrons, and~muons only. It is seen that for an NS of 2.0~$M_{\odot}$,
its average density scaled by that in an NS of 1.4~$M_{\odot}$ is significantly different in the three cases. In~particular, in Case 2,
the scaled density increases by about 50\% from 1.4 to 2.0~$M_{\odot}$. The~recent report by Miller~et~al. indicated that NICER observations favor a mass--radius relation between Case 2 and Case 3~\cite{Miller21}.
As mentioned earlier, NICER found the mass of $1.44^{+0.15}_{-0.14}$$M_{\odot}$ and $R=13.02^{+1.24}_{-1.06}$ km for PSR J0030+0451~\cite{Miller19} and $R=12.2-16.3$ km for PSR J0740+6620 with an updated mass of $2.08\pm 0.07$~$M_{\odot}$~\cite{Miller21}.

\textls[-11]{A systematic Bayesian inference of the EOS parameters using the mocked mass--radius data discussed above was carried out in~\cite{Xie20}. The~resulting PDFs of the EOS parameters are compared in Figure~\ref{Xie-PDF}. In Case 2, where the radius is independent of the mass, the~resulting PDFs of all six EOS parameters are not much different from those of the reference (a single canonical NS with a radius of $R_{1.4}=11.9\pm 1.4$ km), although the average density increases by about 50\% from 1.4 to 2.0~$M_{\odot}$. Nevertheless, comparing the PDFs from Case 1 and Case 3, all PDFs change significantly.
There are also some secondary bumps and dips due to the correlations among the EOS parameters, as discussed in~\cite{Xie20}. The~resulting 68\% confidence boundaries of the symmetry energy (bottom) and SNM EOS (top) are shown in Figure~\ref{Xie-Esym} in comparison with the results of the reference case. Most interestingly, while the confidence boundaries of SNM EOSs are not much different in all three mock cases and the reference case as all EOSs are required to support the same NS minimum mass of 1.97~$M_{\odot}$, the~symmetry energy boundaries especially at high densities are quite different from Case 1 to Case 3. This indicates that the radii of massive NSs will help constrain the symmetry energy at densities above $2\rho_0$, where it is most uncertain, with~little influence from the remaining uncertainties of high-density SNM EOS. We noticed again that the lower boundary of symmetry energy at high densities had appreciable dependence on the NS minimum maximum
mass M$_{\rm{max}}$ used. In~the results shown, M$_{\rm{max}}=1.97$~$M_{\odot}$ was used. If~it were replaced by 2.01~$M_{\odot}$ or even higher, the~lower boundaries were expected to move~up.}

\begin{figure}[H]
%\widefigure
\includegraphics[width=1.\linewidth,angle=0]{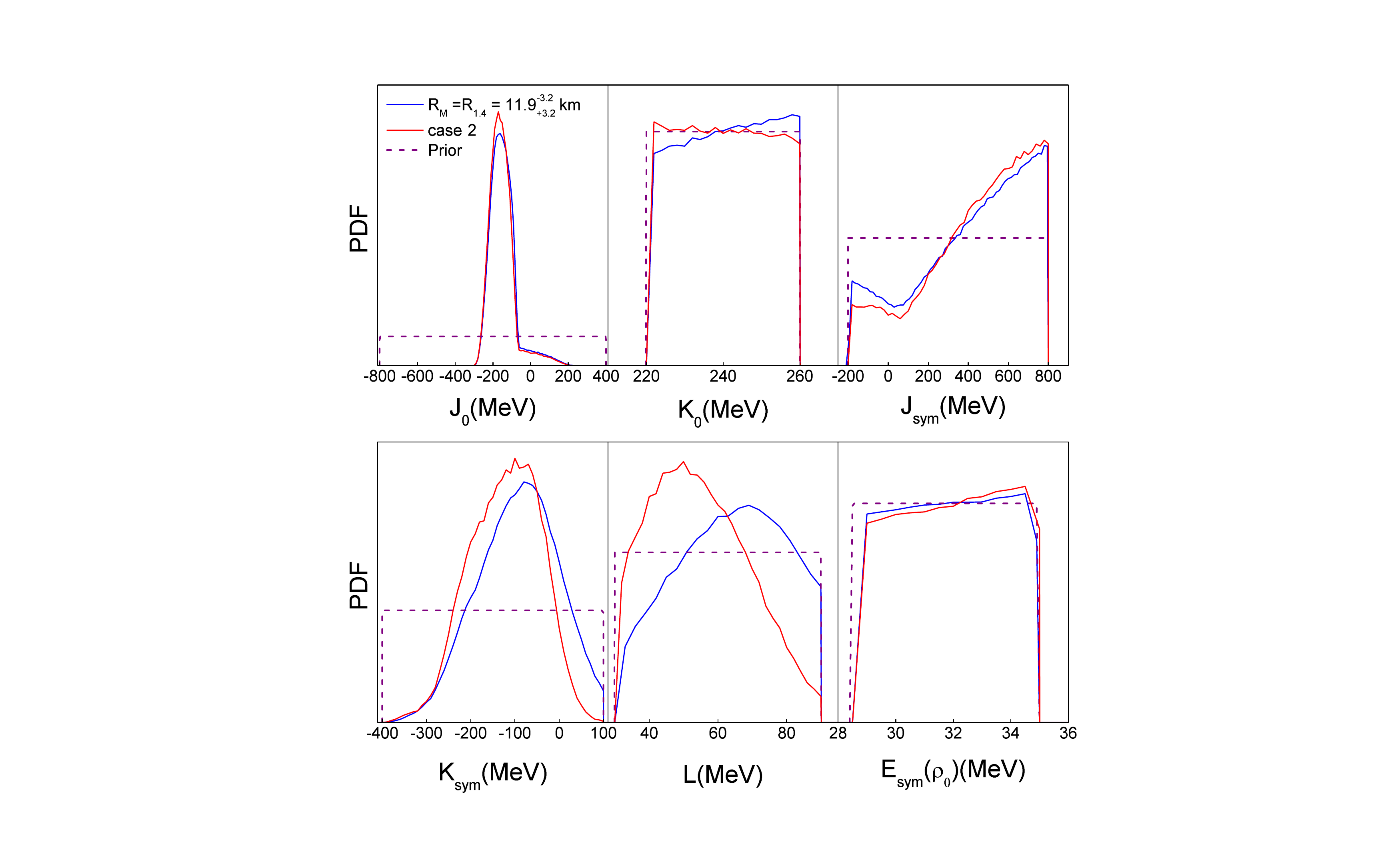}
%MDPI: Please change hyphen to minus sign in the image.
\caption{Prior and posterior probability distribution functions of EOS parameters assuming all neutron stars have the same radius of $R_{\mathrm{M}}=R_{1.4}=11.9 \pm 3.2$ km in comparison with the results from Case 2, which has the same mean radius of 11.9 km, but an error bar of 1.4 km at a 90\% confidence level (Red curves shown in Figure~\ref{Xie-PDF}). Taken from~\cite{Xie20}.}
%MDPI: Please make sure that permission has been obtained and there is no copyright issue.
\label{newradius}
\end{figure}

How important is the precision of radius measurements? What happens if the precision is not high enough
to distinguish the three possible mass--radius correlations shown in Figure~\ref{MR-model}?
To answer these questions, shown in Figure~\ref{newradius} are comparisons of the PDFs of the six EOS parameters calculated for Case 2 discussed above and a new calculation using a constant radius $R_{\mathrm{M}}=R_{1.4}=11.9 \pm 3.2$ km at a 90\% confidence level for all NSs considered. Thus, the~two data sets have the same mean radius, but the new case has an error bar that is three times that of Case 2. It is so large that one can no longer distinguish the three cases shown in Figure~\ref{MR-model}.
As one would expect, the~posterior PDFs of all three symmetry energy parameters, especially for the slope $L$ and curvature $K_{\mathrm{sym}}$, became significantly wider. In~fact, the~$L$ parameter can now go higher than its prior upper limit used. On~the other hand,
it was seen that the PDFs of the SNM EOS parameters were basically the same as the masses, and the mean radii of the NSs considered were kept the same.
Therefore, the~precision of the radius measurement for massive neutron stars can significantly affect the accuracy of constraining the symmetry energy, but not greatly the EOS of~SNM.

\section{Effects of Hadron--Quark Phase Transition on Extracting Nuclear Symmetry Energy from Neutron Star~Observables}

Whether Quark Matter (QM) exists and how big its mass or volume may be are among the longstanding and interesting issues regarding the structure of neutron stars~\cite{Alford}. In~the forward modeling of neutron stars, often, one assumes a hadron--quark phase transition density around several times the saturation density of nuclear matter. To~the best of knowledge, there is still a hadron--quark duality in understanding NS observables, namely there is so far no unique signature for the existence of quark matter in NSs, while high-frequency post-merger gravitational waves from collisions of NSs were predicted to provide more clear signatures of hadron--quark phase transition in neutron star matter~\cite{GSI,FRA}.
Both the purely hadronic matter and the hybrid hadron--quark mixture can equally describe all available NS observations. In~the literature, most of the studies on nuclear symmetry energy using NS observables were carried out using the purely hadronic picture. Therefore, how does the consideration of possible hadron--quark phase transition in NSs affect the extraction of high-density symmetry energy? This question was studied recently in~\cite{Xie-QM}. Here, we summarize the main findings from this~study.

\subsection{Bayesian Inference of Hadronic and Quark Matter EOS Parameters from Observations of Canonical Neutron~Stars}
In~\cite{Xie-QM}, a~nine-parameter metamodel was used to generate both hadronic and quark matter EOSs in the Markov chain Monte Carlo sampling process within the Bayesian statistical framework. For~the hadronic phase, the~six-parameter EOS was constructed using Equations~(5) and (6). It was connected through the Maxwell construction to the Constant Speed of Sound (CSS) model for quark matter developed by Alford, Han, and Prakash~\cite{CSS}. More specifically, the~pressure can be written as:
\begin{equation}
\ep(p) = \left\{\!
\begin{array}{ll}
\ep_{\rm HM}(p) & p<p_{t} \\
\ep_{\rm HM}(p_{t})+\De\ep+c_{\rm QM}^{-2} (p-p_{t}) & p>p_{t}
\end{array}
\right.\ ,
\label{eqn:EoSqm1}
\end{equation}
where $\ep_{\rm HM}(p)$ is the Hadronic Matter (HM) EOS below the hadron--quark transition pressure $p_{t}$, $\De\ep$ is the discontinuity in energy density $\ep$ at the transition, and~$c_{\rm QM}$ is the QM speed of~sound.

In the Bayesian analyses using the metamodel EOS described above, besides~the causality condition and the requirement that all EOSs have to be stiff enough to support NSs at least as massive as 2.14~$M_{\odot}$, the~following radii of canonical NSs were used as independent data: (1) $R_{1.4}=11.9\pm 1.4$ km from GW170817 by the LIGO/VIRGO Collaborations~\cite{LIGO18}, (2) $R_{1.4}=10.8^{+2.1}_{-1.6}$ extracted independently also from GW170817 by De~et~al.~\cite{De18}, (3) $R_{1.4}=11.7^{+1.1}_{-1.1}$ from earlier analysis of quiescent low-mass X-ray binaries~\cite{Lattimer14}, and~(4)
$R=13.02^{+1.24}_{-1.06}$ km with mass $M=1.44^{+0.15}_{-0.14}$~$M_{\odot}$~\cite{Miller19} or $R=12.71^{+1.83}_{-1.85}$ km with mass $M=1.34\pm0.24$~$M_{\odot}$~\cite{Riley19} for PSR J0030+0451 from the NICER Collaboration. The~errors quoted are at the 90\% confidence~level.

\begin{figure}[H]
%\widefigure
\includegraphics[width=.77\linewidth,angle=90]{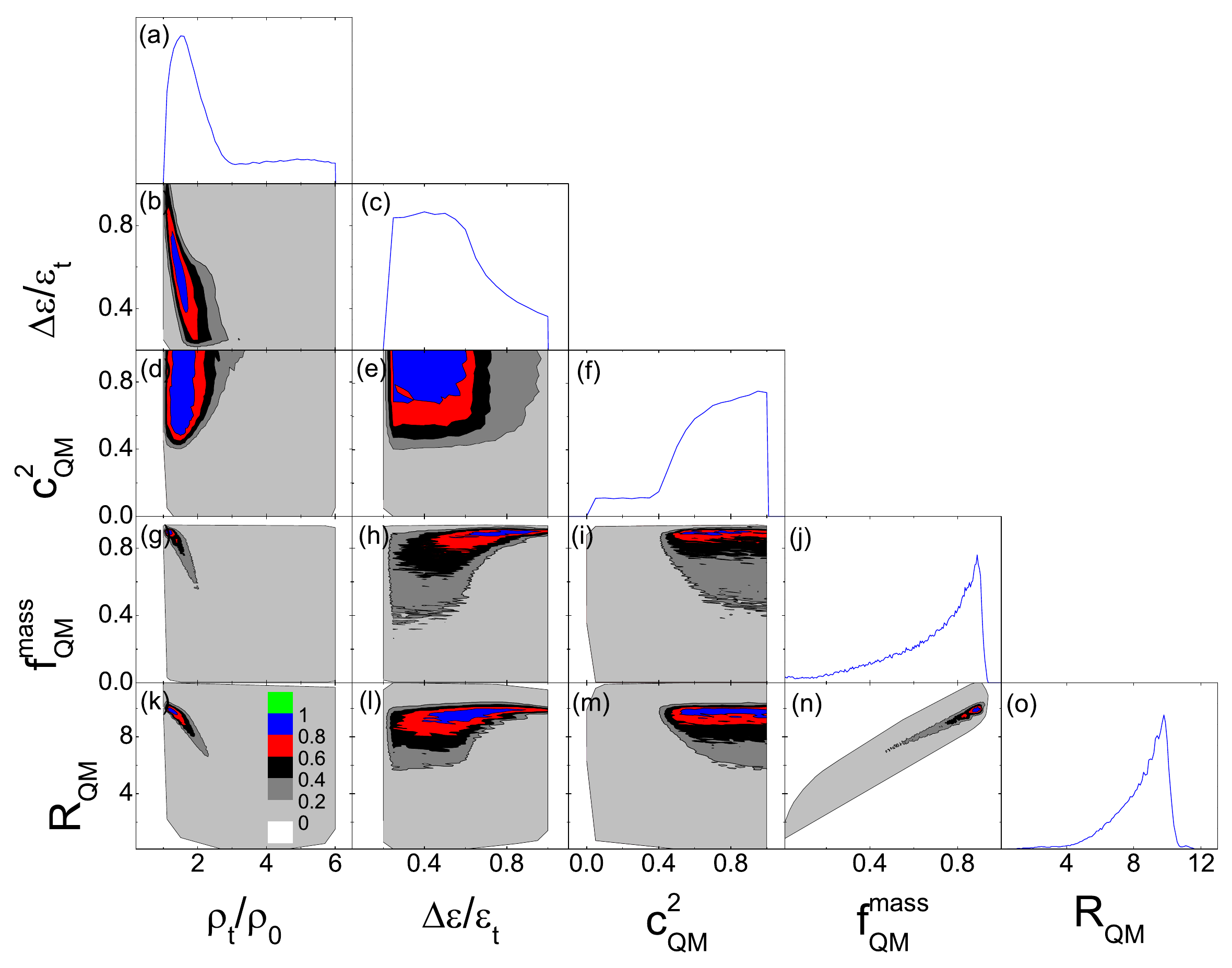}
%MDPI: There is no explanation for the subfigure in the figure, please add.
\caption{The subfigures along the diagonal are the posterior probability distribution functions (in arbitrary units) of the five quantities labeled (the three quark matter EOS parameters, as well as the fraction and radius of quark matter in hybrid neutron stars of mass 1.4~$M_{\odot}$). See text for the definitions of each quantity shown.  All off-diagonal subfigures are the correlations among the 5 quantities labeled. Taken from~\cite{Xie-QM}.}
\label{QM}
%MDPI: Please make sure that permission has been obtained and there is no copyright issue.
\end{figure}

Shown in Figure~\ref{QM} are the resulting posterior PDFs and correlations of quark matter EOS parameters $\rho_t/\rho_0$, $\De\ep/\ep_t$, and~$\cQMsq/c^2$, as~well as the fraction $f^{\rm mass}_{\rm QM}$ (defined as the ratio of QM mass over the total NS mass) and the radius of quark matter $R_{\rm QM}$. Interestingly, the~most probable hadron--quark transition density $\rho_t/\rho_0=1.6^{+1.2}_{-0.4}$ was found to be rather low, while the transition strength $\De\ep/\ep_t=0.4^{+0.20}_{-0.15}$ was modest and the speed of sound in QM \mbox{$\cQMsq/c^2=0.95^{+0.05}_{-0.35}$} very high. The~latter is understandable as the most probable transition density is very low, so the~ stiffness of QM EOS represented by its $\cQMsq$ value has to be high enough to support the neutron star with a large QM core. In fact, the strong anti-correlation between the transition density and speed of sound in QM is clearly shown in their correlation function shown. Since the average baryon density in a canonical NS with a 12 km radius is about $2\rho_0$ (compatible with the low transition density), the QM fraction and its radius are both quite high. We noticed that the rather low hadron--quark transition density found in~\cite{Xie-QM} was also found very recently in two independent Bayesian analyses using similar NS data~\cite{Tang21,AngLi21}. Thus, based on these Bayesian analyses, a large volume of QM exists even in canonical NSs. However, it was found in~\cite{Ann20} that it is unlikely to form a quark core in canonical neutron~stars. 

\subsection{Comparing Symmetry Energy Parameters from Analyzing Neutron Star Observables with and without Considering the Hadron-Quark Phase~Transition}
\begin{figure}[H]
%\widefigure
\includegraphics[width=.9\linewidth]{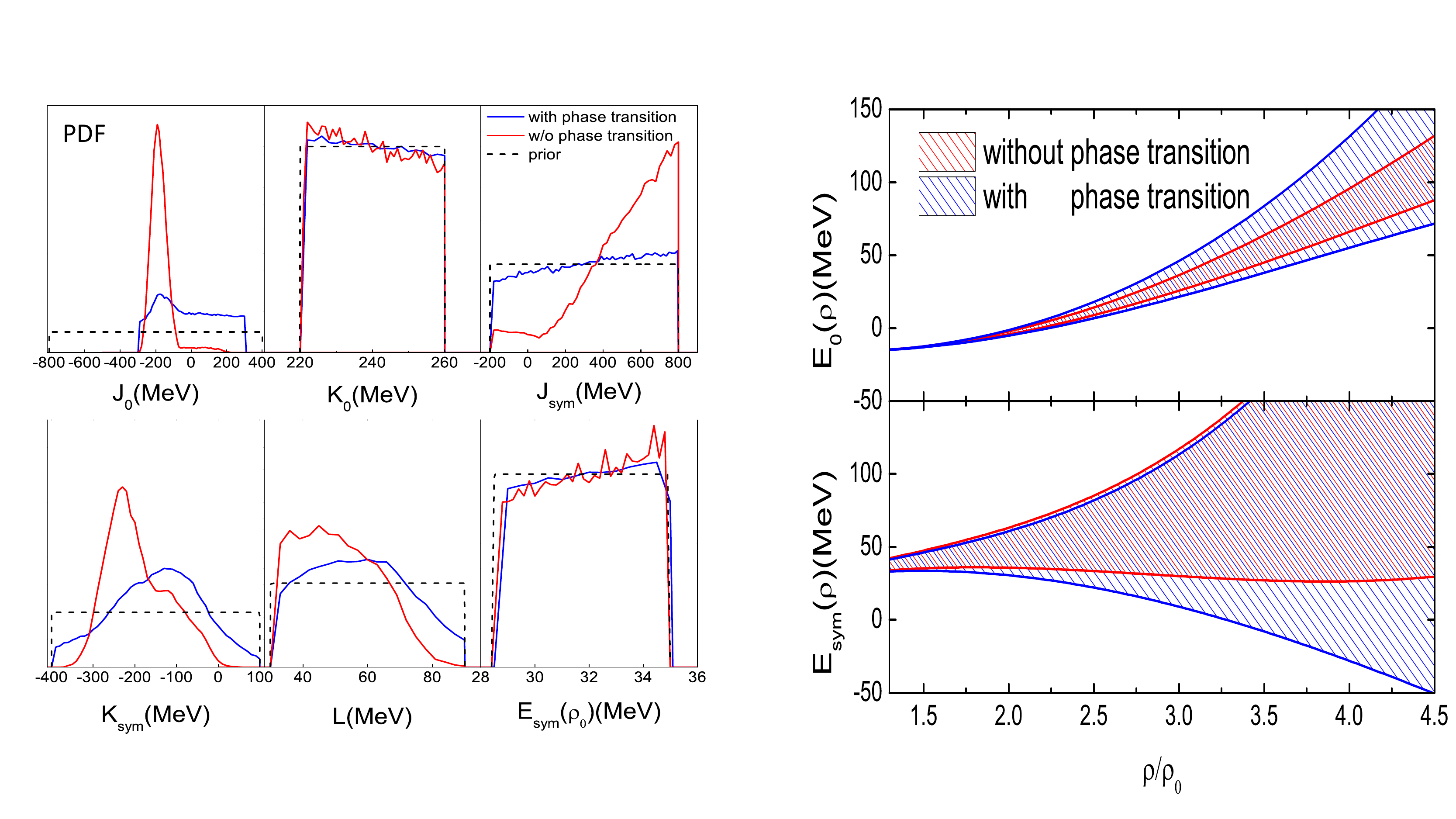}
%MDPI: Please change hyphen to minus sign in the image.
\caption{({Left}) The posterior probability distribution functions (in arbitrary units) of nuclear matter EOS parameters inferred from Bayesian analyses with (thick blue curves) and without (thin red curves) considering the hadron--quark phase transition in neutron stars in comparison with their uniform priors (dashed curves). ({Right}) The corresponding 68\% confidence boundaries of SNM EOS ({top}) and symmetry energy ({bottom}). Taken from~\cite{Xie-QM}. }
%MDPI: Please make sure that permission has been obtained and there is no copyright issue.
\label{NMEOS}
\end{figure}

How does the consideration of the hadron--quark phase transition affect the extraction of nuclear symmetry energy from NS observables? To answer this question, shown in the left window of Figure~\ref{NMEOS} are comparisons of the PDFS of six hadronic EOS parameters with and without considering the hadron--quark phase transition, while in the right are comparisons of the corresponding 68\% confidence boundaries of the SNM EOS (top) and symmetry energy (bottom). 

It is seen that while the incompressibility $K_0$ and symmetry energy $E_{\rm{sym}}(\rho_0)$ at saturation density $\rho_0$ are not much different, as one expects, the~PDF of $J_0$ characterizing the stiffness of SNM at suprasaturation densities shifts towards higher values as the hadron--quark phase transition softens the EOS, unless $\cQMsq/c^2$ is very high. The~hadronic EOS needs to be stiffened to support the same NSs. Furthermore, with~the hadron--quark phase transition, three additional parameters in the CSS model are used. The~68\% confidence boundaries of the SNM EOS become wider, as shown in the upper right window.
Thus, with~the PDF of $\rho_t/\rho_0$ peaks at $1.6^{+1.2}_{-0.4}$, in~the model with the hadron--quark phase transition, the~posterior PDFs of $L$ and $K_{\rm sym}$ shift significantly higher to reproduce the same radius data, while the poster PDF of $J_{\rm sym}$ is not much different from its prior PDF. The~latter is, however, significantly different from the PDF in the case without considering the hadron--quark phase transition. Thus, as~shown in the lower right window, the~high-density behavior of nuclear symmetry energy extracted from NS observables is quite different with or without considering the hadron--quark phase transition. With~three more parameters introduced in the CSS model, the~uncertainty of the high-density symmetry energy becomes larger, similar to the SNM EOS. 
Nevertheless, the~symmetry energy extracted is about the same at densities less than about $2\rho_0$, as one expects from the most probable transition density given~above.

% start a new page without indent 4.6cm
%\clearpage
%\end{paracol}
\nointerlineskip
%\begin{paracol}{2}
%\linenumbers
%\switchcolumn
\vspace{-9pt} 

\section{Effects of Symmetry Energy on the Second Component of GW190814 as a Supermassive and Superfast~Pulsar}
\textls[-33]{Recently, the~LIGO/Virgo Collaborations reported the binary merger event of GW190814:} the coalescence of a (22.2--24.3)~$M_\odot$ black hole with a (2.50--2.67)~$M_\odot$ compact object~\cite{LIGOGW190814}. This event generated much excitement and interest partially because: (1) The mass of the secondary falls into the mass gap range ($\sim$2--5~$M_\odot$) where $\sim$5~$M_\odot$ is the smallest observed/predicted mass of black holes (see, e.g.,~\cite{Ozel2010,Farr2011}) and $\sim$2~$M_\odot$ is the largest mass observed/predicted of neutron stars (i.e., 2.14$^{+0.11}_{-0.10}$~$M_\odot$ for PSR J0740+6620~\citep{Mmax} or its recently revised value of 2.03$^{+0.10}_{-0.08}$~$M_{\odot}$~\cite{Farr2020} when analyzed using a population-informed prior). (2) The highly asymmetric mass ratio and large merger rate of the GW190814-type class of binaries are hard to explain. Different from GW170817~\cite{LIGOGW170817}, no tidal effects in the signal and electromagnetic counterpart to the gravitational waves have been measured/identified. Therefore, whether the secondary is a massive neutron star, low-mass black hole, or~an exotic object is still under hot debate. Determining the nature of GW190814's secondary can potentially help identify the mass boundary between neutron star and black hole and update models of stellar evolution (see, e.g.,~\cite{Vattis2020,Zevin2020,Safarzadeh2020}).

\subsection{Is GW190814's Secondary a Superfast and Supermassive Neutron Star or Something Else?}
Many possible mechanisms leading to the GW190814 event and the related natures of its secondary component have been proposed in a flood of interesting papers in the literature recently. It is certainly beyond our knowledge range to review all of these interesting works. To~the best of our knowledge, most analyses indicated that GW190814's secondary is a neutron star, while many other works suggested it as a black hole or an exotic object. For~example, \citet{LIGOGW190814} initially explained it as a black hole with a probability larger than 97\% using the GW170817-informed spectral EOS samples~\cite{LIGO18}. Applying Bayesian analyses, \citet{Essick2020} and \citet{Tews2021} found that GW190814's secondary is a binary black hole merger with a probability of $> 94\%$ and $> 99.9\%$, respectively. Moreover, \citet{Fattoyev2020,Das2020} found that the requirement of a very stiff EOS to support 2.5~$M_\odot$ neutron stars is inconsistent with either constraints obtained from analyzing energetic heavy-ion collisions or the low deformability of medium-mass neutron stars. While \citet{Sedrakian2020,LiJJ2020} considered the $\Delta$-resonance and hyperons, they found that their results were inconsistent with a stellar nature interpretation of GW190814's secondary, implying that this event involved two black holes rather than a neutron star and a black~hole.

On the other hand, the~possibility for GW190814's secondary as a neutron star can be accomplished by: (1) choosing/constructing stiff EOSs having a maximum mass larger than 2.5~$M_{\odot}$~\cite{Huang20,Biswal2020,Bombaci2020,Goncalves2020, Roupas2020,Tan2020,Godzieba2021,Lim2021,Lourenco2021, Rather2021,Wu2021}; (2) considering the effects of fast rotations, which can increase the maximum mass by about 20\% when a star rotates at the Kepler frequency (the maximum frequency at which the gravitational attraction is still sufficient to keep matter bound to the pulsar surface)~\cite{Most2020,Zhang20b,LiJJ2020,Lim2021,Sedrakian2020,Tsokaros2020,Demircik2021,Dexheimer2021,Khadkikar2021,Riahi2021,Zhou21,Stone21}; (3) considering other effects/models that can modify the maximum mass of a neutron star, such as the magnetic field~\cite{Rather2021}, twin star~\cite{Christian2021}, two families of compact stars~\cite{Bombaci2020}, finite temperature~\cite{Khadkikar2021}, antikaon condensation~\cite{Thapa2020},~net electric charge~\cite{Goncalves2020}, etc.

Since fast rotation is among the easiest mechanisms to increase the masses of NSs, rotational effects have been considered in several model frameworks for investigating the nature of GW190814's secondary. For~example, adopting a maximum mass M$_{\rm TOV}=2.3$~$M_{\odot}$ for non-rotating NSs, \citet{Most2020} found that GW190814's secondary does not need to be an {ab initio} black hole, nor an exotic object. Rather, it can be a fast pulsar collapsed to a rotating BH at some point before the merger. \citet{Riahi2021} calculated the maximum mass at Kepler frequency with four DDRMF EOSs. Two of them can support pulsars heavier than 2.5~$M_{\odot}$ after considering rotation.
\citet{Dexheimer2021} discussed rotating hybrid stars within a Chiral Mean Field (CMF) model. They can generate stellar masses that approach, and~in some cases surpass, 2.5~$M_{\odot}$. It was shown that in such cases, fast rotation does not necessarily suppress exotic degrees of freedom due to changes in stellar central density, but~requires a larger amount of baryons than what is allowed in the non-rotating stars. This is not the case for pure quark stars, which can easily reach 2.5~$M_{\odot}$ and still possess approximately the same amount of baryons as stable non-rotating stars. \citet{Khadkikar2021} selected 11 EOSs from relativistic density functional theories, Skyrme functionals, and~an empirical extension of a variational microscopic model. Hyperons were also included in their discussions. It was shown that rotation can easily increase the maximum mass to larger than 2.5~$M_\odot$. {On the other hand, \citet{Tsokaros2020} showed that one can find EoSs that both satisfy the constraints of GW170817 and at the same time imply that GW190814's secondary component is a non-rotating NS.}

Among the studies supporting GW190814's secondary component as a supermassive and superfast pulsar, the~required rotational frequencies were found to be rather high. In~fact, most of the studies found that the minimum frequencies are significantly higher than the fastest known
pulsar PSR J1748-2446ad with a frequency of 716 Hz~\cite{2006Sci}, thus making GW190814's secondary the most massive and fastest known pulsar if its nature is confirmed. For~example, \citet{Most2020} found that a rotation
frequency of 1210~Hz is needed assuming GW190814's secondary has a typical radius of 12.5 km and a \mbox{$M_{\rm TOV}$ = 2.08~$M_{\odot}$} when it is not rotating. \citet{Zhang20b} found a minimum rotation frequency of 971 Hz and an equatorial radius of 11.9 km using a model EOS that predicts a~$M_{\rm TOV}$ = 2.39$M_{\odot}$. More recently, \citet{Bis2020} derived a minimum frequency of $1143^{+194}_{-155}$ Hz and an equatorial radius $R_e = 15.7^{+1.0}_{-1.7}$ km at the 90\% confidence level assuming~$M_{\rm TOV}$ = 2.14$M_{\odot}$. In~another study, \citet{Demircik2021} investigated a hybrid star based on the APR EOS for hadronic matter and the V-QCD model for quark matter. The~maximum mass can reach about 2.9~$M_\odot$. However, even with the stiff EOS model, high rotation frequencies $\geq$ 1 kHz are required to reach 2.5~$M_\odot$.

\subsection{Is GW190814's Secondary r-Mode Stable If It Is Really a Superfast Pulsar?}
It is well known that fast pulsars could be r-model unstable (see, e.g.,~\cite{r-review} for a review), leading to the exponential growth of Gravitational Wave (GW) emission through the Chandrasekhar--Friedman--Schutz mechanism~\cite{1970PhRvL..24..611C,1978ApJ...222..281F} if the GW growth rate is faster than its damping rate. It is known that a rigid crust provides the strongest r-mode damping, and it can well explain the stability of all observed Low-Mass X-ray Binaries (LMXBs)~\cite{Zhou21}, 
albeit that some people may consider this extreme mechanism unrealistic.

\begin{figure}[H]
%\widefigure
\includegraphics[width=.5\linewidth,angle=0]{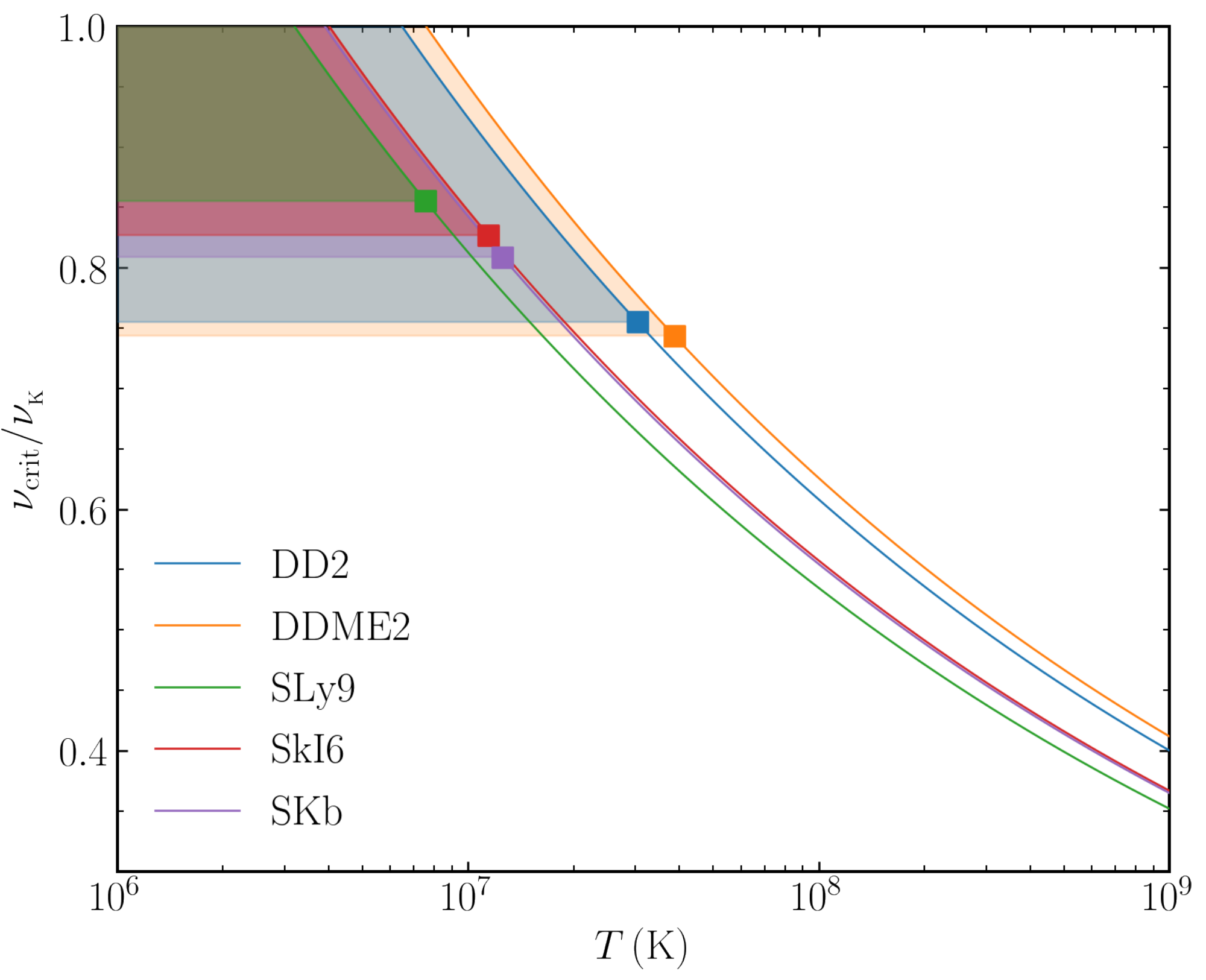}
\vspace{-0.2cm}
\caption{The minimum frequency $\nu_{\rm min}$ to support a neutron star of mass $2.50~M_{\odot}$ is plotted as a horizontal line in the frequency--temperature plane where the r-mode stability boundary for each EOS is also shown. Their cross point is marked with a diamond indicating the maximum temperatures $T_{\rm max}$ below which neutron stars remain r-mode stable. Taken from~\cite{Zhou21}.}
%MDPI: Please make sure that permission has been obtained and there is no copyright issue.
\label{ANG}
\end{figure}

Is GW190814's secondary r-mode stable? This question was recently addressed by \citet{Zhou21} using the six EOSs shown in Figure~\ref{Zhou} that meet all current constraints from both astrophysics and nuclear physics and the rigid crust damping mechanism. It was found that five of them can support pulsars with masses higher than 2.5~$M_\odot$ if they rotate faster than the minimum frequencies $\nu_{\rm min}$ listed in Table~\ref{ANG-table}. Shown also in the table are the maximum mass of non-rotating neutron stars~$M_{\rm TOV}$ and Kepler frequency $\nu_{\rm K}$ for the \mbox{five EOSs} used.~$\nu_{\rm min}$ is plotted as a horizontal line in Figure~\ref{ANG}, where the r-mode stability boundary for each EOS is shown in the frequency--temperature plane. Their cross point is marked with a diamond, indicating the maximum temperatures $T_{\rm max}$ below which neutron stars remain r-mode stable. The~values of $T_{\rm max}$ are listed as the fifth column of Table~\ref{ANG-table}.
It is seen that GW190814's secondary is r-mode stable as long as its temperature is sufficiently low, e.g.,~lower than about $3.9\times 10^7$ K when it is rotating at 1169.6 Hz (0.744 times its Kepler frequency). Because~this temperature is about an order of magnitude higher than that of some known old neutron stars, it was thus concluded that GW190814's secondary can be r-mode stable depending on its temperature~\cite{Zhou21}.

\begin{table}%[tb]
\caption{The maximum mass of non-rotating neutron stars $M_{\rm TOV}$, Kepler frequency $\nu_{\rm K}$, and the minimum frequency $\nu_{\rm min}$ to support a neutron star of mass $2.50 M_{\odot}$ for the 5 EOSs used. $T_{\rm max}$ is the maximum temperature for the $m_2$ to remain r-mode stable.
Taken from Reference \cite{Zhou21}.}\label{ANG-table}
\setlength{\tabcolsep}{1.1pt}
\renewcommand\arraystretch{0.95}
%\begin{ruledtabular}
\center\begin{tabular}{ccccc}
\hline
Model&$M_{\rm TOV}$&$\nu_{\rm K}$ & ~~$\nu_{\rm min}$ & $T_{\rm max}$\\
&$(\Msun)$&(Hz)& ($\nu_{\rm K}$) & ($10^7$ K) \\
 \hline\hline
DD2  & $2.42$  & $1197$  &~~$0.76$ &$3.0$\\
DDME2 & $2.48$ & $1170$  &~~$0.74$ & $3.9$\\
SKb  &$2.20$  &$1447$  &~~$0.81$ & $1.3$\\
SkI6 & $2.20$  & $1433$  &~~$0.83$ & $1.1$\\
SLy9 & $2.16$  & $1515$  &~~$0.86$ & $0.75$\\
\hline\hline
\end{tabular}
\label{tab:vmin}
%\end{ruledtabular}
\end{table}

\vspace{-0.2cm}

\subsection{Effects of Symmetry Energy on the Mass, Radius, and~Minimum Rotation Frequency of GW190814's Secondary Component as a Supermassive and Superfast~Pulsar}

\begin{figure}[H]
%\widefigure
\includegraphics[width=.98\linewidth,angle=0]{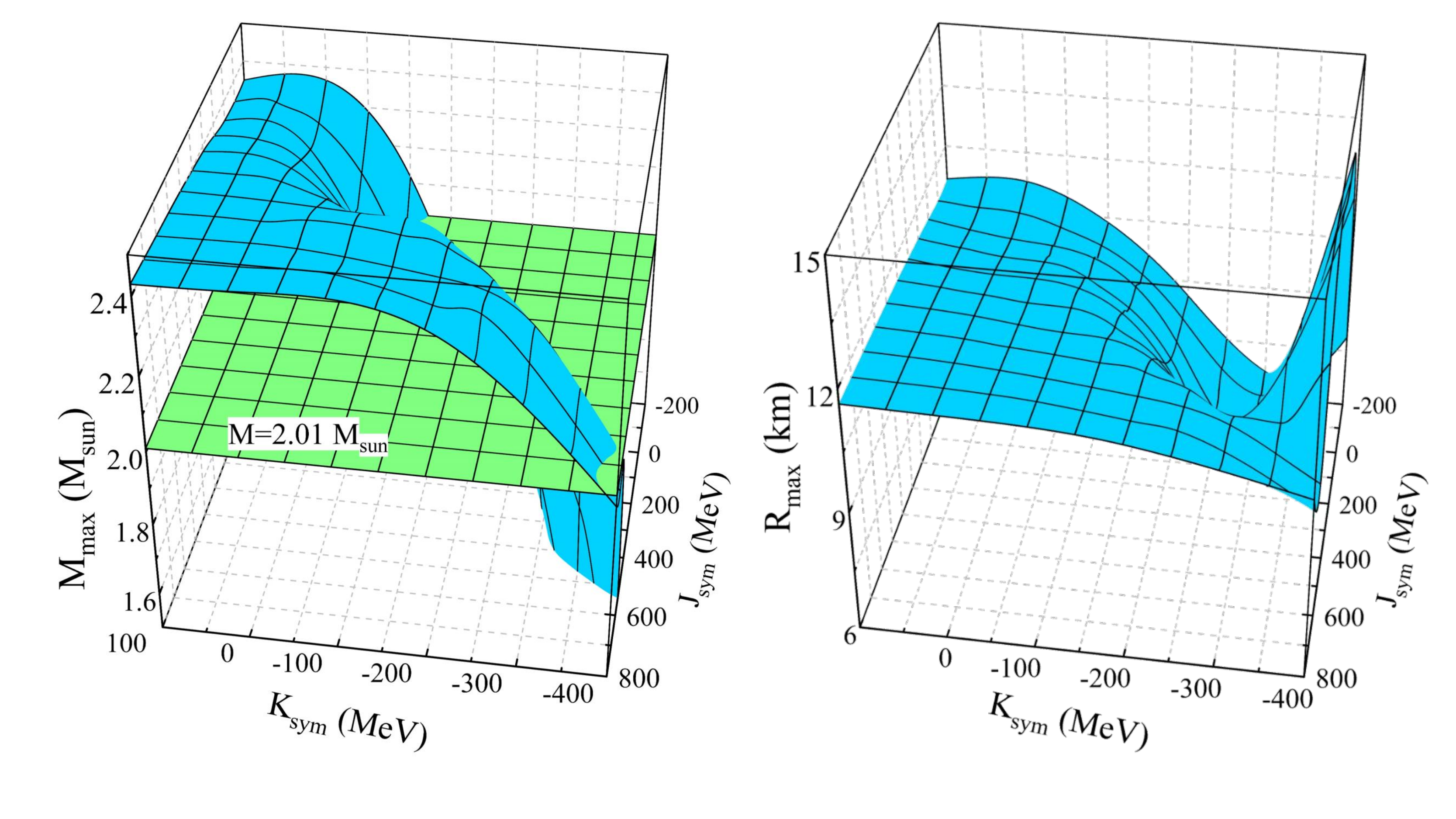}
%MDPI: Please change hyphen to minus sign in the image.
\caption{\textls[-15]The maximum masses~$M_{\rm TOV}$ (left) and corresponding radii (right) of non-rotating NSs on the causality surface as functions of the curvature and skewness of nuclear symmetry energy. Taken from~\cite{Zhang19epj}.}
%MDPI: Please make sure that permission has been obtained and there is no copyright issue.
%MDPI: please add the explanation of the left and right parts. 
\label{MR-max}
\end{figure}

Why is the high-density nuclear symmetry energy important for determining the nature of GW190814's secondary component besides its know effects on the structure of neutron stars as we discussed earlier? Basically, the~answer lies in the fact that the maximum mass of non-rotating neutron stars~$M_{\rm TOV}$~\cite{Zhang19epj}, the~Kepler frequency $\nu_{\rm K}$~\cite{Krastev07,Worley08}, and~the r-mode stability boundaries~\cite{Wen11,Vidana12} are all known to depend sensitively on the high-density behavior of nuclear symmetry energy; for~a review, see, e.g.,~\cite{BALI19}. For~example, shown in the left window of Figure~\ref{MR-max}
are the maximum masses~$M_{\rm TOV}$ while that in the right window are the corresponding radii $R_{\rm{max}}$ of non-rotating NSs. {They were obtained by solving TOV equations for the EOS parameter sets on the causality surface (the blue surface in the left panel of \mbox{Figure~\ref{Constraints}}) as functions of the curvature $K_{\rm sym}$ and skewness $J_{\rm sym}$ of nuclear symmetry energy. All other EOS parameters used to calculate the~$M_{\rm TOV}$ and $R_{\rm{max}}$ surfaces are the same as those used in Figure~\ref{Constraints}, i.e.,~$E_{\rm{sym}}(\rho_0)=31.6$ MeV and $L=58.9$ MeV.} 
The~$M_{\rm TOV}$ on the causality surface represents the absolutely maximum mass allowed for non-rotating neutron stars. For~a comparison, the~2.01~$M_{\odot}$ mass plane is also shown in the left window. {The space below this surface \mbox{is excluded}.}

As we discussed earlier, large positive (negative) $K_{\rm{sym}}$ and $J_{\rm{sym}}$ represent stiff (soft) high-density symmetry energies. Correspondingly, the~interiors of neutron stars are less (more) neutron-rich from the energy consideration or due to the so-called isospin fractionation~\cite{BALI19,LCK08}. Thus, for~the stiff high-density symmetry energy, the~pressure is dominated by that of SNM EOS. It is thus seen that~$M_{\rm TOV}$ flattens towards an asymptotic value of about 2.39~$M_{\odot}$, determined mostly by the $J_0$ of SNM EOS when the high-density symmetry energy becomes very stiff. The~corresponding radius reaches a constant of about 12 km. On~the other hand, when the high-density symmetry energy becomes very soft, the~interior of neutron stars becomes very neutron-rich. Then, the contribution to the nuclear pressure from the high-density symmetry energy becomes more important, but~can be even negative, thus reducing the total nuclear pressure. Consequently,~$M_{\rm TOV}$ becomes even smaller than 2.01~$M_{\odot}$. The~corresponding radius also becomes~smaller.

\begin{figure}[H]
%\widefigure
\includegraphics[width=.98\linewidth]{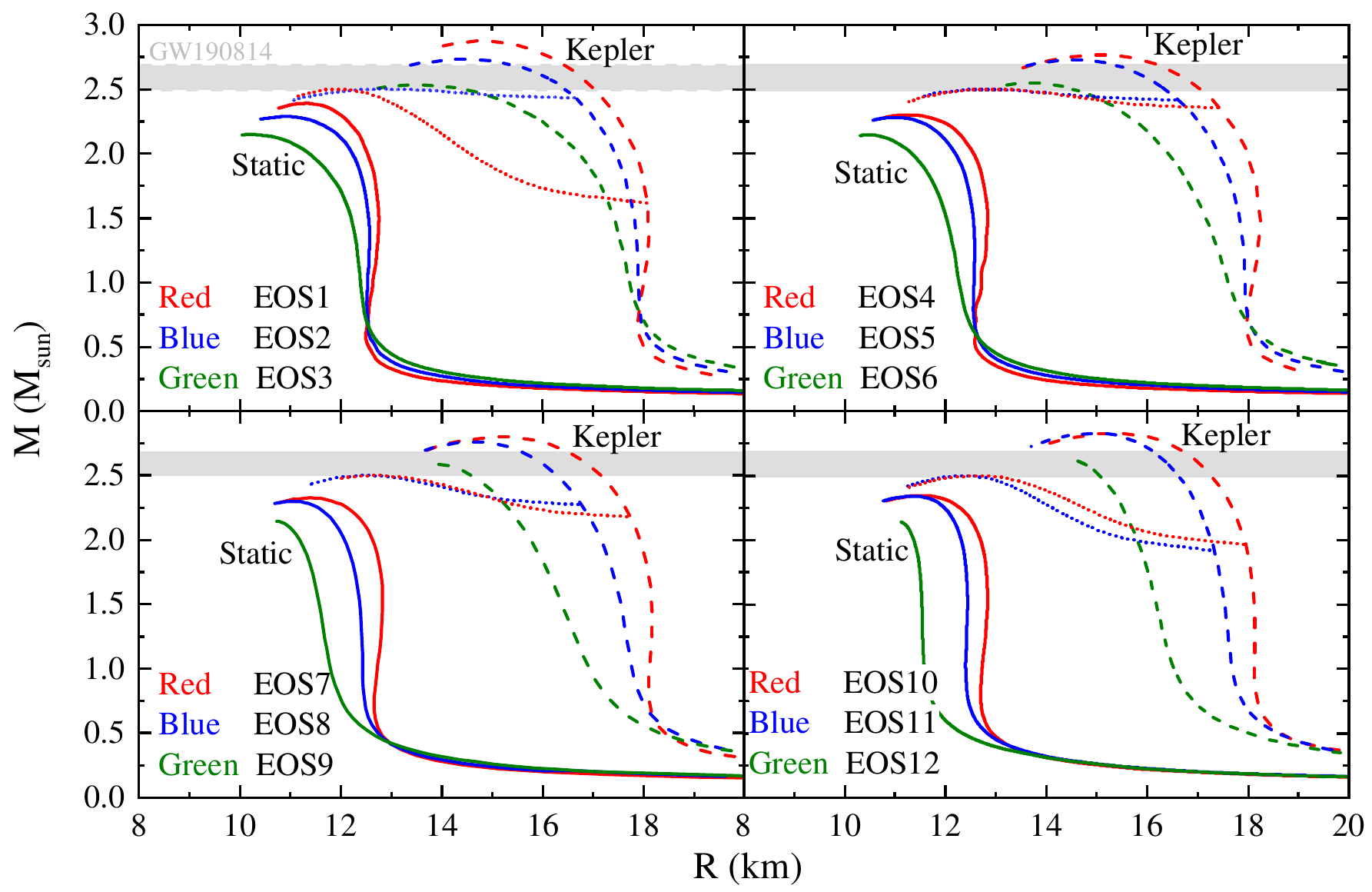}
\caption{Mass--radius relations of both static (solid lines) and rotating neutron stars with the selected EOSs along the symmetry energy boundaries shown in the right window of Figure~\ref{Constraints}. {The NSs rotating at the Kepler frequencies and minimum frequency that can rotationally support a NS with mass 2.5$M_\odot$ are shown as dashed and dotted lines, respectively.} Taken from~\cite{Zhang20b}.}
%MDPI: Please make sure that permission has been obtained and there is no copyright issue.
\label{MR-rotation}
\end{figure}

Depending on what~$M_{\rm TOV}$ one uses for non-rotating neutron stars, the~minimum frequency necessary to rotationally support neutron stars heavier than 2.50~$M_{\odot}$ will thus be different. As~an example, shown in Figure~\ref{MR-rotation} are the mass--radius relations of both static (solid curves) and rotating neutron stars at Kepler frequencies (dashed curves) with the 12 selected EOSs along and/or inside the symmetry energy boundaries shown in the right window of Figure~\ref{Constraints}. We note that~$M_{\rm TOV}$ on the bounded causality surface is between 2.14 and 2.39~$M_\odot$. Neutron stars rotating at the minimum frequency $f_{2.5}$ that can rotationally support a neutron star with mass $2.50$~$M_\odot$ are shown with dotted curves. The~reported mass 2.50--2.67~$M_\odot$ of GW190814's secondary component is shown as gray~bands.

It is interesting to see that while the M$_{\rm TOV}$ of the 12 EOSs are between 2.14 and 2.39~$M_\odot$, pulsars at their respective Kepler frequencies can easily sustain masses heavier than $2.50$~$M_\odot$. For~example, with~the stiffest EOS possible (EOS1 with~$M_{\rm TOV}=2.39$~$M_\odot$), the~maximum rotating mass is 2.87~$M_\odot$, while with the soft EOSs including EOS3, EOS6, EOS9, and~EOS12 on the right boundary of the allowed EOS space shown in \mbox{Figure~\ref{Constraints}}, the~maximum rotating masses are slightly higher than 2.5~$M_\odot$, but less than 2.67~$M_\odot$. {As the maximum mass of these EOSs is close to 2.5~$M_\odot$,~$f_{2.5}$ is slightly lower than the Kepler frequency; thus, the RNS code~\cite{RNS} does not output the $f_{2.5}$ for EOS3, EOS6, EOS9, and~EOS12.} 

\begin{figure}[H]
%\widefigure
\includegraphics[width=.98\linewidth,angle=0]{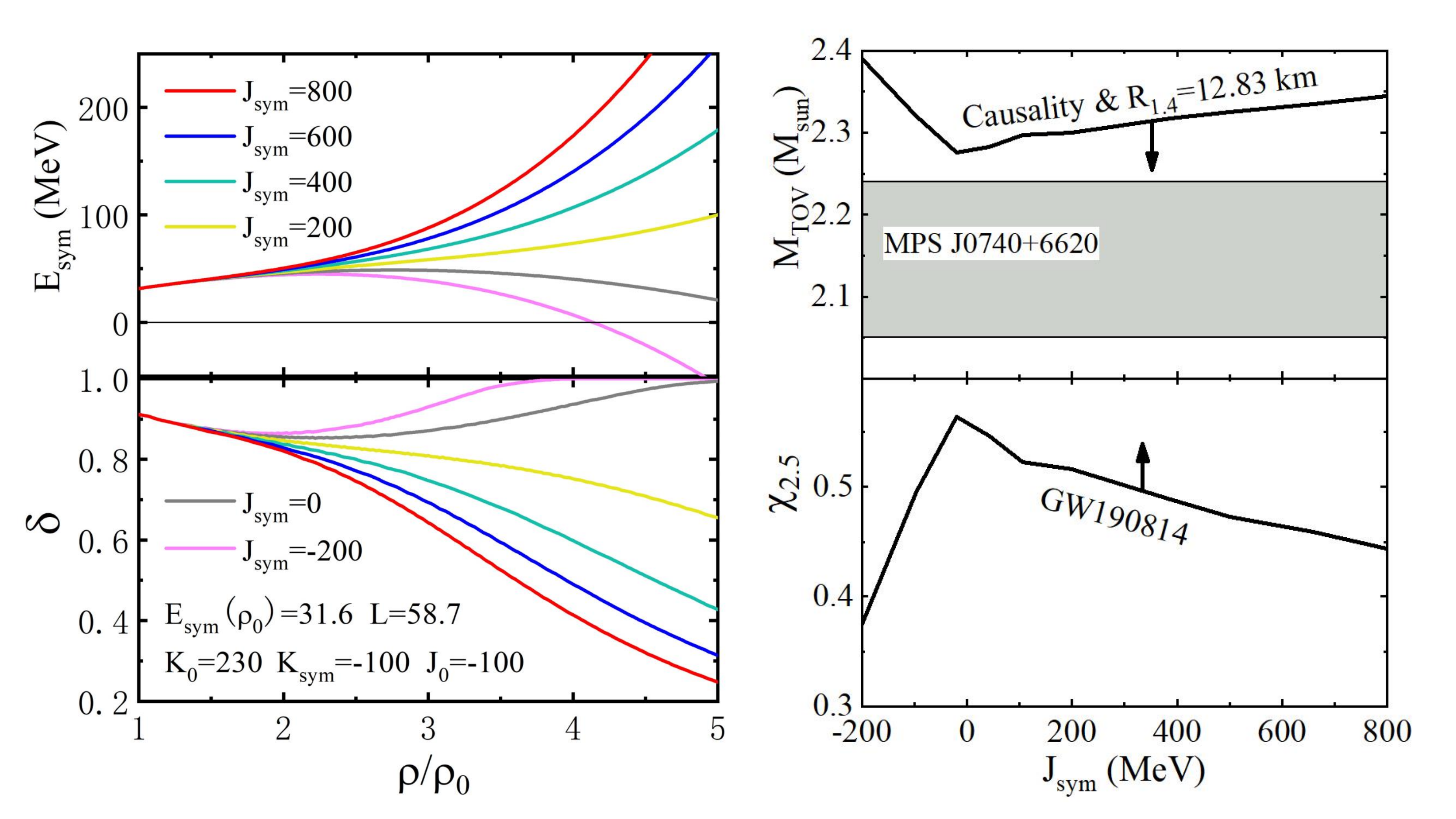}
%MDPI: Please change hyphen to minus sign in the image.
\caption{({Left}) The density dependence of nuclear symmetry energy (upper) and the corresponding isospin asymmetry (lower) in neutron stars at $\beta$ equilibrium with different skewness $J_{\rm sym}$ parameters. ({Right}) The maximum mass~$M_{\rm TOV}$ of non-rotating NSs (upper window) and the minimum spin parameter $\chi_{2.5}$ of pulsars with the frequency $f_{2.5}$ (lower window) as functions of the skewness parameter of nuclear symmetry energy parameter. The~{previously} observed NS maximum mass $M=2.14^{+0.10}_{-0.09}$~$M_\odot$ (68\% confidence level) of MSR J0740+6620 is shown in the upper panel. The~arrows indicate the conditions for GW190814's secondary to be a superfast pulsar. Taken from~\cite{Zhang20b}.}
%MDPI: Please make sure that permission has been obtained and there is no copyright issue.
\label{Mchi25}
\end{figure}

One can measure the minimum frequency necessary to support the pulsar with mass 2.50~$M_\odot$ with the dimensionless spin parameter $\chi_{2.5}=J/M^2$ where $J$ is the angular momentum of the pulsar of mass M = 2.50~$M_\odot$.
Shown in the left window of Figure~\ref{Mchi25} is the density dependence of nuclear symmetry energy (upper) with different skewness $J_{\rm sym}$ parameters and the corresponding isospin asymmetry $\delta$ of neutron star matter at $\beta$-equilibrium (lower). The~latter clearly shows how the high-density symmetry energy affects the composition of neutron stars at $\beta$-equilibrium, i.e.,~their neutron~richness.

{As discussed earlier, the~upper limit of nuclear symmetry is determined by the crossline between the causality surface and the $R_{1.4}=12.83$ km surface (or the nearby $\Lambda_{1.4}=580$ surface) before NICER's observation for the radius of PSR J0740+6620. This crossline was projected to the $K_{\rm sym}-J_{\rm sym}$ plane and shown as the left boundary in the right panel of Figure~\ref{Constraints}. In~Figure~\ref{MR-rotation}, only four selected EOSs (EOS1, EOS4, EOS7, and~EOS10) along this boundary are used. To~show the~$M_{\rm TOV}$ of non-rotating NSs along this boundary more clearly,~$M_{\rm TOV}$ as a continuous function of $J_{\rm sym}$ is now shown in the upper-right window of Figure~\ref{Constraints}.}
For a comparison, the~{previously reported} mass $M=2.14^{+0.10}_{-0.09}$~$M_\odot$ of PSR J0740+6620 is also shown in the upper panel. The~resulting EOSs of neutron star matter along the boundary discussed above are the stiffest while being consistent with all known astrophysical and nuclear constraints. The~resulting minimum spin parameter $\chi_{2.5}$ is shown in the lower right window as a function of the skewness parameter $J_{\rm sym}$. The~arrows indicate the conditions for GW190814's secondary to be a superfast pulsar. Clearly, as~we mentioned earlier, the~minimum spin parameter $\chi_{2.5}$ depends on~$M_{\rm TOV}$, and~both of them depend on $J_{\rm sym}$ characterizing the high-density behavior of nuclear symmetry~energy.

\section{An Auxiliary Function Approach for Predicting the High-Density Behavior of Nuclear Symmetry Energy Based on Its Slope \boldmath{$L$, Curvature $K_{\textbf{sym}}$, and Skewness \mbox{$J_{\textbf{sym}}$ at $\rho_0$}}}
\def\Ts{\Theta_{\rm{sym}}}
\def\Ps{\Pi_{\rm{sym}}(\chi,\Ts)}
\def\Es{E_{\rm{sym}}}
\def\Pso{\Pi_{\rm{sym}}(0,\Ts)}

Assuming now that the community has finally reached a consensus about the characteristics of nuclear symmetry energy at $\rho_0$, i.e., $L,K_{\rm{sym}}$, and~$J_{\rm{sym}}$ are all well determined using various approaches, how can we use these quantities to predict the symmetry energy at suprasaturation densities? While one may use the same theory/model used in extracting these parameters at $\rho_0$ to predict the high-density symmetry energy, it is well known that most of the theories/models are unreliable at high densities above about $2\rho_0$. One may also try to extrapolate the symmetry energy at $\rho_0$ to high densities by using the conventional Taylor expansion, namely $E_{\rm{sym}}(\rho)\approx S+L\chi+2^{-1}K_{\rm{sym}}\chi^2+6^{-1}J_{\rm{sym}}\chi^3+\cdots$. Unfortunately, the~latter breaks down eventually as the density $\rho$ increases above the certainty limit because $\chi$ is not always small enough for small-quantity expansions to work properly. For example, the~RMF model with the FSUGold parameters predicts exactly the symmetry energy $E_{\rm{sym}}(\rho)$ as a function of density in a broad density range. One can extract from the prediction the characteristics $L, K_{\rm{sym}}$, and~$J_{\rm{sym}}$ of the symmetry energy at $\rho_0$. It was shown quantitatively in~\cite{CL21} that using these characteristics in the standard Taylor expansion up to the $\chi^3$ term, the~resulting symmetry energy at densities above about $2\rho_0$ deviates significantly from the exact $E_{\rm{sym}}(\rho)$ function predicted by the RMF model, indicating the non-convergence of the Taylor expansion at high densities. Moreover, as~we shall show, %MDPI: Figures should be cited in sequential numerical order. Please confirm if it’s invalid citation, if no, please revise the order.
Taylor expansions up to the $\chi^2$ or $\chi^3$ term give significantly different predictions for $E_{\rm{sym}}(\rho)$ at densities above $2\rho_0$, indicating again the non-convergence of the Taylor expansion at high densities.

Therefore, is there a better way to predict the symmetry energy at $2-3\rho_0$ using its characteristics at $\rho_0$? This question was recently studied in~\cite{CL21} using an auxiliary function approach. Here, we summarize the main idea and~results.

\subsection{Theoretical~Framework}
To predict accurately the symmetry energy at suprasaturation densities based on its known characteristics at $\rho_0$, such as those we surveyed in Section~\ref{ASTR}, the~first task is to find an appropriate auxiliary function that naturally reproduces the first several terms given by the conventional expansion when it is expanded around $\chi=0$. Secondly, certain higher $\chi$-order contributions should also be effectively encapsulated in the auxiliary function using still only the first few characteristics of $E_{\rm{sym}}(\rho)$ at $\rho_0$, i.e.,~$L, K_{\rm{sym}}$, and~$J_{\rm{sym}}$. Mathematically, this was performed by introducing the auxiliary function $\Pi_{\rm{sym}}(\chi,\Ts)$, which itself is a function of the density $\rho$ (or equivalently of the $\chi$) and depends on a new parameter $\Ts$. In~expanding the symmetry energy, one has the following replacement,
\begin{equation}
\frac{\d^nE_{\rm{sym}}}{\d\rho^n}(\rho-\rho_0)^n\rightarrow
\frac{\d^nE_{\rm{sym}}}{\d\Pi_{\rm{sym}}^n}\widetilde{\nu}_{\rm{sym}}^n(\chi,\Ts),\end{equation}
\textls[-15]{where
$\widetilde{\nu}_{\rm{sym}}(\chi,\Ts)=\Pi_{\rm{sym}}(\chi,\Ts)-\Pi_{\rm{sym}}(0,\Ts)$ corresponds to the dimensionless quantity $\chi$.
Once a model $\Pi_{\rm{sym}}(\chi,\Theta_{\rm{sym}})$ is adopted/selected, the~parameter $\Theta_{\rm{sym}}$ can be determined by the symmetry energy at a density where it is well determined~experimentally.}

Given the four characteristic parameters $S\equiv E_{\rm{sym}}(\rho_{0}), L, K_{\rm{sym}}$, and~$J_{\rm{sym}}$ of $E_{\rm{sym}}(\rho)$ at $\rho_0$, the~symmetry energy $E_{\rm{sym}}(\rho)$ can be expanded around $\Pi_{\rm{sym}}(\chi,\Theta_{\rm{sym}})=\Pi_{\rm{sym}}(0,\Theta_{\rm{sym}})$ to order $\nu_{\rm{sym}}^3(\chi,\Ts)$ as:
\begin{align}
E_{\rm{sym}}(\rho)\approx&S+L\nu_{\rm{sym}}(\chi,\Theta_{\rm{sym}})
+\frac{1}{2}K_{\rm{sym}}\Phi\nu^2_{\rm{sym}}(\chi,\Theta_{\rm{sym}})
+\frac{1}{6}J_{\rm{sym}}\Psi\nu^3_{\rm{sym}}(\chi,\Theta_{\rm{sym}}),\label{EXPEsymNEW}
\end{align}
where:
\begin{align}
\Phi=&1
+\left.\frac{L}{K_{\rm{sym}}}\left({\displaystyle\frac{1}{3\rho}\frac{\partial^2\rho}{\partial\Pi_{\rm{sym}}^2}}\right)\right/{\displaystyle\left(\frac{1}{3\rho}\frac{\partial\rho}{\partial\Pi_{\rm{sym}}}\right)^2}_{\chi=0},\label{def_PHI}\\
\Psi=&1+\left.\frac{K_{\rm{sym}}}{J_{\rm{sym}}}
\left({\displaystyle\frac{1}{3\rho^2}\frac{\partial\rho}{\partial\Pi_{\rm{sym}}}\frac{\partial^2\rho}{\partial\Pi_{\rm{sym}}^2}}\right)\right/{\displaystyle\left(\frac{1}{3\rho}\frac{\partial\rho}{\partial\Pi_{\rm{sym}}}\right)^3}_{\chi=0}\notag\\
&+\left.\frac{L}{J_{\rm{sym}}}\left({\displaystyle\frac{1}{3\rho}\frac{\partial^3\rho}{\partial\Pi_{\rm{sym}}^3}}\right)\right/{\displaystyle\left(\frac{1}{3\rho}\frac{\partial\rho}{\partial\Pi_{\rm{sym}}}\right)^3}_{\chi=0},\label{def_PSI}
\end{align}
and,
\begin{equation}\label{def_nusymkk}
\nu_{\rm{sym}}(\chi,\Theta_{\rm{sym}})\equiv \left[\frac{1}{3\rho}\frac{\partial\rho}{\partial\Pi_{\rm{sym}}(\chi,\Theta_{\rm{sym}})}\right]_{\chi=0}\cdot\widetilde{\nu}_{\rm{sym}}(\chi,\Theta_{\rm{sym}}),
\end{equation}

It was shown that the conventional Taylor expansion corresponds to the special case of selecting $\Pi_{\rm{sym}}(\chi,\Ts)\propto \chi$.
Moreover, terms higher than $\chi^3$ are effectively included in Equation~(\ref{EXPEsymNEW}), although it is truncated at order $\nu_{\rm{sym}}^3(\chi,\Ts)$, since the latter itself encapsulates the higher order effects in $\chi$.

In\,\cite{CL21}, an~exponential (abbreviated as ``exp'') and an algebraic (abbreviated as ``alge'') auxiliary function were used. These functions and the corresponding expansions of symmetry energy can be written as:
\vspace{12pt}
\begingroup\makeatletter\def\f@size{8.8}\check@mathfonts
\def\maketag@@@#1{\hbox{\m@th\normalsize\normalfont#1}}%
\begin{align}
\rm{exp}:~~&\nu_{\rm{sym}}(\chi,\Theta_{\rm{sym}})=\frac{1}{3\Theta_{\rm{sym}}}
\left(1-e^{-3\chi\Theta_{\rm{sym}}}\right),\label{nnn-k}\\
&E_{\rm{sym}}(\rho)\approx S+L\nu_{\rm{sym}}(\chi,\Theta_{\rm{sym}})
+\frac{1}{2}K_{\rm{sym}}\left(1+\frac{3L}{K_{\rm{sym}}}\Ts\right)\nu_{\rm{sym}}^2(\chi,\Theta_{\rm{sym}})\notag\\
&\hspace{2cm}+\frac{1}{6}J_{\rm{sym}}\left(1+\frac{9K_{\rm{sym}}}{J_{\rm{sym}}}\Ts
+\frac{18L}{J_{\rm{sym}}}\Ts^2
\right)\nu_{\rm{sym}}^3(\chi,\Theta_{\rm{sym}})\label{tt1},\\
\rm{alge}:~~&\nu_{\rm{sym}}(\chi,\Theta_{\rm{sym}})=\chi\frac{1+\Theta^{-1}_{\rm{sym}}}{1+3\chi+\Theta^{-1}_{\rm{sym}}},\label{nnn-1}\\
&E_{\rm{sym}}(\rho)\approx S+L\nu_{\rm{sym}}(\chi,\Theta_{\rm{sym}})
+\frac{1}{2}K_{\rm{sym}}\left(1+\frac{6L}{K_{\rm{sym}}}\frac{1}{1+\Theta^{-1}_{\rm{sym}}}\right)\nu_{\rm{sym}}^2(\chi,\Theta_{\rm{sym}})\notag\\
&\hspace{2cm}+\frac{1}{6}J_{\rm{sym}}\left[1+\frac{18K_{\rm{sym}}}{J_{\rm{sym}}}\frac{1}{1+\Theta^{-1}_{\rm{sym}}}+\frac{54L}{J_{\rm{sym}}}\left(\frac{1}{1+\Theta^{-1}_{\rm{sym}}}\right)^2
\right]\nu_{\rm{sym}}^3(\chi,\Theta_{\rm{sym}})\label{tt2}.
\end{align}
\endgroup

\begin{figure}[H]
%\widefigure
\includegraphics[width=.9\linewidth,angle=0]{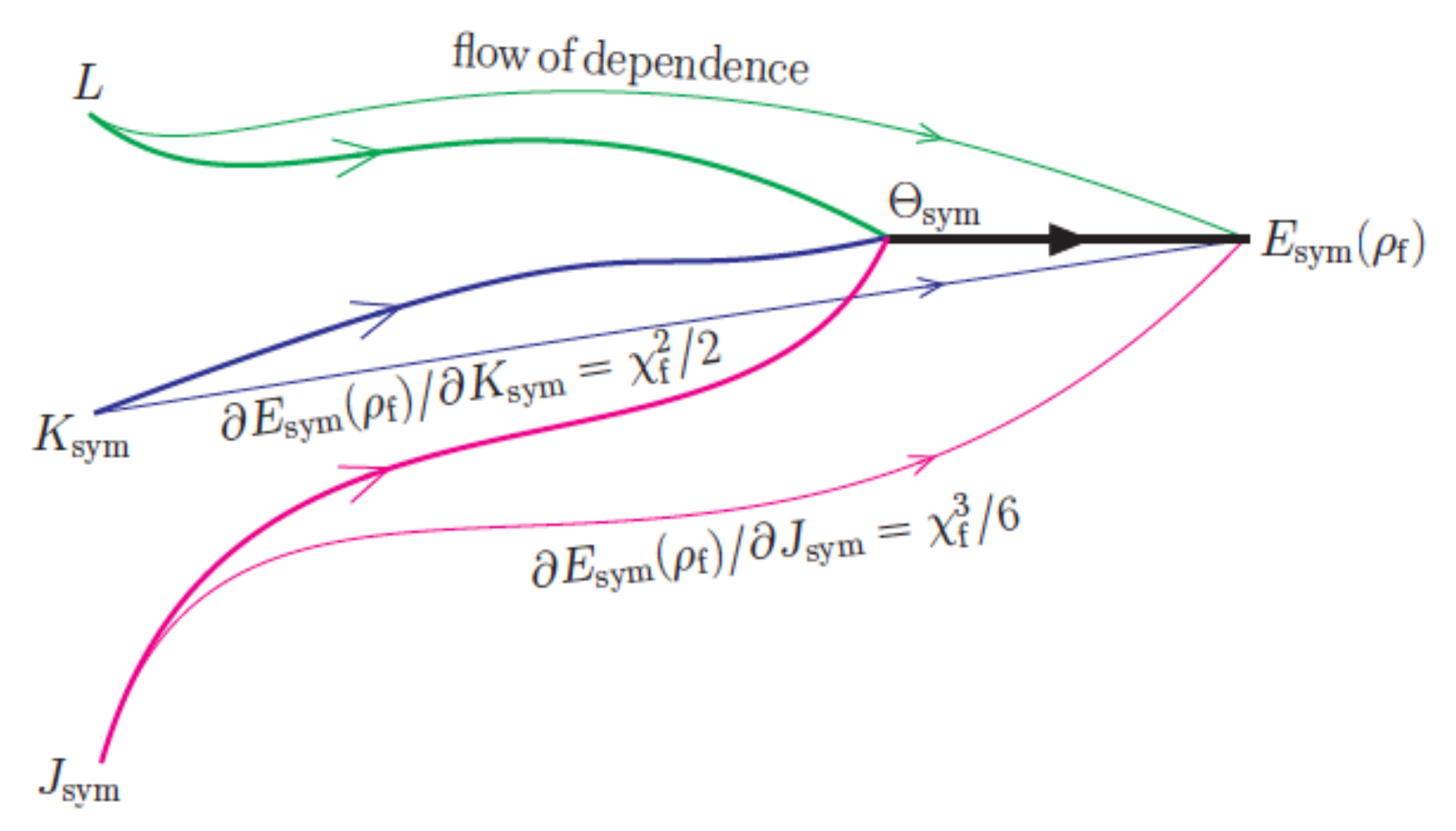}
\caption{Illustration of how the characteristics $L,K_{\rm{sym}}$, and~$J_{\rm{sym}}$ of symmetry energy at $\rho_0$ affect both directly and indirectly through the parameter $\Ts$ the high-density symmetry energy at $\rho_{\rm{f}}$.}
\label{fig_Ts1}
\end{figure}

\textls[-15]{Take the exponential model as an example. Some new features emerge in \mbox{Equation~(\ref{tt1})}}. Besides~the conventional term $2^{-1}K_{\rm{sym}}
$, a~new term $3\Ts L/K_{\rm{sym}}$ (normalized by $2^{-1}K_{\rm{sym}}$) contributes at order $\nu_{\rm{sym}}^2(\chi,\Ts)$. This term is generally sizable and cannot be thought as a perturbation.
For small $\chi$, e.g.,~$\rho_0\lesssim\rho\lesssim3\rho_0$, one has:
\begin{align}\label{dk}
\nu_{\rm{sym}}(\chi,\Ts)\approx&\chi-\frac{3}{2}\Ts\chi^2+\frac{3}{2}\Ts^2\chi^3-\frac{9}{8}\Ts^3\chi^4
+\frac{27}{40}\Ts^4\chi^5-\cdots,
\end{align}
which approaches zero as $\chi\to0$. It is clear that the
effects of $\chi^4$ or $\chi^5$ are effectively generated with the help
of the function $\Pi_{\rm{sym}}(\chi,
\Theta_{\rm{sym}})$. Shown in Figure~\ref{fig_Ts1} is an illustration of how the characteristics $L,K_{\rm{sym}}$, and~$J_{\rm{sym}}$ of symmetry energy at $\rho_0$ affect both directly and indirectly through the parameter $\Ts$ the high-density symmetry energy at density $\rho_{\rm{f}}$.
The flow indicates the dependence. The~conventional dependence of $E_{\rm{sym}}(\rho_{\rm{f}})$ on $K_{\rm{sym}}$ and $J_{\rm{sym}}$, namely $\chi_{\rm{f}}^2/2$ and $\chi_{\rm{f}}^3/6$ are indicated, where $\chi_{\rm{f}}=(\rho_{\rm{f}}-\rho_0)/3\rho_0$.

Several possible ways of determining the parameter $\Ts$ were given in~\cite{CL21}. One approach is to fix it by using the experimentally known symmetry energy at some reference density. In~particular, one can use the empirical constraint:
\begin{equation}\label{elow}
E_{\rm{sym}}(\rho_{\rm{low}}\approx0.05~\,\rm{fm}^{-3})\approx16.4\pm0.5~\,\rm{MeV}.
\end{equation}

This is from analyzing both nuclear structures and reactions, e.g.,~isobaric analog states of several nuclei, the~centroid energy of isovector giant dipole resonance, and electrical dipole polarizability of $^{208}$Pb~\cite{Dan14,Zhang15,JXu20}.
%MDPI: Figures should be cited in sequential numerical order. Please confirm if it’s invalid citation, if no, please revise the order.

By adopting the auxiliary-function-based reconstruction, one can easily find, e.g.,~in the exponential model, that the dependence of the symmetry energy at some density $\rho_{\rm{f}}$ on the characteristics $K_{\rm{sym}}$ and $J_{\rm{sym}}$ as,
\begingroup\makeatletter\def\f@size{9.4}\check@mathfonts
\def\maketag@@@#1{\hbox{\m@th\normalsize\normalfont#1}}%
\begin{align}
\frac{\partial E_{\rm{sym}}(\rho_{\rm{f}})}{\partial K_{\rm{sym}}}=&
\frac{1}{2}\nu_{\rm{sym}}^{\rm{f},2}\left(1+3\Ts\nu_{\rm{sym}}^{\rm{f}}\right)
\times\left[1-\left(\frac{\nu_{\rm{sym}}^{\rm{low}}}{\nu_{\rm{sym}}^{\rm{f}}}\right)^2\left(\frac{1+3\Ts\nu_{\rm{sym}}^{\rm{low}}}{1+3\Ts\nu_{\rm{sym}}^{\rm{f}}}\right)\left(\frac{\Upsilon_{\rm{f}}}{\Upsilon_{\rm{low}}}\right)\right]
,\label{Ek}\\
\frac{\partial E_{\rm{sym}}(\rho_{\rm{f}})}{\partial J_{\rm{sym}}}=&\frac{1}{6}\nu_{\rm{sym}}^{\rm{f},3}\times
\left[1-\left(\frac{\nu_{\rm{sym}}^{\rm{low}}}{\nu_{\rm{sym}}^{\rm{f}}}\right)^3\left(\frac{\Upsilon_{\rm{f}}}{\Upsilon_{\rm{low}}}\right)\right],\label{Ej}
\end{align}
\endgroup
where the superscripts/subscripts ``f'' and ``low'' are for $\rho_{\rm{f}}$ and $\rho_{\rm{low}}\approx0.05\,\rm{fm}^{-3}$, respectively.
The function $\Upsilon(\rho)$ in the above two equations is given by:
\begingroup\makeatletter\def\f@size{9}\check@mathfonts
\def\maketag@@@#1{\hbox{\m@th\normalsize\normalfont#1}}%
\begin{align}
\Upsilon(\rho)
=&\frac{3}{2}L\nu_{\rm{sym}}^2+\frac{3}{2}\left(K_{\rm{sym}}+4L\Ts\right)\nu_{\rm{sym}}^3\notag\\
&+\frac{\partial\nu_{\rm{sym}}}{\partial\Ts}\left[L+\left(K_{\rm{sym}}+3L\Ts\right)\nu_{\rm{sym}}+\frac{1}{2}\left(J_{\rm{sym}}+9K_{\rm{sym}}\Ts+18L\Ts^2\right)\nu_{\rm{sym}}^2
\right].\end{align}
\endgroup

\textls[-11]{The corrections in the square brackets in Equations~(\ref{Ek}) and~(\ref{Ej}) come from the dependence of the $\Ts$ parameter on the curvature $K_{\rm{sym}}$ and the skewness $J_{\rm{sym}}$ of the symmetry energy, i.e.,~$\partial\Ts/\partial K_{\rm{sym}}$ and $\partial\Ts/\partial J_{\rm{sym}}$, as~sketched in Figure\,\ref{fig_Ts1}.
In the conventional expansion, one has:}
\begin{equation}\label{Ekj}
\frac{\partial E_{\rm{sym}}(\rho_{\rm{f}})}{\partial K_{\rm{sym}}}=\frac{1}{2}\chi_{\rm{f}}^2,~~\frac{\partial E_{\rm{sym}}(\rho_{\rm{f}})}{\partial J_{\rm{sym}}}=\frac{1}{6}\chi_{\rm{f}}^3.
\end{equation}

\textls[-15]{By comparing Equation~(\ref{Ek}) with the first relation of Equation~(\ref{Ekj}) or Equation~(\ref{Ej})} with the second relation of Equation~(\ref{Ekj}), one can immediately find that if the new expansion element $\nu_{\rm{sym}}$ is constructed reasonably, the~dependence of $E_{\rm{sym}}(\rho_{\rm{f}})$ on $K_{\rm{sym}}$ or $J_{\rm{sym}}$ at high densities should be reduced, as~compared with the conventional approach, where the expansion element $\chi$ is unbounded from~above.

\subsection{An Example of~Applications}
\begin{figure}[H]
%\widefigure
\includegraphics[width=.93\linewidth]{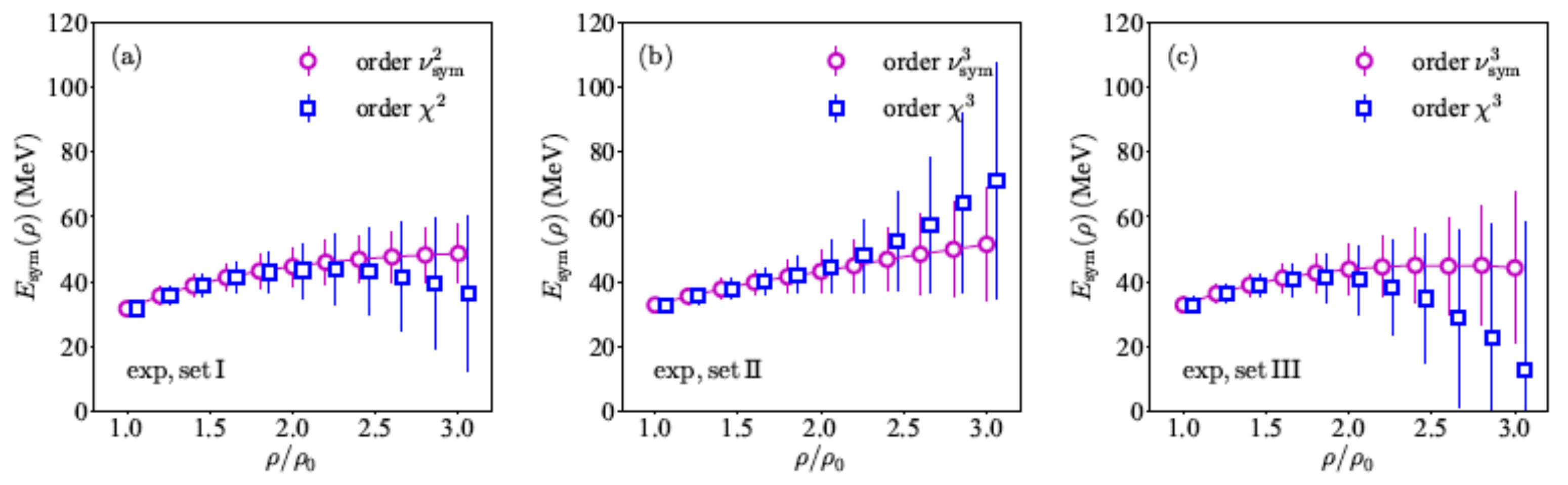}
\caption{Density dependence of nuclear symmetry energy from simulations adopting the exponential model for the auxiliary function (purple) and the conventional Taylor expansion (blue) for Test Set I (\textbf{a}), II (\textbf{b}), and~III (\textbf{c}), respectively. Taken from~\cite{CL21}.}
%MDPI: Please make sure that permission has been obtained and there is no copyright issue.
\label{fig_Esymu-set-ABC}
\end{figure}

\textls[-11]{To illustrate the advantages of the auxiliary function approach over the Taylor expansion in predicting the symmetry energy at high densities,
shown in Figure~\ref{fig_Esymu-set-ABC} are comparisons of the symmetry energies from the two approaches from Monte Carlo simulations performed in\,\cite{CL21}. The~three test sets have different characteristic parameters for $E_{\rm{sym}}(\rho)$ at $\rho_0$:
(I) $E_{\rm{sym}}(\rho)$ is expanded to order $\nu_{\rm{sym}}^2$ or order $\chi^2$ with \mbox{$-300\,\rm{MeV}\leq $ K$_{\rm{sym}}\leq 0\,\rm{MeV}$~\cite{Xie20}};
(II) $E_{\rm{sym}}(\rho)$ is expanded to order $\nu_{\rm{sym}}^3$ or order $\chi^3$ with \mbox{$-300\,\rm{MeV}\leq $ K$_{\rm{sym}}\leq 0\,\rm{MeV}$}, \mbox{$0\,\rm{MeV}\leq $ J$_{\rm{sym}}\leq2000\,\rm{MeV}$};
(III) $E_{\rm{sym}}(\rho)$ is expanded to order $\nu_{\rm{sym}}^3$ or order $\chi^3$ with $K_{\rm{sym}}$ and $J_{\rm{sym}}$ given by the intrinsic relations imposed by the unbound nature of PNM~\cite{CL21-i},}
\begin{align}
K_{\rm{sym}}\approx&K_0\left(1-\frac{1}{3}\frac{K_0}{L}+\frac{1}{2}\frac{J_0}{K_0}\frac{L}{K_0}\right),\label{IC1}\\
J_{\rm{sym}}\approx&\frac{2K_0^3}{3L^2}\left(1-\frac{3L}{K_0}\right)+\frac{I_0L}{3K_0}+\left(\frac{2K_0K_{\rm{sym}}}{L}-J_0\right)
\left(1+\frac{J_0L}{K_0^2}-\frac{K_{\rm{sym}}}{K_0}\right).\label{IC2}
\end{align}
For these demonstrations, the~magnitude $S\approx32\pm4\,\rm{MeV}$ and slope $L\approx60\pm30\,\rm{MeV}$ of symmetry energy at $\rho_0$\,\cite{LiBA13}, the~incompressibility $K_0\approx240\pm40\,\rm{MeV}$\,\cite{Garg18,You99,Shl06,Che12,Col14}, skewness $J_0\approx-300\pm200\,\rm{MeV}$\,\cite{Cai17x}, and~kurtosis $I_0\approx0\pm2000\,\rm{MeV}$ of SNM were adopted in~\cite{CL21}. The~$\Ts$ parameter in Set I was found to be about $\Ts\approx 1.67\pm0.56$, while that in Set II (Set III) was found to be about $\Ts\approx 1.41\pm0.88$ ($1.74\pm0.81$) {using the condition of Equation~(\ref{elow})}.

It is clearly seen from Figure~\ref{fig_Esymu-set-ABC} that below about $1.5\rho_0$, the auxiliary-function-based (purple) and the conventional expansions (blue) give almost identical results. However, at~higher densities, changing from Test Set I to Set III, the~result from the auxiliary-function-based approach is stable and always has smaller error bars compared to that from the conventional expansion.
Moreover, the~higher order contributions from $\nu_{\rm{sym}}^3$ are relatively small in the auxiliary-function-based reconstruction, indicating its fast convergence, by~comparing Panel (c) with Panel (a) or Panel (b). As was pointed out in~\cite{CL21}, this is because the function $\nu_{\rm{sym}}$ itself generates higher order terms in $\chi$, e.g.,~$\chi^3$, $\chi^4$, etc., even when the symmetry energy is truncated apparently at order $\nu_{\rm{sym}}^2$.
Consequently, the~reconstructed symmetry energy at suprasaturation densities from the auxiliary-function-based approach either to order $\nu_{\rm{sym}}^2$ or to order $\nu_{\rm{sym}}^3$ looks very~similar.

\begin{figure}[H]
%\widefigure
\includegraphics[width=.57\linewidth,angle=0]{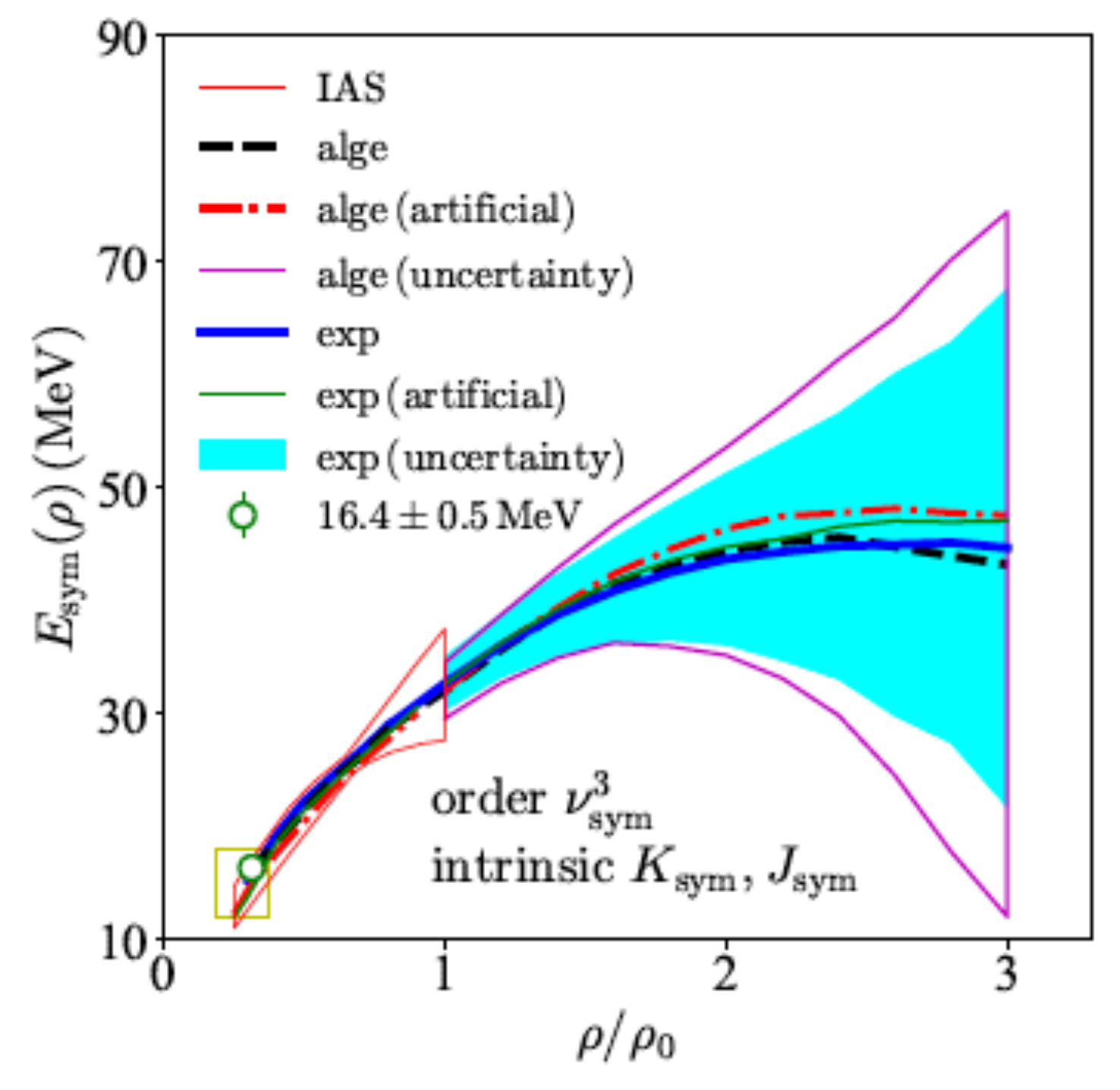}
\caption{Symmetry energy reconstructed via the auxiliary functions in the exponential and algebraic models with Test Set III (adopting the intrinsic correlation between $K_{\rm{sym}}$ and $J_{\rm{sym}}$ imposed by the unbound nature of PNM). Taken from~\cite{CL21}.}
%MDPI: Please make sure that permission has been obtained and there is no copyright issue.
\label{fig_Esym_exp_alge}
\end{figure}

Shown in Figure~\ref{fig_Esym_exp_alge} are the reconstructed
symmetry energy from $0.3\rho_0$ to $3\rho_0$ with the $1\sigma$ uncertainty band using the exponential and algebraic auxiliary function, respectively, adopting Parameter Set III~\cite{CL21}.
The $\Ts$ parameter for the algebraic model was found to be about $\Ts\approx1.91\pm1.80$. Interestingly, the~$E_{\rm{sym}}(\rho)$ obtained from the two models (blue solid and black dashed) behaved very similarly, indicating the approximate independence of the auxiliary function used. Quantitatively, the~symmetry energy at $2\rho$ was found to be $E_{\rm{sym}}^{\rm{exp}}(2\rho_0)\approx44.8\pm8.1\,\rm{MeV}$ and $E_{\rm{sym}}^{\rm{alge}}(2\rho_0)\approx 46.4\pm9.1\,\rm{MeV}$ with the two different auxiliary functions. They were in good agreement with the fiducial value of $E_{\rm{sym}}(2\rho_0)\approx 51\pm 13$ MeV at a 68\% confidence level from the nine different analyses of neutron star observables summarized in Figure~\ref{Esym2} in Section~\ref{ASTR}.

% start a new page without indent 4.6cm
%\clearpage
%\end{paracol}
\nointerlineskip
%\begin{paracol}{2}
%\linenumbers
%\switchcolumn

In short, knowing the characteristics of nuclear symmetry energy at $\rho_0$, regardless of how they were obtained, they can be used in properly chosen auxiliary functions to accurately predict the symmetry energy at suprasaturation densities up to about $3\rho_0$ before the hadron--quark phase transition happens. 
Similar approaches can be developed to predict the SNM EOS or generally the EOS of isospin asymmetric nuclear matter at high densities using their characteristics at $\rho_0$~\cite{CL21}.

\section{Summary and~Outlook}
The detection of the GW170817 event marked the beginning of gravitational wave astronomy. It stimulated many interesting new studies on the EOS of
dense neutron-rich matter. Together with observations of neutron stars using other messengers, as well as constraints from terrestrial nuclear experiments and predictions of nuclear theories, the~tidal deformation of neutron stars from GW170817 has provided more information about the EOS of neutron star matter within various model frameworks. The~nuclear symmetry energy encodes information about the energy cost to make nuclear matter more neutron rich, thus determining the content of neutron star matter. Many analyses of neutron star observables use models that directly construct/model/parameterize the pressure as a function of energy density above $\rho_0$ without considering explicitly the isospin degree of freedom, thus missing information about the symmetry energy at suprasaturation densities. To~extract the latter, one has to model the EOS starting at the level of single-nucleon energy in neutron-rich matter with the explicit isospin degree of freedom. Indeed, many analyses have used such models. These analyses have extracted in various ways the characteristics of nuclear symmetry energy, especially its slope L and, to some extent, curvature $K_{\rm{sym}}$ at $\rho_0$. Very few studies have tried to constrain the entire function of symmetry energy at high densities. Of~course, there is the longstanding and interesting issue: At what density does the hadron--quark phase transition happen in neutron stars? An answer to this question is relevant for the study of high-density nuclear symmetry energy, as the latter will lose its original physical meaning once the phase transition happens. In~turn, the~high-density behavior of nuclear symmetry energy may affect the properties of the hadron--quark phase transition. Thus, ideally, the high-density symmetry energy should be extracted from neutron star observables using models encapsulating the hadron--quark phase transition and the explicit isospin degree of freedom for the hadronic~phase.

\textls[-15]{In this brief review, within~our limited knowledge range, we summarized what the community has extracted about the characteristics of nuclear symmetry energy at $\rho_0$ and some new understandings about its high-density behavior from analyzing neutron star observables since the discovery of GW170817. More specifically, corresponding to the questions listed in the introduction, in~our possibly biased opinion, we learned the following:}
%\newpage
\begin{enumerate}
\item \textls[-11]{{What have we learned about the symmetry energy from the tidal deformation of canonical neutron stars from GW170817, the mass of PSR J0740+6620, and NICER's simultaneous observation of the mass and radius of PSR J0030+0451 and PSR J0740+6620?}}
%MDPI: Is the bold necessary? NOT NECESSARY, but better to have it to emphasize the correspondence with the questions we listed in the introduction. 
\begin{itemize}
\item
The average value of the slope parameter L of nuclear symmetry energy from 24 new analyses of neutron star observables was about $L\approx 57.7\pm 19$ MeV at a 68\% confidence level, while the average value of the curvature $K_{\rm{sym}}$ from 16 new analyses was about $K_{\rm{sym}}\approx -107\pm 88$ MeV, and the magnitude of nuclear symmetry energy at $2\rho_0$ was found to be $E_{\rm{sym}}(2\rho_0)\approx 51\pm 13$ MeV from nine new~analyses;

\item While the available data from canonical neutron stars did not provide tight constraints on nuclear symmetry energy at densities above about $2\rho_0$, the~lower radius boundary $R_{2.01}=12.2$ km from NICER's very recent observation of PSR J0740+6620 having a mass of $2.08\pm 0.07$~$M_{\odot}$ and radius $R$ = 12.2--16.3~km at a 68\% confidence level set a tighter lower limit for nuclear symmetry energy at densities above $2\rho_0$ compared to what we knew before from analyzing \mbox{earlier data};
\end{itemize}

\item {How do the symmetry energy parameters extracted from recent observations of neutron stars compare with what we knew before the discovery of GW170817?}
Before GW170817, there were surveys of symmetry energy parameters based on over \mbox{50 analyses} of various terrestrial nuclear experiments and some astrophysical observations of neutron stars. The~newly extracted average value of L was in good agreement with the earlier fiducial value within the error bars. There was little information about the curvature $K_{\rm{sym}}$ and $E_{\rm{sym}}(2\rho_0)$ before GW170817. The~latter two quantities characterizing the symmetry energy from $\rho_0 \sim 2\rho_0$ were mostly from analyzing the new data of neutron star~observations;

\item {What can we learn about the high-density symmetry energy from future, more precise radius measurement of massive neutron stars? }
Using characteristically different mock mass--radius data up to 2~$M_{\odot}$ within Bayesian analyses, it was found that the radius of massive neutron stars can constrain more tightly the lower boundary of high-density symmetry energy without much influence of the remaining uncertainties of SNM EOS. Indeed, as~mentioned above, NICER's very recent observation of PSR J0740+6620 made this real. Moreover, the~radii of massive neutron stars may help identify twin stars, the size of quark cores, and the nature of the hadron--quark phase transition~\cite{David,Sophia,Jerome};

\item {What are the effects of hadron--quark phase transition on extracting the symmetry energy from neutron star observables? How does the symmetry energy affect the fraction and size of quark cores in hybrid stars?}
Bayesian inferences of nuclear symmetry energy using models encapsulating a first-order hadron--quark phase transition from observables of canonical neutron stars indicated that the phase transition shifts appreciably both L and $K_{\rm{sym}}$ to higher values, but with larger uncertainties compared to analyses assuming no such phase transition.
It was also found that the available astrophysical data prefer the formation of a large volume of quark matter even in canonical NSs. The~correlations among the symmetry energy parameters and the hadron--quark phase transition density, as well as the quark matter fraction were found to be weak. Moreover, the~symmetry energy parameters extracted with or without considering the hadron--quark phase transition in neutron stars were all consistent with their known constraints within the still relatively large uncertainties. Thus, more precise constraints on the high-density symmetry energy are~needed;

\item {What are the effects of symmetry energy on the nature of GW190814's second component of mass (2.50--2.67) $M_{\odot}$?}
The high-density behavior of nuclear symmetry energy significantly affects the minimum rotational frequency of GW190814's secondary component of mass (2.50--2.67) $M_{\odot}$ as a superfast pulsar.
It also affects the r-mode stability boundary of GW190814's secondary in the frequency--temperature plane. Moreover, its equatorial radius and Kepler frequency also depend strongly on the high-density behavior of nuclear symmetry~energy;

\item \textls[-5]{{If all the characteristics of nuclear symmetry energy at saturation density $\rho_0$, e.g.,~its slope L, curvature $K_{\rm{sym}}$, and skewness $J_{\rm{sym}}$, are precisely determined by the astrophysical observations and/or terrestrial experiments, how do we use them to predict the symmetry energy at suprasaturation densities, such as $2-3 \rho_0$?} It was found very recently that by expanding $E_{\rm{sym}}(\rho)$ in terms of a properly chosen auxiliary function $\Pi_{\rm{sym}}(\chi,\Theta_{\rm{sym}})$ with a parameter $\Theta_{\rm{sym}}$ fixed accurately by an experimental $E_{\rm{sym}}(\rho_{\rm{r}})$ value at a reference density $\rho_{\rm{r}}$, the~shortcomings of the conventional $\chi$-expansion can be completely removed or significantly reduced in determining the high-density behavior of $E_{\rm{sym}}(\rho)$.}
\end{enumerate}

Thus, thanks to the historical GW170817 event and the hard work of many people in both astrophysics and nuclear physics, some interesting new knowledge about nuclear symmetry energy besides many other fundamental physics have been obtained from analyzing neutron star observables.
Many interesting issues especially about the high-density behavior of nuclear symmetry energy remain to be resolved. More precise radius measurements of massive neutron stars are particularly useful for addressing these issues. Comprehensive analyses of combined multi-messengers from various astrophysical observatories, laboratory experiments, and nuclear theories
will hopefully help us soon realize the ultimate goal of determining precisely the EOS of super-dense neutron-rich~matter.

	%%%%%%%%%%%%%%%%%%%%%%%%%%%%%%%%%%%%%%%%%%
	\vspace{6pt}
	
	%%%%%%%%%%%%%%%%%%%%%%%%%%%%%%%%%%%%%%%%%%
	\authorcontributions{Conceptualization, B.-A.L., B.-J.C., W.-J.X. and N.-B.Z.; methodology, B.-A.L.,
B.-J.C., W.-J.X. and N.-B.Z.; software, B.-A.L., B.-J.C., W.-J.X. and N.-B.Z.; validation, B.-A.L., B.-
J.C.,W.-J.X. and N.-B.Z.; formal analysis, B.-A.L., B.-J.C.,W.-J.X. and N.-B.Z.; investigation, B.-A.L.,
B.-J.C., W.-J.X. and N.-B.Z.; resources, B.-A.L., B.-J.C., W.-J.X. and N.-B.Z.; data curation, B.-A.L.,
B.-J.C.,W.-J.X. and N.-B.Z.; writing?original draft preparation, B.-A.L., B.-J.C.,W.-J.X. and N.-B.Z.;
writing?review and editing, B.-A.L.,W.-J.X. and N.-B.Z.; visualization, B.-A.L., B.-J.C.,W.-J.X. and
N.-B.Z.; supervision, B.-A.L.; project administration, B.-A.L.; funding acquisition, B.-A.L., B.-J.C.,
W.-J.X. and N.-B.Z. All authors have read and agreed to the published version of the manuscript.} %please complete
	%MDPI: For research articles with several authors, the following statements should be used “Conceptualization, X.X. and Y.Y.; methodology, X.X.; software, X.X.; validation, X.X., Y.Y. and Z.Z.; formal analysis, X.X.; investigation, X.X.; resources, X.X.; data curation, X.X.; writing—original draft preparation, X.X.; writing—review and editing, X.X.; visualization, X.X.; supervision, X.X.; project administration, X.X.; funding acquisition, Y.Y. All authors have read and agreed to the published version of the manuscript.”   THIS IS A REVIEW ARTICLE. It seems unnecessary to have the above statements. 

	%%%%%%%%%%%%%%%%%%%%%%%%%%%%%%%%%%%%%%%%%%
	
	\funding{This work was supported in part by the U.S. Department of Energy, Office of Science, under~Award Number DE-SC0013702, the~CUSTIPEN (China-U.S. Theory Institute for Physics with Exotic Nuclei) under the U.S. Department of Energy Grant No. DE-SC0009971, the~Yuncheng University Research Project under Grant No. YQ-2017005, the~National Natural Science Foundation of China under Grant No. 11505150 and No. 12005118, the~Scientific and Technological Innovation Programs of Higher Education Institutions in Shanxi under Grant No. 2020L0550, and~the Shandong Provincial Natural Science Foundation under Grant No. ZR2020QA085.}
	%please complete the Funding section
	
	%%%%%%%%%%%%%%%%%%%%%%%%%%%%%%%%%%%%%%%%%%
	
%
%\institutionalreview {Not applicable.}
%MDPI: Please add “The study was conducted according to the guidelines of the Declaration of Helsinki and approved by the Institutional Review Board (or Ethics Committee) of NAME OF INSTITUTE (protocol code XXX and date of approval).” OR “Ethical review and approval were waived for this study due to REASON (please provide a detailed justification).” OR “Not applicable” for studies not involving human or animals.

%
%\informedconsent {Not applicable.}
%MDPI: Please add “Informed consent was obtained from all subjects involved in the study.” OR “Patient consent was waived due to REASON (please provide a detailed justification).” OR “Not applicable” for studies not involving human.

%
%\dataavailability {The data presented in this study are available on request from the corresponding author.}
%MDPI: Please refer to suggested Data Availability Statement in section “MDPI Research Data Policies” at https://www.mdpi.com/ethics.
	
	\acknowledgments{We would like to thank Lie-Wen Chen, Wei-Zhou Jiang, Ang Li, De-Hua Wen, Jun Xu, and Xia Zhou for collaborations and discussion on some of the issues reviewed~here.}
	\conflictsofinterest{The authors declare no conflict of~interest.}%please complete it. Declare conflicts of interest or state ``The authors declare no conflict of~interest.''
	
%\newpage	

%\end{paracol}
%\newpage
\reftitle{References}

\end{document}